\documentclass [11pt,a4paper]{article}
\pdfoutput=1
\usepackage{jcappub}

\usepackage{graphicx,rotating,slashed,amsmath,xcolor,amssymb,amsfonts,colortbl,subfigure,float}
\usepackage{upgreek}
\usepackage{latexsym}     
\usepackage{graphics}
\usepackage{graphicx} %Loading the package
\usepackage[latin1,utf8]{inputenc}
\usepackage[OT2,OT1]{fontenc}

\newcommand{\nn}{\nonumber}

\newcommand{\iBox}{\Box^{-1}}

\renewcommand\({\left(}
\renewcommand\){\right)}
\renewcommand\[{\left[}
\renewcommand\]{\right]}

\newcommand{\ra}{\rightarrow}

\def\lsim{\raise 0.4ex\hbox{$<$}\kern -0.8em\lower 0.62
ex\hbox{$\sim$}}

\def\gsim{\raise 0.4ex\hbox{$>$}\kern -0.7em\lower 0.62
ex\hbox{$\sim$}}

\def\lbar{{\hbox{$\lambda$}\kern -0.7em\raise 0.6ex
\hbox{$-$}}}

\newcommand\eq[1]{eq.~(\ref{#1})}
\newcommand\eqs[2]{eqs.~(\ref{#1}) and (\ref{#2})}

\newcommand\eqst[2]{eqs.~(\ref{#1})--(\ref{#2})}

\newcommand\pa{\partial}
\newcommand\p{\partial}

\newcommand\ee{\end{equation}}
\newcommand\be{\begin{equation}}
\def\bea{\begin{array}}
\def\eea{\end{array}}\def\ea{\end{array}}
\newcommand\ees{\end{eqnarray}}
\newcommand\bees{\begin{eqnarray}}
\def\nn{\nonumber}

\def\a{\alpha}

\def\g{\gamma}

\def\dslash{\hspace{-1mm}\not{\hbox{\kern-2pt $\partial$}}}
\def\Dslash{\not{\hbox{\kern-2pt $D$}}}
\def\pslash{\not{\hbox{\kern-2.1pt $p$}}}
\def\kslash{\not{\hbox{\kern-2.3pt $k$}}}
\def\qslash{\not{\hbox{\kern-2.3pt $q$}}}

\newcommand{\vk}{{\bf k}}

\def\p1{{\bf p}_1}
\def\p2{{\bf p}_2}
\def\k1{{\bf k}_1}
\def\k2{{\bf k}_2}

\newcommand{\gmn}{g_{\mu\nu}}

\newcommand{\gMN}{g^{\mu\nu}}

\newcommand{\hmn}{h_{\mu\nu}}

\newcommand{\pam}{\pa_{\mu}}

\newcommand{\pan}{\pa_{\nu}}

\newcommand{\Gmn}{G_{\mu\nu}}

\newcommand{\Tmn}{T_{\mu\nu}}
\newcommand{\Smn}{S_{\mu\nu}}

\newcommand{\dddM}{\kern 0.2em \raise 1.9ex\hbox{$...$}\kern -1.0em \hbox{$M$}}
\newcommand{\dddQ}{\kern 0.2em \raise 1.9ex\hbox{$...$}\kern -1.0em \hbox{$Q$}}
\newcommand{\dddI}{\kern 0.2em \raise 1.9ex\hbox{$...$}\kern -1.0em\hbox{$I$}}
\newcommand{\dddJ}{\kern 0.2em \raise 1.9ex\hbox{$...$}\kern-1.0em
\hbox{$J$}}
\newcommand{\dddcalJ}{\kern 0.2em \raise 1.9ex\hbox{$...$}\kern-1.0em
\hbox{${\cal J}$}}

\newcommand{\dddO}{\kern 0.2em \raise 1.9ex\hbox{$...$}\kern -1.0em
\hbox{${\cal O}$}}
\def\dddz{\raise 1.5ex\hbox{$...$}\kern -0.8em \hbox{$z$}}
\def\dddd{\raise 1.8ex\hbox{$...$}\kern -0.8em \hbox{$d$}}
\def\dddbd{\raise 1.8ex\hbox{$...$}\kern -0.8em \hbox{${\bf d}$}}
\def\ddbd{\raise 1.8ex\hbox{$..$}\kern -0.8em \hbox{${\bf d}$}}
\def\dddx{\raise 1.6ex\hbox{$...$}\kern -0.8em \hbox{$x$}}

\newcommand{\gt}{\g_{\rm t}}

\newcommand{\mpl}{M_{\rm Pl}}
\newcommand{\mplr}{m_{\rm Pl}}

\newcommand{\ode}{\Omega_{\rm DE}}
\newcommand{\oma}{\Omega_{M}}
\newcommand{\ora}{\Omega_{R}}

\newcommand{\ola}{\Omega_{\Lambda}}

\newcommand{\rde}{\rho_{\rm DE}}
\newcommand{\wde}{w_{\rm DE}}

\def\be{\begin{equation}}
\def\ee{\end{equation}}
\def\bea{\begin{eqnarray}}
\def\eea{\end{eqnarray}}

\def\dd{{\rm d}}

\newcommand{\bpm}{\begin{pmatrix}}
\newcommand{\epm}{\end{pmatrix}}
\newcommand{\lpar}{\left(}
\newcommand{\rpar}{\right)}
\newcommand{\lbra}{\left[}
\newcommand{\rbra}{\right]}
\newcommand{\Hc}{\mathcal{H}}%Conformal Hubble rate
\newcommand{\nM}{\hat{\nu}} %friction matrix
\newcommand{\cM}{\hat{C}} %friction matrix
\newcommand{\mM}{\hat{M}} %mass matrix
\newcommand{\eM}{\hat{E}} %eigenvectors matrix
\newcommand{\gM}{\hat{G}} %exponential phase matrix
\newcommand{\tM}{\hat{\theta}} %exponential phase matrix
\newcommand{\iM}{\hat{I}} %identity matrix

\newcommand{\bphi}{\bar{\Phi}}
\newcommand\tg{\theta_g}%mixing angle

\def\dh{d_{\rm H}}

\def\p{\partial}

\def\cM{\mathcal{M}}

\def\ds{d_{\rm S}}

\def\bs{\begin{subequations}}
\def\es{\end{subequations}}
\def\a{\alpha}

\def\g{\gamma}

\def\k{\kappa}

\def\ve{\varepsilon}

\def\cM{\mathcal{M}}

\def\ds{d_{\rm S}}
\def\dh{d_{\rm H}}

\def\p{\partial}

\newcommand{\Eq}[1]{(\ref{#1})}

\def\Pl{{\rm Pl}}
\def\lp{\ell_\Pl}

\def\rme{e}

\def\leq{\leqslant}

\numberwithin{equation}{section}
\definecolor{verde}{rgb}{0,0.5,0}

\def\gt{\color{black}}

\definecolor{MyBlue}{rgb}{0.15,0.15,0.70}
\hypersetup{
colorlinks=true,
citecolor=MyBlue,
linkcolor=MyBlue,
urlcolor=MyBlue
}

\input epsf
\usepackage{hyperref}

\begin{document}

\begin{figure}
\hskip13.cm \href{https://lisa.pages.in2p3.fr/consortium-userguide/wg_cosmo.html}{\includegraphics[width = 0.19 \textwidth]{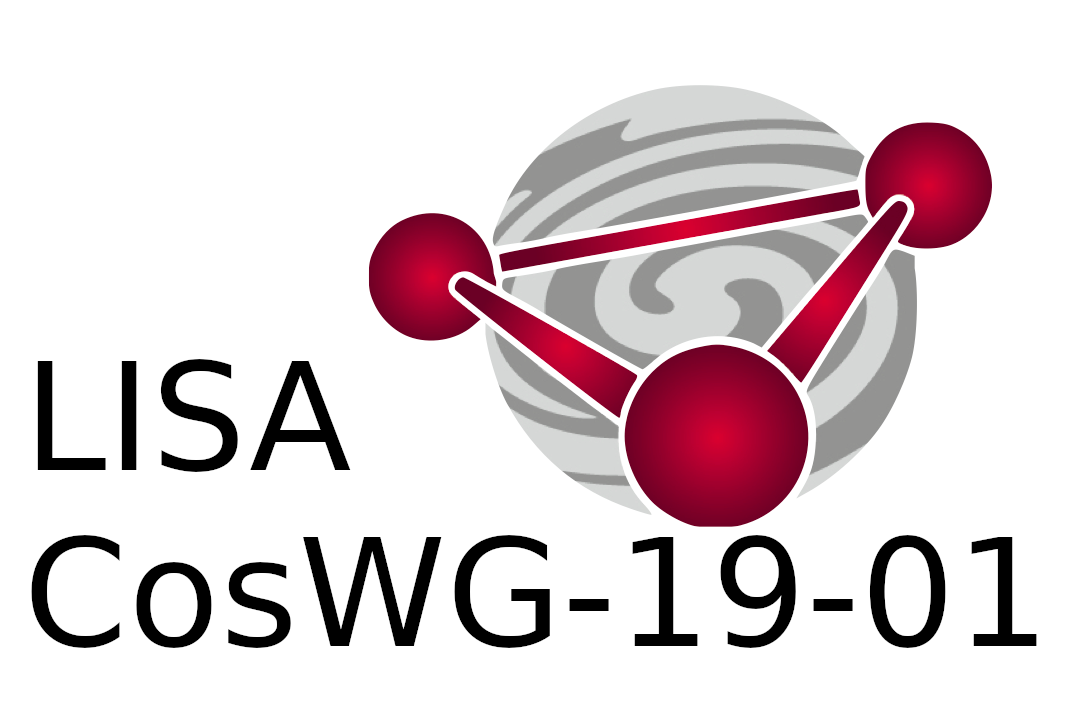}}
\end{figure}
\title{Testing modified gravity at cosmological distances with LISA standard sirens}

\author{\large Enis Belgacem$^a$~\footnote{Corresponding author. E-mail: Enis.Belgacem@unige.ch}, Gianluca Calcagni$^b$, Marco Crisostomi$^{c,d,e}$, Charles Dalang$^{a}$, Yves Dirian$^{a,f}$,  Jose Mar\'ia Ezquiaga$^g$,  Matteo Fasiello$^h$, Stefano Foffa$^a$, Alexander Ganz$^i$, Juan Garc\'ia-Bellido$^g$, Lucas Lombriser$^a$, Michele Maggiore$^a$, Nicola Tamanini$^j$, Gianmassimo Tasinato$^k$~\footnote{Project coordinator. E-mail: g.tasinato2208@gmail.com\\ \vskip -0.25cm% 
 \hskip 0.22cm 
 The authors' affiliations can be found at the end of the article. }, Miguel Zumalac\'arregui$^{l,c}$, Enrico Barausse$^{m,n}$, Nicola Bartolo$^{i,o}$, Daniele Bertacca$^i$, Antoine Klein$^{m,r}$, Sabino Matarrese$^{i,o,p}$, Mairi Sakellariadou$^q$}

\abstract{Modifications of General Relativity leave their imprint both on the cosmic expansion history through a non-trivial dark energy equation of state, and on the evolution  of cosmological perturbations in the  scalar and in the tensor sectors. In particular, the modification in the  tensor sector gives rise to a notion of  gravitational-wave (GW) luminosity distance, different from the standard electromagnetic luminosity distance, that can be studied with standard sirens at  GW detectors such as LISA or third-generation ground based experiments. We discuss the predictions for modified GW  propagation from some of the  best  studied theories of modified gravity, such as Horndeski or the more general   degenerate higher order scalar-tensor (DHOST) theories, non-local infrared modifications of gravity,  bigravity theories  and the corresponding phenomenon of GW oscillation, as well as theories with extra or varying dimensions. We show that modified GW  propagation is a completely generic phenomenon in modified gravity. We then use a simple parametrization of the effect in terms of two parameters $(\Xi_0,n)$, that is shown to fit well the results from a large class of models, to study the prospects of observing modified GW propagation using supermassive black hole binaries as standard sirens with LISA. We construct  mock source catalogs and perform detailed   Markov Chain Monte Carlo studies of the likelihood obtained from LISA standard sirens alone, as well as by combining them with CMB, BAO and SNe data to reduce the degeneracies between cosmological parameters. We find that the combination of LISA with the other cosmological datasets allows one to measure the parameter $\Xi_0$ that characterizes modified GW  propagation to  the percent level accuracy,  sufficient to test several modified gravity theories.  LISA standard sirens can also  improve constraints on GW oscillations  induced by extra field content by about three orders of magnitude relative to the current capability of ground detectors. We also update the forecasts on the accuracy on $H_0$ and on the dark-energy equation of state using more recent estimates for the LISA sensitivity.}

\maketitle

\section{Introduction}
\label{sec:general}

The physics of gravitational waves (GWs) provides important opportunities to improve our understanding of gravitational interactions, and
to test theories of gravity alternative to General Relativity (GR). 
For example, the recent simultaneous detection of a gravitational  and an
electromagnetic signal from GW170817 \cite{TheLIGOScientific:2017qsa,Goldstein:2017mmi,Savchenko:2017ffs,Monitor:2017mdv}, and the follow-up studies of the electromagnetic counterpart 
\cite{GBM:2017lvd} impose  severe constraints on theories predicting a different speed for gravitational and electromagnetic waves~\cite{Lombriser:2015sxa,Creminelli:2017sry,Sakstein:2017xjx,Ezquiaga:2017ekz,Baker:2017hug}. 

The next generation of GW experiments, and in particular the LISA mission~\cite{Audley:2017drz}, has the potential of performing stringent tests of other aspects of modified gravity theories, by studying the propagation of GWs across cosmological distances. In GR,
the linearised evolution equation for GWs traveling on an FRW background in four dimensional space-time 
is 
\be \label{gen-ev-GR}
 h''_A+ 2{ \cal H}\, h'_A+k^2h_A\,=\,\Pi_A\,,
\ee 
where the primes indicate 
derivatives  with respect to conformal time $\eta$, related to the physical time through the usual relation  
$d \eta=d t/a(t)$, with $a(t)$ the scale
factor, $A=+,\times$ labels the two polarizations, and $\Pi_A$ is the source term, related to the anisotropic stress tensor. In a generic modified gravity model the above equation is modified into
\be \label{gen-ev-eq1}
 h''_A+ 2\left[1-\delta(\eta) \right]{\cal H}\, h'_A+\[ c_T^2(\eta)\,{k^2}+m_T^2(\eta) \] h_A\,=\,\Pi_A
\,.
\ee 
More generally,  
$\delta(\eta)$ could be a function also of wavenumber $k$, and the expression $(c_T^2(\eta)\,{k^2}+m_T^2(\eta) )$ could be replaced by an expression with a more complicated $k$ dependence, corresponding to a non-trivial dispersion relation (parity-violating theories could also a priori introduce a dependence of
these quantities on the polarization index $A$).
Each of the quantities $\delta(\eta), c_T(\eta)$ and $m_T(\eta)$ appearing in the equation \eqref{gen-ev-eq1}  can in principle be tested with GW observations. With respect to GR, the GW evolution equation \eqref{gen-ev-eq1} contains several new ingredients:
\begin{enumerate}
\item The cosmological homogeneous background  controlling the Hubble parameter ${\cal H}$ is in general  distinct from
$\Lambda$CDM, given the different  background solutions in modified gravity. This, as we shall see,  {must}   be taken into
due  account when investigating
 the dynamics of  gravitational waves propagating through cosmological distances. 
\item The function $\delta(\eta)$ modifies the friction term in the propagation equation. As we will recall below, this affects the amplitude of a GW propagating across cosmological distances,
giving rise to a notion of ``gravitational-wave luminosity distance"~\cite{Deffayet:2007kf,Saltas:2014dha,Lombriser:2015sxa,Nishizawa:2017nef,Belgacem:2017ihm,Belgacem:2018lbp}.\footnote{Notice that  $\delta$ is also indicated with other names
in the literature, such as
$\nu$ and $\alpha_M$, related to $\delta$  by 
 $\nu\,=\,\alpha_M\,=\,-2 \delta$. See for
example \cite{Gleyzes:2014rba}.} 
In several modified gravity theories the function $\delta(\eta)$ is related to the time dependence of the effective 
Planck mass $M_{\rm eff}(\eta)$ (that could arise because of  non-minimal coupling of tensor modes with other fields),  via the relation 
\be \label{deltaMstar}
\delta(\eta)=-\frac{d\ln M_{\rm eff}}{d\ln a}\,.
\ee
In GR, $\delta=0$. Notice however that (as discussed in \cite{Belgacem:2018lbp} and as we will recall below), the relation (\ref{deltaMstar}) is not universally valid in modified gravity theories, and is not obeyed in some interesting models; thus, one should not {\it a priori} identify  a non-vanishing $\delta(\eta)$ with a time-dependent  effective Planck mass.

\item The tensor velocity $c_T$ can be in general time (and scale) dependent. In GR, $c_T=c$, with $c$
the light speed (that we have set to one). 
\item While in GR the tensor modes are massless, in theories of modified gravity the tensor mode can be massive, with $m_T$  its mass. This can occur in theories such as massive gravity or bigravity. 
\item  In the presence of anisotropic stress, or in theories where tensors couple with additional fields already at linearised level (as in theories
breaking spatial diffeomorphisms), the tensor evolution equation contains
 a ``source term'' $\Pi_A$ in the right hand side of eq.~\eqref{gen-ev-eq1}.
 In  absence of anisotropic stress, and in cosmological scenarios where spatial diffeomorphisms are preserved,   we have $\Pi_A=0$.
\end{enumerate}
The physical consequences of these parameters have been discussed at length in the literature (see \cite{Ezquiaga:2018btd} for a review on their implications for GW astronomy). 
In this paper we   investigate how they affect a specific observable, the  GW luminosity distance, which can be probed
by LISA standard sirens. 

The space-based interferometer  LISA can qualitatively and quantitatively  improve our tests on the 
propagation of gravitational waves in theories of  modified
gravity. 
     LISA can probe signals from standard sirens of supermassive black hole mergers (MBHs) at  redshifts $z\sim {\cal O}(1-10)$, much larger than the redshifts $z\sim {\cal O}(10^{-1})$ of typical sources detectable from second-generation
ground-based interferometers. This implies that LISA can  test the possible time dependence of the
parameters controlling deviations from GR or the standard $\Lambda$CDM model, since GWs travel large cosmological distances before reaching the observer. Moreover, as we will review in section~\ref{sect:catalogs}, LISA can measure the luminosity distance to 
MBHs with remarkable precision, thereby reaching an accuracy not possible for second-generation ground-based detectors.
 
It is also interesting to observe that  LISA can probe GWs in the frequency range in the milli-Hz regime (more precisely, in the interval $10^{-4}-10^{0}$ Hz),  much smaller than the typical frequency interval of ground-based detectors, $10^{1}-10^{3}$ Hz. This is a theoretically interesting range to explore since several theories of modified gravity designed to explain dark energy, such as Horndeski, {\gt degenerate higher order scalar-tensor (DHOST) theories} or massive gravity,
have a  low UV cutoff, typically of order 
%\be \label{cutoff1}
%\Lambda_{\text{cutoff}}\sim \left( H_0^2\,M_{\text{Pl}}\right)^{1/3}\,\sim\,10^2\,{\text{Hz}}\,.
%\ee
$\Lambda_{\text{cutoff}}\sim \left( H_0^2\,M_{\text{Pl}}\right)^{1/3}\,\sim\,10^2\,{\text{Hz}}$.
This cutoff is  within  the frequency regime probed by LIGO, making a comparison
between modified gravity predictions and GW observations delicate  
 \cite{deRham:2018red}. The frequency range  tested by LISA, instead, is well below this
 cutoff, hence it lies within the range of validity of the theories under consideration.   

The paper is organized as follows. In section~\ref{sect:dLgw} we recall the notion of modified GW propagation and GW luminosity distance, that emerges generically in modified theories of gravity. In section~\ref{sect:models} we discuss the prediction on modified GW propagation of some of the best studied modified-gravity theories: scalar-tensor theories  (with particular emphasis on Horndeski and DHOST theories), infrared non-local modifications of gravity, bigravity, and theories with extra {\gt and varying} dimensions. We then turn to the study of the capability of LISA to detect modified GW propagation using the coalescence of supermassive black hole  binaries as standard sirens. In section~\ref{sect:catalogs} we discuss the construction of mock catalogs of events with LISA, and in section \ref{sect:MCMC} we present the results, obtained by running a series of Markov Chain Monte Carlo (MCMC). We present our conclusions in section~\ref{sect:conclusions}.

\section{The  gravitational-wave  luminosity distance}\label{sect:dLgw}

\subsection{Luminosity distance and standard sirens in  GR}

The standard luminosity distance associated to electromagnetic signals, $d_L$, is defined in terms
of the energy flux ${\cal F}$  measured in the observer frame, and of the intrinsic luminosity ${\cal L}$
measured at the source frame, as 
\be\label{fluxluminosity}
{\cal F}\,\equiv\,
 \frac{\cal L}{4 \pi d_L^2}\,.
\ee
From the propagation of electromagnetic signals over an (unperturbed)  FRW background, one  finds the standard expression for $d_L$ as a function of redshift,
\be\label{dLem}
d_L(z)=(1+z)\int_0^z\, 
\frac{d\tilde{z}}{H(\tilde{z})}\, ,
\ee
where
\be\label{E(z)}
H(z)=H_0\sqrt{\oma (1+z)^3+\ora (1+z)^4+\ode(z)}\, .
\ee
Here $\oma=\rho_M(t_0)/\rho_0 $ is the present matter density fraction
(where,  as usual, $\rho_0=3H_0^2/(8\pi G)$ is the critical energy density and $t_0$ the present value of cosmic time), $\ora=\rho_R(t_0)/\rho_0$ is the present {\gt radiation} density fraction, and
$\ode(z)=\rde(z)/\rho_0$, where
$\rde(z)$ is the dark energy (DE) density in the cosmological model under consideration. In particular, in 
$\Lambda$CDM, $\ode(z)=\ola$ is constant. We are assuming for simplicity spatial flatness, so
$\oma+\ora+\ode(z=0)=1$. The luminosity distance therefore encodes important information about the cosmological model, and is a prime cosmological observable, that can be measured in particular using type Ia supernovae.

In GR, the  amplitude  of GWs produced by a binary astrophysical system provides yet another measurement of $d_L(z)$. Indeed, introducing
a field $\chi_A(\eta, \vk)$ from
\be\label{4defhchiproofs}
h_A(\eta, \vk)=\frac{1}{a(\eta)}  \chi_A(\eta, \vk)\, ,
\ee
\eq{gen-ev-GR} becomes
\be\label{4propchiproofs1}
\chi''_A+\(k^2-\frac{a''}{a}\) \chi_A=0\, .
\ee
For modes well inside the horizon $a''/a$ is  negligible with respect to $k^2$,\footnote{In principle, the effect of the $a''/a$ term could be included use a WKB approximation, as in \cite{Nishizawa:2017nef}. However,
the relative size of the $a''/a$ and $k^2$  terms is of order of the  square of the 
wavelength $\lambda_{\rm GW}$ of the GW over the size of the horizon, $(\lambda_{\rm GW}/H_0^{-1})^2$,  and this correction is therefore not significant for LISA or for ground-based detectors.} and we get  a solution for $\chi_A$ of the form
\be\label{chiAsin}
\chi_A(\eta, \vk)\simeq {\cal A}_A\sin (k\eta+\varphi_A)\, ,
\ee
with ${\cal A}_A$ the amplitude and $\varphi_A$ a phase. This shows that, overall, $h_A$ scales as $1/a$ in the propagation over cosmological distances. For a coalescing binary, combining this factor with the standard behavior $1/r$ in the near region, as well as with a redshift-dependent factor that arises in  transforming frequencies and masses 
   from the rest-frame  to the observed one, we obtain the standard result $h_A\propto 1/d_L(z)$ (see e.g. section 4.1.4 of \cite{Maggiore:1900zz} for a detailed derivation). More precisely, in the so-called restricted 
post-Newtonian (PN) approximation, where one takes into account the PN corrections to the phase but not to the amplitude, one finds 
\bees
h_+(t)&=&\frac{2(1+\cos^2\iota)}{d_L(z)}\,   (G{\cal M}_c)^{5/3} [\pi f(t)]^{2/3} \cos\Phi(t)\, ,
\label{hplus}\\
h_{\times}(t)&=&\frac{4 \cos\iota}{d_L(z)}\, (G{\cal M}_c)^{5/3} [\pi f(t)]^{2/3} \, 
\sin\Phi(t)\, ,\label{hcross}
\ees
where $\Phi(t)$ is the phase, that in general needs to be computed to a high PN order,
${\cal M}_c=(1+z) (m_1m_2)^{3/5} (m_1+m_2)^{-1/5}$ is the redshifted chirp mass (i.e. the quantity actually observed in the detector frame), $f(t)$ is the observed GW frequency, that sweeps upward  in time, and $\iota$ is the inclination angle of the normal to the orbit with respect to the line of sight.

The chirp mass is accurately determined from the time evolution of $f(t)$ [e.g., to lowest order in the PN expansion, $\dot{f}=(96/5)\pi^{8/3}  (G{\cal M}_c)^{5/3} f^{11/3}$]. Then, 
as first observed  in \cite{Schutz:1986gp},   the amplitude  of GWs from a coalescing compact binary provides an absolute measurement of its luminosity distance and in this sense coalescing compact binaries are the GW analogue  of standard candles, or {\it standard sirens}. As we see from \eqs{hplus}{hcross}, the main uncertainty on the standard siren measurement of $d_L(z)$ comes from the partial degeneracy with  $\cos\iota$. This can be broken in particular if both polarizations can be measured, or if we have informations on the inclination angle, e.g. from the observation of an electromagnetic jet. Much work as been devoted in the literature to studying how standard sirens can be used for cosmology~\cite{Holz:2005df,Dalal:2006qt,MacLeod:2007jd,Nissanke:2009kt,Cutler:2009qv,Sathyaprakash:2009xt,Zhao:2010sz,DelPozzo:2011yh,Nishizawa:2011eq,Taylor:2012db,Camera:2013xfa,Cai:2016sby,Bertacca:2017vod}; see in particular   \cite{Tamanini:2016zlh,Caprini:2016qxs,Congedo:2018wfn} for recent work specifically related to LISA.

\subsection{GW luminosity distance in modified gravity theories}

\subsubsection{GW luminosity distance from modified friction term}

Let us now discuss  how the situation changes in modified gravity.
The expression for the luminosity distance that enters \eqs{hplus}{hcross} now depends on the  equation for
propagation of GWs  in the theory under consideration. In particular, as recognized in
\cite{Saltas:2014dha,Lombriser:2015sxa,Nishizawa:2017nef,Belgacem:2017ihm,Belgacem:2018lbp} (see also Sect.~19.6.3 of \cite{Maggiore:2018zz}), the function $\delta(\eta)$, that modifies the ``friction term''  in \eq{gen-ev-eq1} affects the luminosity distance extracted from the observation of GWs from a coalescing binary. Indeed, consider the equation 
\be\label{prophmodgrav}
h''_A  +2 {\cal H}[1-\delta(\eta)] h'_A+k^2h_A=0\, ,
\ee
where, for the moment, we have only  retained the deviations from GR  induced by the function $\delta(\eta)$ in \eq{gen-ev-eq1}.  In this case, to eliminate the friction term, we must  introduce $\chi_A(\eta, \vk)$ from 
\be\label{4defhchiproofsRR}
h_A(\eta, \vk)=\frac{1}{\tilde{a}(\eta)} \chi_A(\eta, \vk)\, ,
\ee
where 
\be\label{deftildea}
\frac{\tilde{a}'}{\tilde{a}}={\cal H}[1-\delta(\eta)]\, .
\ee
Then we get 
\be
\chi''_A+\(k^2-\frac{\tilde{a}''}{\tilde{a}}\) \chi_A=0\, .
\ee 
Once again, inside the horizon the term $\tilde{a}''/\tilde{a}$ is totally negligible, so   GWs propagate at the speed of light. However, now the amplitude of $\tilde{h}_A$ is proportional to $1/\tilde{a}$ rather than $1/a$. As a result, the GW amplitude measured by coalescing binaries is still given by  \eqs{hplus}{hcross}, where now the standard ``electromagnetic" luminosity distance $d_L(z)$ (that we will henceforth denote by $d_L^{\rm em}$) is replaced by a ``GW luminosity distance" $d_L^{\,\rm gw}$ such that
\be\label{dgwaadem}
d_L^{\,\rm gw}(z)=\frac{a(z)}{\tilde{a}(z)}\, d_L^{\,\rm em}(z)\, .
\ee
By integrating \eq{deftildea} (and choosing, without loss of generality,  the normalizations $\tilde{a}(t_0)=a(t_0)=1$, where $t_0$ is the present value of cosmic time),
this can be nicely rewritten in terms of $\delta(z)$ as~\cite{Belgacem:2017ihm,Belgacem:2018lbp}
\be\label{dLgwdLem}
d_L^{\,\rm gw}(z)=d_L^{\,\rm em}(z)\exp\left\{-\int_0^z \,\frac{dz'}{1+z'}\,\delta(z')\right\}\, .
\ee

\subsubsection{Relation to the time variation of the Planck mass}\label{sect:timevarMpl}

We next discuss the relation between the function $\delta(\eta)$ and the possibility of a time-varying Planck mass. In several explicit models (see section~\ref{sect:models}), it has been found that the two are related by \eq{deltaMstar}. Upon integration, using $d\ln a =-dz/(1+z)$, \eq{deltaMstar} gives
\be\label{MeffMPlintdelta}
\ln\frac{M_{\rm eff}(z)}{M_{\rm eff}(0)}=\int_0^z \,\frac{dz'}{1+z'}\,\delta(z')\, ,
\ee
Comparing with \eq{dLgwdLem} we see that
\be\label{ddmplMstar}
\frac{d_L^{\,\rm gw}(z)}{d_L^{\,\rm em}(z)}=\frac{M_{\rm eff}(0)}{M_{\rm eff}(z)}\, .
\ee
Equivalently, in terms of  
the effective Newton constant $G_{\rm eff}(z)=1/M^2_{{\rm eff}}(z)$, we have
\be\label{dLgwdLemGeff}
\frac{d_L^{\,\rm gw}(z)}{d_L^{\,\rm em}(z)}=\, \sqrt{\frac{G_{\rm eff}(z) }{G_{\rm eff}(0)}}\, .
\ee 
The origin of this relation has been discussed in \cite{Belgacem:2018lbp}, where it has been found that it expresses the conservation of the (comoving) number of gravitons during the propagation. The argument runs as follows. Recall 
first of all how the usual scaling of the GW energy density $\rho_{\rm gw}\propto a^{-4}$ is deduced from the cosmological evolution in GR (we follow Section~19.5.1 of \cite{Maggiore:2018zz}). From \eqs{4defhchiproofs}{chiAsin}  we learn that, for the modes well inside the horizon that we are considering, 
\be\label{tildehAvsa}
\tilde{h}_A(\eta, \vk)= \frac{1}{a(\eta)} {\cal A}_A\sin(k\eta+\varphi_A)\, .
\ee
Therefore
$h'_A(\eta,\vk)\simeq   {\cal A}_Ak\cos(k\eta+\varphi_A)/a(\eta)$,
since the extra term obtained when the time derivative acts on  $1/a(\eta)$ gives a term proportional to $a'/a^2$, which, for modes well inside the horizon, is negligible compared to $k/a$.
Using $\dot{h}_A=(1/a)h'_A$, where the dot is the derivative with respect to cosmic time, we get
\be\label{dothGR}
\dot{h}_A\simeq   \frac{{\cal A}_Ak\cos(k\eta+\varphi_A)}{a^2}\, , 
\ee
so $\dot{h}_A$ scales as $1/a^2$.
The energy density of the GWs is given by the usual expression
\be\label{rhogwGR}
\rho_{\rm gw}=\frac{1}{16\pi G}\sum_{A=+,\times} \langle \dot{h}_A^2\rangle\, ,
\ee
where the angular bracket denotes an average over several periods. Inserting here \eq{dothGR}, we find that the term $\cos^2(k\eta+\varphi_A)$, averaged over several periods, gives a factor $1/2$, and it then 
follows that $\rho_{\rm gw}\propto a^{-4}$. This  
expresses the fact that the physical number density of gravitons scales as $1/a^3$ (i.e. the comoving number density is conserved), and that the energy of each graviton redshifts as $1/a$, leading to the overall $1/a^4$ scaling.  

In a modified gravity theory in which the comoving number of gravitons remains conserved, the physical number density still scales as $1/a^3$. If, furthermore, the  graviton dispersion relation remains $E=k$,  its energy still scales as $1/a$, so we have again $\rho_{\rm gw}\propto a^{-4}$. However,
we have seen that $h_A$ now scales as $1/\tilde{a}$, with $\tilde{a}$ in general different from $a$ and,
repeating the argument in \eqs{tildehAvsa}{dothGR},  $\dot{h}_A$ scales as $1/(a\tilde{a})$ (since the extra factor of $a$ comes from the transformation between cosmic time and conformal time).
On the other hand, $G$ in \eq{rhogwGR} must be replaced by $G_{\rm eff}(t)$.  In order to get
$\rho_{\rm gw}\propto a^{-4}$, we must then have
\be
\frac{1}{G_{\rm eff}(z)}\, \frac{1}{ (a\tilde{a})^2}=\frac{1}{G_{\rm eff}(z=0)}\,\frac{1}{a^4}\, ,
\ee
which, upon use of \eq{dgwaadem}, is equivalent to \eq{dLgwdLemGeff}. Thus, \eq{dLgwdLemGeff} is valid in any modified gravity theory in which the graviton number is conserved and the graviton dispersion relation is not modified, and therefore in a very broad class of models. In particular, it can be shown~\cite{Belgacem:2018lbp} that this relation holds  in any 
modified gravity theory described by an action of the form
\be
S=\frac{1}{8 \pi G}\int d^4x\, \sqrt{-g}\,  A(\phi) R + \dots\,, \label{eq:AtimesR}
\ee
which is minimally coupled to matter, where $A$ can be a nontrivial functional of extra fields in the gravitational  sector, here denoted collectively as $\phi$, and the dots denote other possible gravitational interaction terms, that can depend on $\phi$ but do not contain terms  purely quadratic in the gravitational field $\hmn$ nor interactions with ordinary matter.

It is important to observe that the time variation of the effective Newton constant given in \eq{dLgwdLemGeff} [or, equivalently, the time variation of the Planck mass given by \eq{ddmplMstar}] can only hold on cosmological scales. At such scales, typical cosmological models predict a time dependence such that, today, $|\dot{G}/G|\, \simeq\, H_0$, with a coefficient in general of order one.
A scenario  in which this result  holds down to solar system or    Earth-Moon scales  would be ruled out by Lunar Laser Ranging experiments, that 
by now impose a bound $|\dot{G}/G|\, \lsim\, 10^{-3}H_0$~\cite{Hofmann:2018myc}. If a model predicts a non-trivial value for the ratio $d_L^{\,\rm gw}(z)/d_L^{\,\rm em}(z)$ then, to be observationally viable,
it must either have a screening mechanism at short scales, so that the time-dependent effective Newton constant predicted by \eq{dLgwdLemGeff} cannot be extrapolated down to the Earth-Moon scale 
(this is the case, for instance, for Hordenski and DHOST theories\footnote{Very recently it has been pointed out \cite{Tsujikawa:2019pih} that, for the special case of the cubic Galileon, the screening is not sufficient to guarantee an observable effect in the measurement of \eqref{dLgwdLemGeff} for the redshifts probed by LISA standard sirens. However, a more general form for the cubic Horndeski term, or the screening provided by DHOST theories, can still provide sizable effects of  \eqref{dLgwdLemGeff} within the reach of LISA as we will see in Section \ref{sub_dhost}.}, and bigravity) or else must violate the relation (\ref{dLgwdLemGeff}) and predict $G_{\rm eff}=G$ for all modes well inside the horizon, even 
if it still predicts a non-trivial result for $d_L^{\,\rm gw}(z)/d_L^{\,\rm em}(z)$. As we will see in section~\ref{Sec:nonlocal}, the latter option is realized in the so-called RT non-local gravity model.

Note also that the  relation between the effective Newton constant $G_{\rm eff}$ that governs the gravitational dynamics at cosmological scales and the Newton constant $G$ observed at laboratory or solar-system scales can be model-dependent. In the absence of screening mechanisms, the effective Newton constant at cosmological scales is the same as that at laboratory or solar-system scales, and then 
$G=G_{\rm eff}(z=0)$. In the presence of screening mechanisms, however, the relation can be more complex.

\subsubsection{GW luminosity distance from a time-dependent speed of GWs}

We now discuss how the above results for the luminosity distance  change if we also include a non-trivial function $c_T(\eta)$ in \eq{gen-ev-eq1}. In that case we still define $\chi$ from \eq{4defhchiproofsRR}, with $\tilde{a}$  given by \eq{deftildea}. Then the ``friction term'' in \eq{deftildea} is again eliminated, and $\chi_A$ now satisfies the equation
\be
\chi''_A+\(k^2c^2_T(\eta)-\frac{\tilde{a}''}{\tilde{a}}\) \chi_A=0\, .
\ee 
For modes inside the horizon (defined now by the condition $|a''/a| \ll k^2c^2_T$) we can neglect the term 
$\tilde{a}''/\tilde{a}$.  Assuming that the frequency at which $c_T(\eta)$ changes is much smaller than the frequency $f=k/(2\pi)$ of the GW [which is an extremely good approximation for GWs detectable with LISA or with ground based interferometers, given that the typical scale of change of $c_T(\eta)$ is expected to be fixed by $H(\eta)$], the resulting equation has the WKB-like  solution 
\be\label{chiAsinmod}
\chi_A(\eta, \vk)\simeq {\cal A}_A\, \sqrt{\frac{c_T(\eta_0)}{c_T(\eta)} }\, \sin \[k\int^{\eta} d\eta' c_T(\eta')+\varphi_A\]\, ,
\ee
where we  have normalized the amplitude ${\cal A}_A$ so that, for $c_T(\eta)$ constant and equal to the present value
$c_T(\eta_0)$, we recover \eq{chiAsin}.
Thus, a non-trivial speed of GWs, $c_T(\eta)$,  affects both the phase and the amplitude of the signal from a coalescing binary. Therefore it also affects the GW luminosity distance, so the GR relation $d_L(z)=(1+z)a(t_0)r_{\rm com}(z)$, where $r_{\rm com}$ is the comoving distance to the source, is now modified into
\be
d_L^{\rm gw}(z)=\sqrt{\frac{c_T(z)}{c_T(0)} }\frac{a(z)}{\tilde{a}(z)} \, (1+z) r_{\rm com}(z)\, ,
\ee
where $c_T(0)\equiv c_T(z=0)$ and we used $a(t_0)=\tilde{a}(t_0)=1$.
Furthermore, the expression of the comoving distance measured with GWs as a function of redshift  is also affected by a non-trivial speed of GWs, and
is now given by 
\be
r_{\rm com}(z)=\int_0^z\, d\tilde{z}\, \frac{c_T(\tilde{z}) }{H(\tilde{z})}\, .
\ee
Thus, in the end,
\be\label{dLgwdLemcT}
d_L^{\,\rm gw}(z)=\sqrt{\frac{c_T(z)}{c_T(0)} }\,\exp\left\{-\int_0^z \,\frac{dz'}{1+z'}\,\delta(z')\right\}\, (1+z)
\int_0^z\, d\tilde{z}\, \frac{c_T(\tilde{z}) }{H(\tilde{z})}\, ,
\ee
to be compared with
\be
d_L^{\,\rm em}(z)=(1+z)
\int_0^z\, d\tilde{z}\, \frac{1 }{H(\tilde{z})}\, .
\ee
In App.~\ref{app-A} we show that the same relation emerges if we define the GW luminosity distance from \eq{fluxluminosity}, rather than from  the $1/d_L^{\rm gw}$ behavior of the GW amplitude from a localized source, as we have done here.
Also note that the effect can be related to Eq.~\eqref{eq:AtimesR} with the conformal factor for scalar-tensor theories corresponding to $M_{\rm eff}^2 c_T$~\cite{Lombriser:2015sxa,Gleyzes:2015pma}.

The observation of  GW170817/GRB~170817A
puts a limit   $|c_T-c|/c< O(10^{-15})$, but this limit only holds for redshifts smaller than the redshift of that source, i.e. for $z\,\lsim\, 0.01$.  LISA can measure $c_T(z)$ to much higher redshifts. In particular, the simultaneous observation of a GW event and of its electromagnetic counterpart would put extremely strong limits on $|c_T(z)-c|/c$ up to the redshift of the source, 
just as for GW170817/GRB~170817A.

Given the strong observational constraint from  GW170817/GRB~170817A, and the lack of   explicit models where $c_T(z)$ evolves from a value equal to $c$ within 15 digits at $z<0.01$, to a sensibly different value at higher redshift, in the following we will limit our analysis to the case $c_T(z)=c$. Note also that, if at higher redshift $c_T(z)$ should be sensibly different from $c$, with LISA one would simply not see an electromagnetic counterpart even if it existed, since the time delay of the electromagnetic and gravitational signal, over such distances, would be huge. In that case the analysis of the present paper, that assumes standard sirens with electromagnetic counterpart, would not be applicable, and one would have to resort to statistical methods.\footnote{At low-$z$, an alternative way to test an anomalous GW speed at LISA frequencies, $c_T(k_{_{\rm LISA}})\neq c$, is to measure the phase lag between GW and EM signals of continuous sources such as the LISA verification binaries. This test can constrain the graviton mass \cite{Larson:1999kg,Cutler:2002ef} as well as the propagation speed \cite{Finn:2013toa,Bettoni:2016mij}.}

\subsubsection{Phenomenological parametrization of \texorpdfstring{$d_L^{\,\rm gw}(z)/d_L^{\,\rm em}(z)$}{}}\label{sect:phenopar}

In general, in  modified gravity, both the cosmological background evolution and the cosmological perturbations are 
different with respect to  GR. It is obviously useful to have phenomenological parametrizations of these effects, that encompass a large class of theories.  In modified gravity, the deviation of the background evolution from   $\Lambda$CDM is determined by the DE density $\rde(z)$ or, equivalently, by the DE equation of state $\wde(z)$. In principle one could try to reconstruct  the whole function $\wde(z)$ from  cosmological observations, but  current results are unavoidably not very accurate (see e.g. fig.~5 of \cite{Ade:2015rim}). The standard approach is rather to use a parametrization for this function, that catches the qualitative features of a large class of models. The most common  is the
Chevallier--Polarski--Linder  parametrization~\cite{Chevallier:2000qy,Linder:2002et},  which makes use of two parameters $(w_0,w_a)$,
\be
w_{DE}(a)=w_0+w_a (1-a)\, ,
\ee
corresponding to the value and the slope of the function at the present time.
In terms of redshift, 
\be\label{w0wa}
w_{\rm DE}(z)= w_0+\frac{z}{1+z} w_a\, .
\ee 
One can then analyze the cosmological data adding $(w_0,w_a)$ to the standard set of cosmological parameters.
Similarly, some standard parametrizations are used for describing the modification from GR in the scalar perturbation sector, in order to compare with structure formation and weak lensing, see e.g.  \cite{Daniel:2010ky,Amendola:2007rr}. Here we are interested in  tensor perturbations, where  the effect is encoded in the non-trivial function  $d_L^{\,\rm gw}(z)/d_L^{\,\rm em}(z)$. Again, rather than trying to reconstruct this whole function from the data, it is more convenient to look for a simple parametrization that catches the main features of a large class of models
in terms of a small number of parameters. We shall adopt 
the 2-parameter parameterization proposed in Ref.~\cite{Belgacem:2018lbp},
\be\label{eq:param}
\Xi(z)\equiv \frac{d_L^{\,\rm gw}(z)}{d_L^{\,\rm em}(z)}
=\Xi_0 +\frac{1-\Xi_0}{(1+z)^n}
\, ,
\ee
which depends on the parameters $\Xi_0$ and $n$, both taken to be positive.
In terms of the scale factor $a=1/(1+z)$ corresponding to the redshift of the source,
\be\label{eq:paramvsa}
\frac{d_L^{\,\rm gw}(a)}{d_L^{\,\rm em}(a)}
=\Xi_0 +a^n (1-\Xi_0)
\, .
\ee
The value $\Xi_0\,=\,1$ corresponds to GR. This parameterization is designed to smoothly interpolate between a unit value 
\be
\Xi(z\ll1)\,=\,1 \label{Chias1}\,,
\ee
 at small redshifts -- where
 cumulative effects of modified gravity  wave propagation have not  sufficient time to accumulate  differences
 with respect to GR, see \eq{dLgwdLem} -- 
 to a constant value $\Xi_0$
\be
\Xi(z\gg1)\,=\,\Xi_0\,,
\label{Chias2}
\ee
at large redshift.  Indeed, in the large redshift regime we expect that the effects of modified gravity ``turn-off'' and
$|\delta(z\gg1)|\ll 1$, since modified
gravity should mainly   affect  late-time evolution (also  for 
ensuring compatibility with CMB observations), in which case the quantity $\Xi(z)$ approaches a constant. 
This parametrization was originally proposed in \cite{Belgacem:2018lbp}, inspired by the fact that it fits extremely well the prediction for
$\Xi(z)$ obtained from a nonlocal modification of gravity (see   section~\ref{Sec:nonlocal} and~ref.~\cite{Belgacem:2017cqo} for review), but it was then realized that its features are very general, so that it is expected to fit the predictions from a large class of models. Indeed, in section~\ref{sect:models} we will compare this fitting formula to the explicit predictions of several modified gravity models, and confirm that it is  appropriate in many situations.\footnote{Recently, an analysis of modified GW propagation for ground-based advanced detectors, including ET, has been presented in 
\cite{Nishizawa:2019rra}. That work uses a parametrization for modified GW propagation  which corresponds to setting the function $\delta(z)$ in \eq{dLgwdLem} to  a constant value. As discussed in \cite{Belgacem:2018lbp}, this is a special case of  the parametrization (\ref{eq:param}), with $\Xi_0=0$ and $n=\delta$. As we will see in section~\ref{sect:models}, typical models actually predict variation of $\Xi_0$ of only a few percent from the GR value $\Xi_0=1$. In any case, the conclusion of \cite{Nishizawa:2019rra} that modified GW propagation can be tested at the $1\%$ level at future ground-based detectors such as ET broadly agrees with the finding in \cite{Belgacem:2018lbp}. Ref.~\cite{Lagos:2019kds} recently studied the bound on modified GW propagation from the single event GW170817. Again, the parametrization of modified GW propagation is different, with $\delta(z)$ taken to be proportional to $\ode(z)$, but the results are consistent with those presented earlier in  \cite{Belgacem:2018lbp}, where the limit from
GW170817 was also computed.}

\begin{figure}[t]
\begin{center}
\includegraphics[width = 0.44 \textwidth]{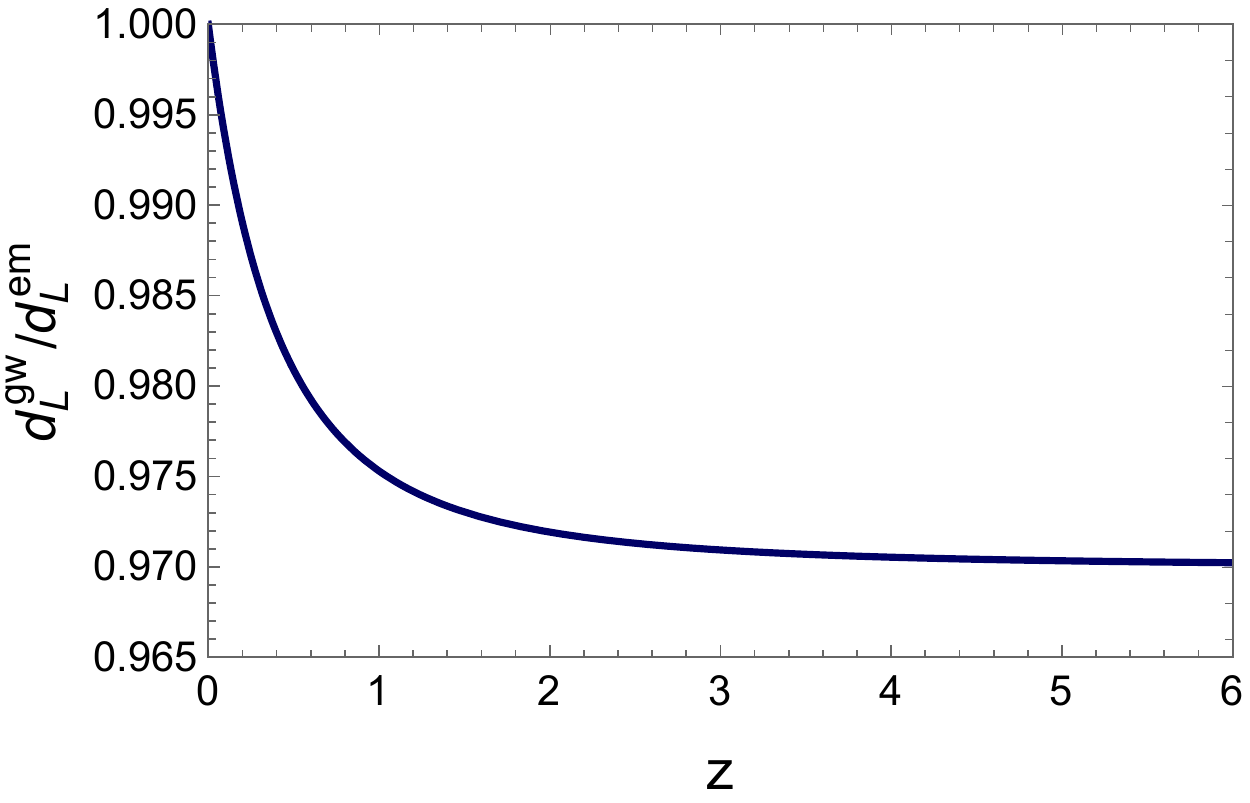}
\includegraphics[width = 0.44 \textwidth]{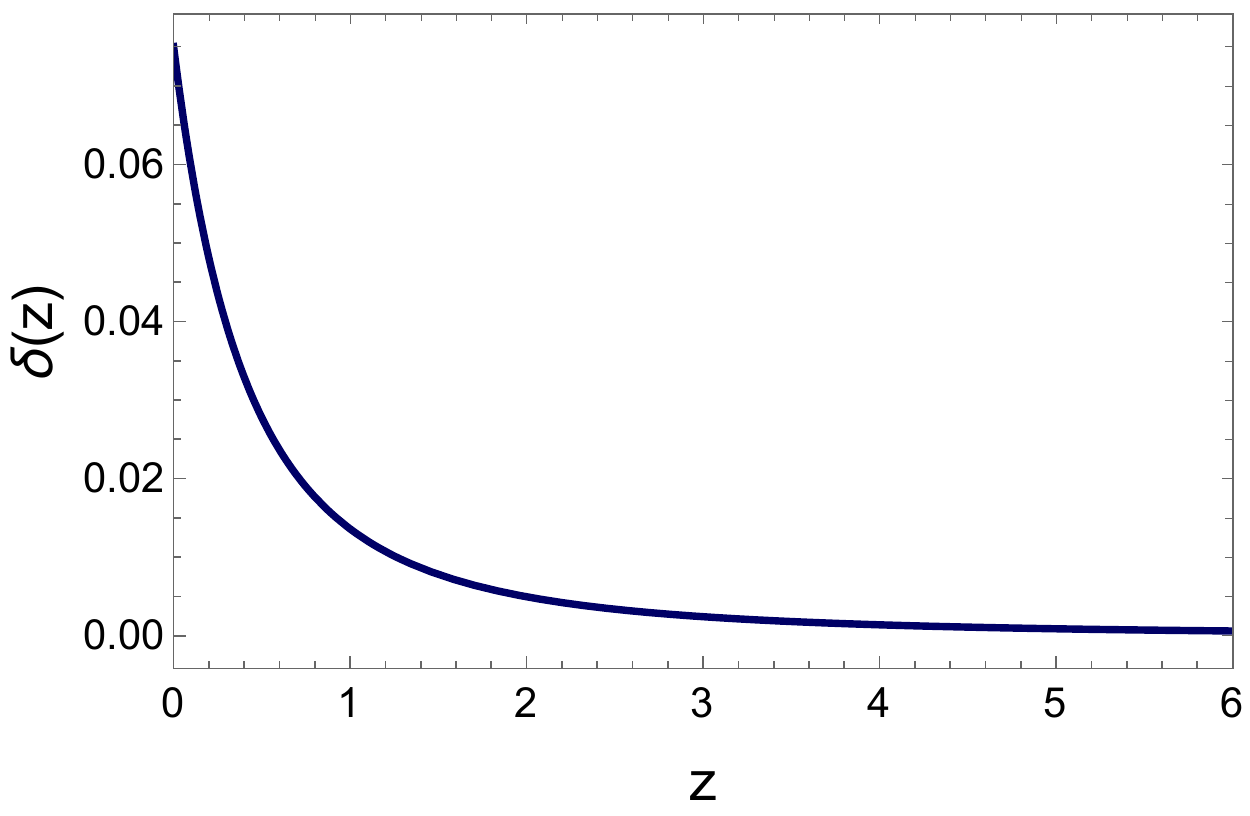}
\includegraphics[width = 0.44 \textwidth]{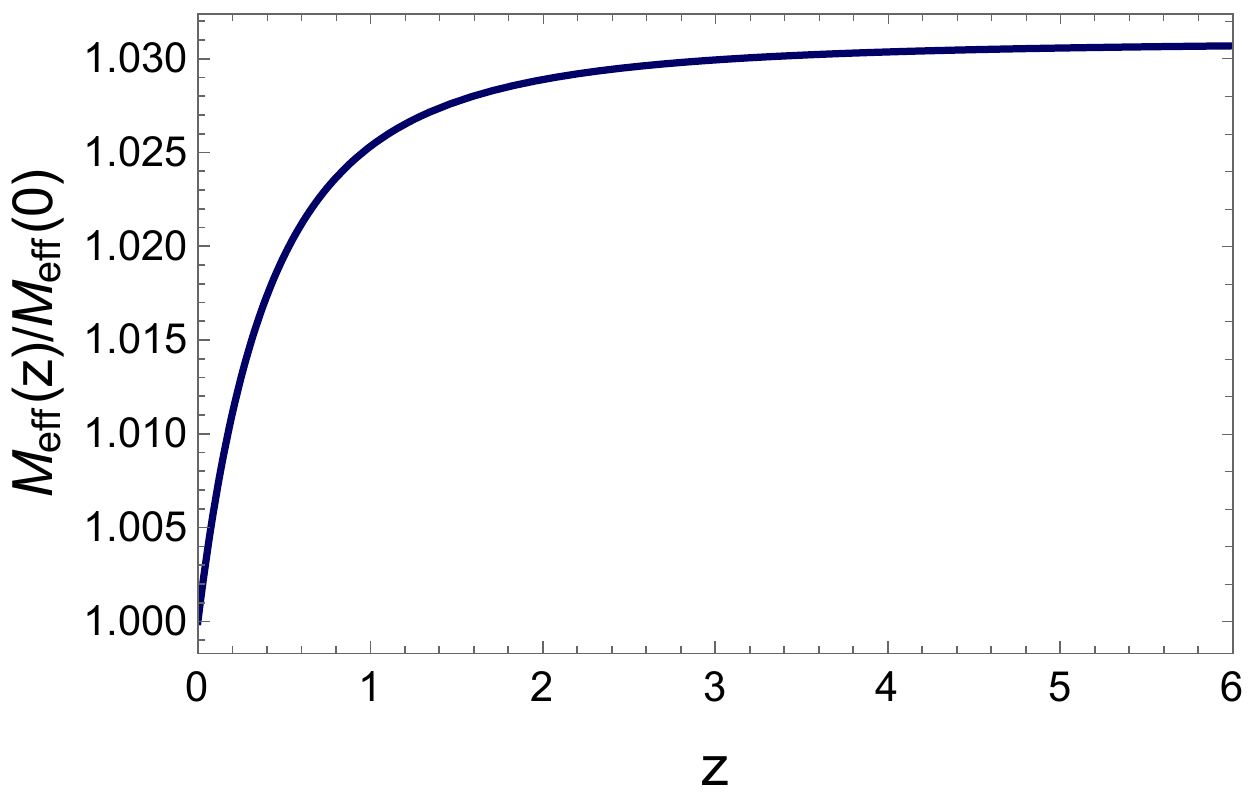}
\includegraphics[width = 0.44 \textwidth]{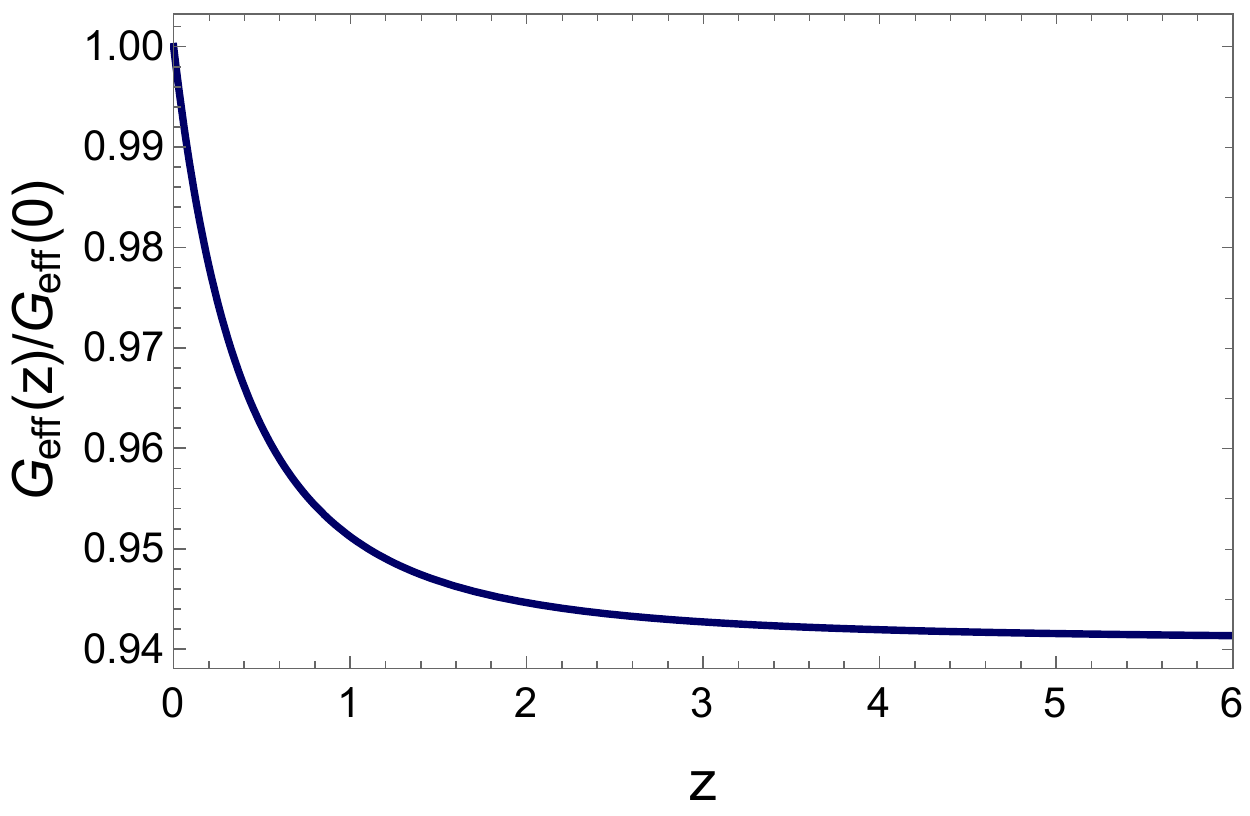} 
 \caption{ Plots of the quantities  $\Xi(z)$, $\delta(z)$, $M_{\rm eff}(z)/\mpl$ and 
 $G_{\rm eff}(z)/G$ as function of the redshift.   Here we choose the values $n=2.5$, $\Xi_0=0.97$, corresponding to one of the models discussed in section~\ref{Sec:nonlocal}.}
\label{fig:HDcl}
\end{center} 
\end{figure}
For theories for which \eqs{ddmplMstar}{dLgwdLemGeff} hold (see the discussion in section~\ref{sect:timevarMpl}),  from
\eq{eq:param} we can obtain a corresponding parametrization for the time variation of the effective Planck mass or of the effective Newton constant,
\bees
M_{\rm eff}(z) &=&\mpl\: \Xi^{-1}(z) \,, \\
G_{\rm eff}(z) &=&G\: \Xi^{2}(z) \, .
\ees
Furthermore, from \eq{dLgwdLem}, we have
\bees\label{nuprof}
\delta(z) &=& - \frac{\,d \ln \Xi(z)}{d\,\ln (1+z)}\nn\\
&=&\frac{n\left(1-\Xi_0\right)}{1-\Xi_0+\Xi_0 (1+z)^n} 
\,.
\ees
In this parametrization the quantity $\delta(z)$ indeed goes to zero at large redshifts,  as desired: at early times gravity propagates as in GR. At late times, instead,
$\delta(z\ll1)\,=\, n\,(1-\Xi_0 )$. Fig.~\ref{fig:HDcl} shows $\Xi(z)$, $\delta(z)$, $M_{\rm eff}(z)/M_{\rm eff}(0)$ and 
 $G_{\rm eff}(z)/G_{\rm eff}(0)$ as function of the redshift, setting for definiteness $n=2.5$ and $\Xi_0=0.97$.

It should also be observed that the parametrization (\ref{eq:param}) of $d_L^{\,\rm gw}(z)/d_L^{\,\rm em}(z)$ is more robust than the corresponding parametrization (\ref{nuprof}) of $\delta(z)$. Indeed, even if $\delta(z)$ should have some non-trivial features as a function of redshift, such as a peak, still these features will  be smoothed out by the integral in \eq{dLgwdLem}. Since anyhow  $d_L^{\,\rm gw}(z)/d_L^{\,\rm em}(z)$ must go to one as $z\ra 0$ and we expect that in most models it will go asymptotically to a constant at large $z$, in general the fit (\ref{eq:param}) to $d_L^{\,\rm gw}(z)/d_L^{\,\rm em}(z)$ will work reasonably well even in cases where the corresponding fit (\ref{nuprof}) is not too  good. We will see an explicit example of this behavior in section~\ref{sub_dhost}. Since, for standard sirens, the observable quantity is
$d_L^{\,\rm gw}(z)$, rather than the function $\delta(z)$ itself, this means that the parametrization (\ref{eq:param}) is quite robust for observational purposes.

Of course, in general, over a large interval of redshifts the simple parametrization (\ref{eq:param}) might no longer be accurate, and in particular the same index $n$ might not fit well  the whole range of redshifts of interest. An extreme case are the bigravity theories that we shall discuss in section \ref{Sec:bigravity}, where the effects of graviton oscillations lead to 
peculiar behaviors for the GW luminosity distance as a function of redshift, see e.g. Fig.~\ref{fig:bigravity_dl}.
Similarly to what is usually done for the parametrization of the DE equation of state, one could further improve the accuracy of   the parametrization (\ref{eq:param})  by fixing $n$ at a pivot redshift $z_p$ where the observational errors are the lowest,
\be         
 n(z_p)  =  \frac{\ln( 1 - \Xi_0) - \ln\left[\Xi(z_p) - \Xi_0 \right]}{\ln (1+z_p)} \,.
\ee
Another possibility is to specify the slope of the evolution of $\Xi(z)$ at $z=0$, i.e., 
\begin{equation}
 n \equiv n(0) = \frac{\delta(0)}{1-\Xi_0} \,.
\end{equation}
However, in considering LISA  standard sirens  at sufficiently large redshift, where $\Xi(z)$ approaches its asymptotic value, the precise value of $n$ will be of limited importance, and the crucial parameter in the parametrization (\ref{eq:param}) will rather be $\Xi_0$.

An alternative parametrization where  
$d_L^{\,\rm gw}(z)/d_L^{\,\rm   em}(z)$ differs from unit  by  ${\cal O}(z^m)$ at small redshifts, rather than by ${\cal O}(z)$, can be obtained by modifying \Eq{eq:param} into
\be\label{eq:paramzm}
\Xi(z) =\Xi_0 +\frac{1-\Xi_0}{1+z^m}\, . 
\ee 
In the
low-$z$ limit, the expression on the right-hand side approaches $1+(\Xi_0-1)z^m$, and still goes to $\Xi_0$
at large $z$. We will discuss in sections \ref{sub_dhost} and~\ref{sec:QG}  examples of models where a $1+ {\cal O}(z^m)$ behavior could emerge, as well as other possible parametrizations of this type. Note however that, for a model in which  at low redshift $\Xi(z)=1+{\cal O}(z^m)$, a two-parameter parametrization might not be adequate, since it would connect the index $m$ that determines the behavior at $z\ll 1$  to the index describing the decrease at large $z$, that in general will be given by a different power. Thus, in these cases, it might be necessary to rather use a three-parameter expression such as
\be
\Xi(z) =\Xi_0 +\frac{1-\Xi_0}{(1+z^m)^n}\, , 
\ee 
which is obtained by replacing $z\ra z^m$ in \eq{eq:param}. In the
low-$z$ limit it approaches $1+n(\Xi_0-1)z^m$, while at large $z$ it approaches $\Xi_0$ with a different power-like behavior.
 
\vspace{5mm}

To conclude this section, let us emphasize the importance of modified GW propagation, as encoded for instance in the parameters $(\Xi_0,n)$ defined by \eq{eq:param}, for studies of dark energy and modified gravity at advanced GW detectors. This stems from two important considerations:

\begin{enumerate}

\item We have seen that, in general, a modified gravity model induces deviations from $\Lambda$CDM at the background level (through the DE equation of state), in the scalar perturbation sector, and in the tensor perturbation sector. Concerning the background modifications, as encoded for instance in the $(w_0,w_a)$ parameters, several studies~\cite{Sathyaprakash:2009xt,Zhao:2010sz,Tamanini:2016zlh,Belgacem:2018lbp} have shown that the accuracy that LISA, or third-generation ground-based interferometers such as the Einstein Telescope, could reach on  $w_0$  is not really better than the measurement, at the level of a few  percent, that we already  have from {\em Planck} in combination with other cosmological probes such as Baryon Acoustic Oscillations (BAO) and Supernovae (SNe). Our analysis in section~\ref{sect:MCMC} will indeed confirm these findings.
In contrast, $d_L^{\,\rm gw}(z)/d_L^{\,\rm em}(z)$  is an observable that is only accessible thanks to GW observations.

\item On top of this, \label{page:degen} it was realized in \cite{Belgacem:2017ihm,Belgacem:2018lbp} that, in a generic modified gravity theory, in which  the deviation of $d_L^{\,\rm gw}(z)/d_L^{\,\rm em}(z)$ from  1 is of the same order as the deviation of $\wde(z)$ from $-1$, the effect of 
$d_L^{\,\rm gw}(z)/d_L^{\,\rm em}(z)$ on standard sirens dominates over the effect of $\wde(z)$.
This can be understood  from the explicit expression of the standard (``electromagnetic") luminosity distance given in \eqs{dLem}{E(z)}. From this expression one might think, naively, that if one changes the equation of state of DE by, say, $10\%$, this will induce a corresponding relative variation of $d_L$ again of order $10\%$. However, this is not true because the cosmological parameters, such as $H_0$ and $\oma$, that enters in \eq{E(z)}, are not fixed, but must themselves be determined self-consistently within the model, by comparing with cosmological observations and performing Bayesian parameter estimation, and their best-fit values change if we modify the DE content of the model. The fit to cosmological data basically amounts to requiring that the model reproduces some fixed distance indicators at large redshifts, such as the scale  determined by the  peaks of the CMB or that from the BAO. Thus, Bayesian parameter estimation has a compensating effect, changing the luminosity distance in a direction opposite to that induced by a change in $\rde(z)$, in such a way to keep as small as possible the variation of $d_L^{\,\rm em}(z)$ at large redshifts. Thus, after performing Bayesian parameter estimation, a relative change in $\wde$ by, say, $10\%$, would only induce a relative change of order, say, $1\%$ in $d_L(z)$.
In contrast, modified GW propagation is an extra effect, that is not compensated by {\gt degeneracies with other (fitted) cosmological parameters}, 
and therefore dominates over the effect of $\wde$.

This physical argument has been confirmed with an explicit MCMC study for the Einstein Telescope in \cite{Belgacem:2018lbp} where it was found that, assuming $10^3$ standard sirens with an electromagnetic counterpart, ET, in combination with other cosmological probes, could measure $w_0$ with a precision of $3.2\%$ and $\Xi_0$ with a precision $0.8\%$. In this paper we will see, from the explicit MCMC study in section~\ref{sect:MCMC}, that the same pattern holds for LISA, and, using standard sirens in combination with other cosmological probes, $\Xi_0$ can be measured with a significantly better accuracy 
than $w_0$.

Therefore $\Xi_0$, or more generally modified GW propagation, is a prime observable for dark energy studies with advanced GW detectors.

\end{enumerate}

%%%%%
\section{Models of modified gravity}\label{sect:models}

In this section we discuss   the prediction for the ratio $d_L^{\,\rm gw}(z)/d_L^{\,\rm em}(z)$ in 
various scenarios. We will see that modified GW propagation is a  generic phenomenon in modified gravity theories, and  we will  assess how well the parameterization (\ref{eq:param}) 
describes the predictions of various models.   We examine  scalar-tensor theories, non-local models, bigravity and  theories with extra  dimensions.
  In  our discussion we shall make the simplifying  hypothesis that modified gravity affects the propagation, and not the production, of GWs. This
 is a delicate assumption: on the other hand it is known to  hold in specific, interesting examples. For 
 the cubic Galileon model,  a
  scalar-tensor theory of gravity equipped
 with a Vainshtein screening mechanism (see section \ref{sect:scalartens}),  \cite{Dar:2018dra} shown that
 scalar contributions to  
GW radiation   from merging binaries is very suppressed. Also for the non-local RT model discussed in section \ref{Sec:nonlocal}, any effect
of modified gravity is expected to be negligible at the (cosmologically) small scales characterizing astrophysical merging events.
 Moreover, even for systems where our assumption is not strictly valid, we can expect that modified gravity effects
  occurring when  GWs are produced    do not influence the redshift dependence of the ratio 
  $d_L^{\,\rm gw}(z)/d_L^{\,\rm em}(z)$,    and    might be  readsorbed into appropriate rescalings of  (redshift-independent) parameters  entering  into the expression for     the GW luminosity distance.

\subsection{Scalar-tensor theories of gravity}\label{sect:scalartens}

The class of scalar-tensor theories of gravity includes some of the best-studied alternatives to GR aimed at explaining late-time cosmic acceleration.   Besides a massless spin-2 mode, these theories contain a  scalar degree of freedom whose dynamics play  an important role for the evolution of the universe. Scalar-tensor theories 
include the simplest systems that exhibit the following, cosmologically relevant, phenomena:
 \begin{enumerate}
 \item {\it self-acceleration}; it can explain the
   present day acceleration of the universe without the need to invoke a small positive cosmological constant. See \cite{Joyce:2014kja} for a  comprehensive review.
 \item {\it screening mechanisms}; these avoid fifth-force constraints associated with long-range scalar forces by suppressing the effect of the scalar sector on matter at smaller scales. Depending on which screening mechanism is in play, the corresponding suppression relies on specific features of the scalar action and/or its couplings. See \cite{Burrage:2017qrf, Babichev:2013usa} for reviews. \end{enumerate}

In theories where the scalar acquires a non-trivial profile and is non-minimally coupled to gravity, the dynamics of the tensor modes is naturally affected: this may lead to deviations from GR. The physics of GW propagation is a particularly clean probe  of modified gravity, the reason being that the tensor sector is typically the one less affected by screening mechanisms.
In terms of the evolution equation \eqref{gen-ev-eq1}, we can expect that scalar-tensor systems lead to a cosmological
 expansion
 history different from $\Lambda$CDM, and
 the gravitational wave equation \eqref{gen-ev-eq1} 
  can have $\delta \neq 0$, and $c_T\,\neq\,1$. On the other hand,
 $m_T\,=0$ and $\Pi_{ij}\,=\,0$. 
We examine some aspects of  the evolution of tensor modes over cosmological distances in these systems, first by briefly reviewing known  results
from the perspective of  GW physics, then by
delivering new predictions for recent degenerate scalar-tensor theories. In light of the GW170817 event, we focus on the case where $c_T\,=\,1$.

%%%%%%%%%%%%%%%%%%%%%%%%
  \subsubsection{Horndeski theories of gravity}\label{subsec_hd}
%%%%%%%%%%%%%%%%%%%%%%%%  

The most well-known scalar-tensor theories of gravity (Brans--Dicke, $f(R)$, covariant Galileon models)
belong to the wide class of Horndeski theories, the most general covariant scalar-tensor theories of gravity leading
to second-order equations of motion (see e.g.{  \cite{Kobayashi:2019hrl} } for a recent review). They are described by the action
\cite{Horndeski:1974wa,Deffayet:2011gz,Kobayashi:2011nu}
\begin{equation}
 S = \int d^4x \sqrt{-g} \left[ \sum_{i=2}^5 \mathcal{L}_i + \mathcal{L}_{\rm m}(g_{\mu\nu},\psi_{\rm m}) \right] \label{eq:horndeski}
\end{equation}
with the Lagrangian densities
\begin{eqnarray}
\hspace*{-5mm} \mathcal{L}_2 & = & \hspace*{0mm}G_2(\phi,X) \,, \\
\hspace*{-5mm} \mathcal{L}_3 & = & \hspace*{0mm}G_3(\phi,X)\Box \phi \,, \\
\hspace*{-5mm} \mathcal{L}_4 & = &\hspace*{0mm} G_4(\phi,X)R - 2G_{4X}(\phi,X) \left[ (\Box\phi)^2 - (\nabla_{\mu}\nabla_{\nu}\phi)^2 \right] \,, \\
\hspace*{-5mm} \mathcal{L}_5 & = &\hspace*{0mm} G_5(\phi,X) G_{\mu\nu}\nabla^{\mu}\nabla^{\nu}\phi + \frac{1}{3} G_{5X}(\phi,X) \left[ (\Box\phi)^3 - 3\Box\phi(\nabla_{\mu}\nabla_{\nu}\phi)^2 + 2 (\nabla_{\mu}\nabla_{\nu}\phi)^3 \right]\hspace*{-1mm},
\end{eqnarray}
where $X\equiv\partial_{\mu}\phi\partial^{\mu}\phi$; $\phi$ is the scalar;
  $R$ and $G_{\mu\nu}$ denote the Ricci scalar and Einstein tensor of the Jordan frame metric $g_{\mu\nu}$; and the matter fields $\psi_{\rm m}$ in $\mathcal{L}_{\rm m}$ are  minimally coupled with gravity. The $G_i$ are arbitrary functions of $\phi$, $X$, and with $G_{iX}$ we denote derivatives
  along $X$. 
   For convenience, we have set the Planck mass $M_{\rm Pl}$ and the speed of light in vacuum to unity.

\begin{figure}
 \resizebox{0.5075\textwidth}{!}{
 \includegraphics{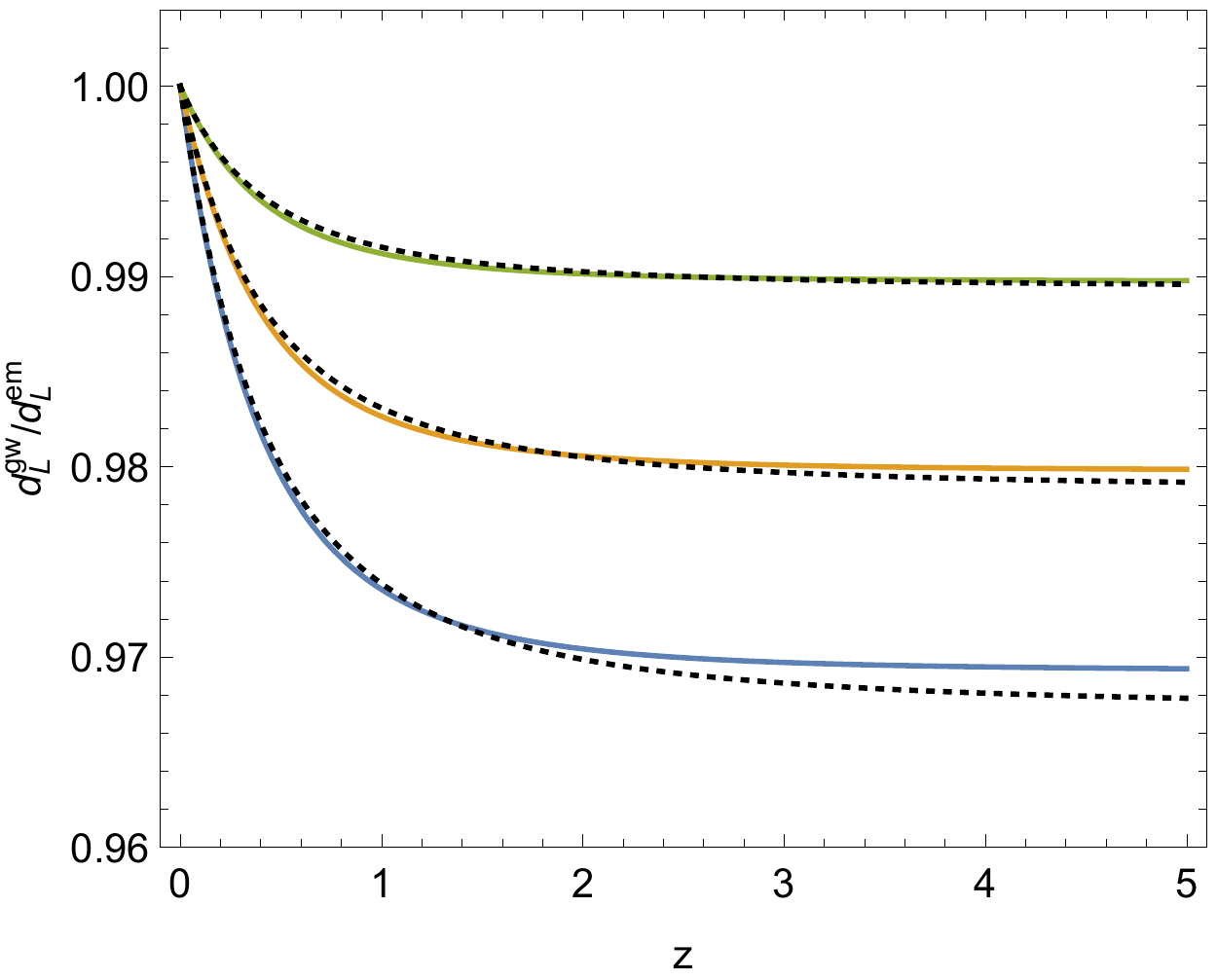}
 }
 \resizebox{0.4925\textwidth}{!}{
 \includegraphics{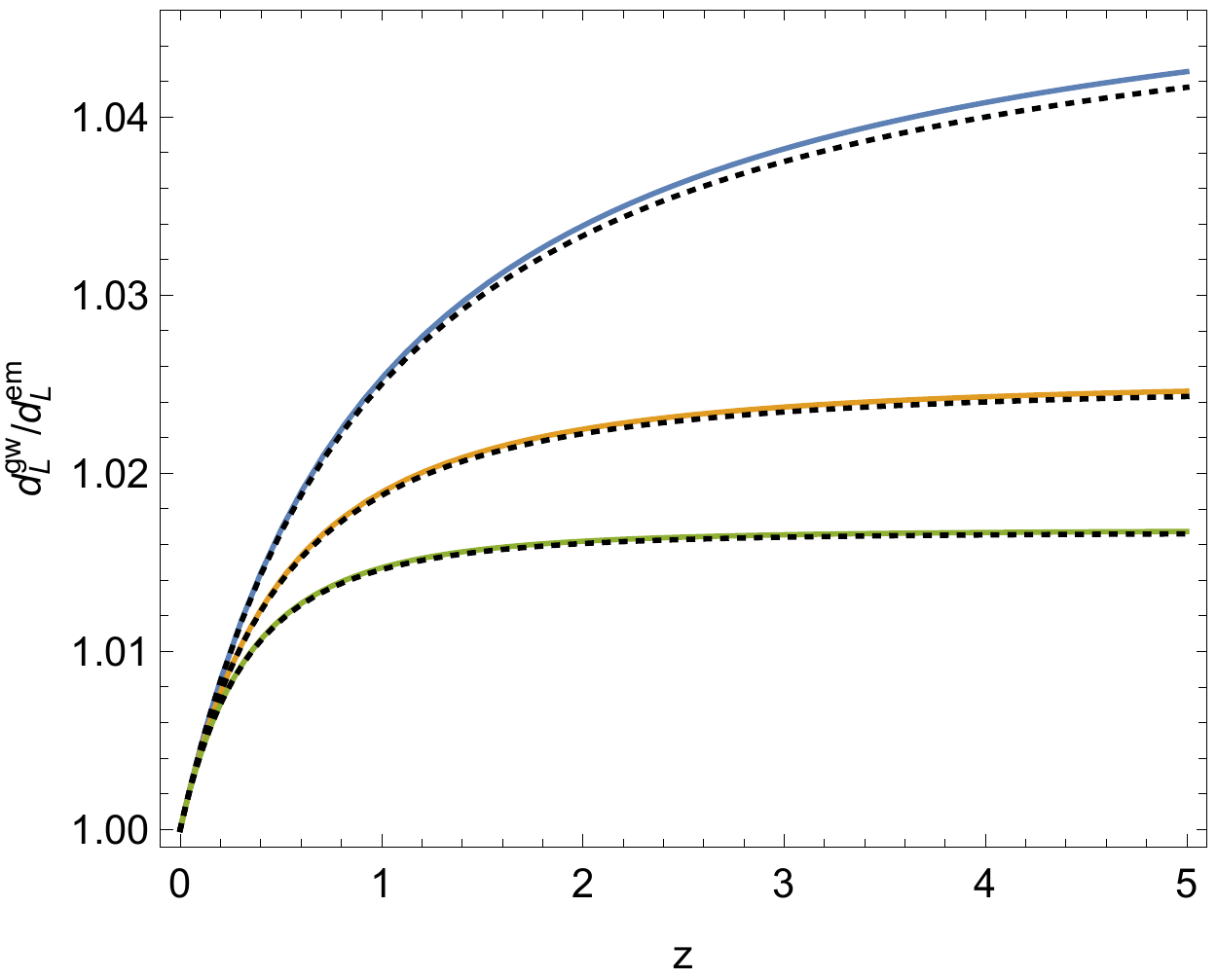}
 }
\caption{
Luminosity distance ratios for designer $f(R)$ gravity (\emph{left panel}) and a power-law modification in $\alpha_M$ (\emph{right panel}).
We adopt {$B_0=  0.34, 0.22, 0.11$}  (bottom to top) and $\alpha_{M0} = 0.01$ with $\tilde{n} = 1, 2, 3$ (top to bottom).
Solid and dashed curves illustrate the model predictions and the corresponding parametrisation~(\ref{eq:param}),
respectively. {\gt The deviations of the parameter $\Xi_0$ from the GR value of unity are of the order of  percent for these
 scenarios.}
%{\rosso [LL: change amplitudes?]}
}
\label{fig:horndeskiparametrisation}
\end{figure}

In these theories, due to the non-minimal couplings between  gravity and the scalar $\phi$, tensor fluctuations
typically propagate with a speed different from light, unless the functions $G_i$ satisfy some constraints. To ensure
$c_T=1$, we require that~\cite{Kimura2011,McManus2016,Bettoni:2016mij}
\begin{equation}
 G_{4X} \simeq 0 \,, \quad G_5 \simeq \textrm{const.}\,,
\end{equation}
where, without loss of generality, $G_5$ can be set to zero as should be clear after performing an integration by parts (see also refs.~\cite{Creminelli:2017sry,Sakstein:2017xjx,Ezquiaga:2017ekz,Baker:2017hug,Copeland:2018yuh}  for discussions on fine-tuned solutions, and \cite{Scomparin:2019ziw} for further aspects of GW propagation in Horndeski theories of gravity). %discussions on specific solutions).
It follows that Eq.~\eqref{eq:horndeski} can no longer provide an observationally viable acceleration by
means of $G_4,\,G_5$ % an $X-$dependent profile for 
 %$G_5$~
  \cite{Lombriser:2015sxa,Lombriser:2016yzn}. The function
 $G_4(\phi)$ is associated with the effective Planck mass, $G_4=M_{\rm eff}^2/2$. As a result, there can be interesting cosmological models where $G_4(\phi)$ depends on time, something that can be tested with GW observations. In this 
 %and similar set-ups [{\rosso meaning the $G_4$ case, correct?}],
  set-up,  $\delta$ is related to the effective Planck mass according to \eq{deltaMstar} (see ref.~\cite{Gleyzes:2014rba}):
\be
\alpha_{\rm M}\,\equiv\,-2 \delta\,=\, \frac{\dd \ln M_{\rm eff}^2}{\dd \ln a}\,,
\ee
as follows from the discussion in section~\ref{sect:timevarMpl}.
In this section we use the notation $\alpha_M$~\cite{Bellini:2014fua}, usually employed in analyzing  scalar-tensor systems. 
The use of GW standard sirens in order to test the time evolution of $G_4$ was already proposed in Ref.~\cite{Saltas:2014dha} and a {\gt preliminary 
forecast}
%first constraint
 at the level
%\be 
 $|M_{\rm eff}^2(z=0)-1|\lesssim 3.5\times10^{-3}
%\ee 
$ 
was estimated in Ref.~\cite{Lombriser:2015sxa} by adapting  forecasts on the accuracy  that LISA can reach on $H_0$.
%%%%
%
We specify the mapping between a range of Horndeski models and the parametrisation in Eq.~(\ref{eq:param}), which will enable us to interpret the constraints on $\Xi_0$ and $n$ for given values of the model parameters.
The mapping for Horndeski scalar-tensor theories can be generally performed\footnote{ Note that one may have to specify $M^2(z\rightarrow\infty)$ whenever it does not reduce to $M_{\rm Pl}^2$ at early times. The early time matching is usually necessary
  for the purpose of recovering GR. In the complementary late-time regime, matching may be due to screening effects in the laboratory at $z=0$.} by  specifying $M^2(0)$ and $\alpha_{M0}$ according to
%From section~\ref{sec:param} 
\begin{eqnarray}
 \Xi_0 & = & \lim_{z\to \infty}\frac{M_{\rm eff}(0)}{M_{\rm eff}(z)} \,, \label{eq:Xi0param} \\
 n & \simeq & \frac{\alpha_{M0}}{2(\Xi_0-1)} \,. \label{eq:nparam}
\end{eqnarray}
%
%Table~\ref{tab:mapping} provides a mapping for a number of frequently studied modified gravity models to the parametrisation in Eq.~(\ref{eq:param}).
This overall ``dictionary" is  summarised in Table~\ref{tab:mapping}.
%
%For completeness we shall provide here derivations of the mappings in Table~\ref{tab:mapping}.
Note that we assume the constraint $|\Xi_0-1|\ll1$ (and $n\sim1$) and that all models recover $M_{\rm eff}(z\rightarrow\infty)=M_{\rm Pl}$, hence, $\Xi_0=M_0$, and we set $M_{\rm Pl}\equiv1$ for convenience.
In Fig.~\ref{fig:horndeskiparametrisation} we illustrate the performance of the fit provided by the parametrisation (\ref{eq:param}), with the values of $\Xi_0$ and $n$ given 
in eqs.~\eqref{eq:Xi0param} and \eqref{eq:nparam}, for two examples embedded in the Horndeski action. We see that the parametrization (\ref{eq:param}) works  well.

%%% TABLE %%%
\begin{table}
 \centering
 \begin{tabular}{lccl}
  Model & $\Xi_0 - 1$ & $n$ & Refs.  \\ 
  \hline\hline
HS $f(R)$ gravity & $\frac{1}{2}f_{R0}$ & $\frac{3(\tilde{n}+1)\Omega_m}{4-3\Omega_m}$ & \cite{Hu:2007nk} \phantom{ \Huge $X$} \vspace*{1mm} \\ 
  \hline
  Designer $f(R)$ gravity & $-0.24\Omega_m^{0.76}B_0$ & $3.1\Omega_m^{0.24}$ & \cite{Song:2006ej} \phantom{ \Huge $X$} \vspace*{1mm} \\ 
  \hline
  %
  %MSG/Palatini $f(R)$ & $(\phi(0)- \phi(z \to \infty))$ &$\frac{\partial_z\phi(0)}{\phi(0)-\phi(z\to \infty)}$ & \cite{Bellini:2014fua,Carroll2006} \\
  %
  Jordan--Brans--Dicke & $\frac{1}{2}\delta\phi_0$ & $\frac{3(\tilde{n}+1)\Omega_m}{4-3\Omega_m}$ & \cite{Brans:1961}  \phantom{ \Huge $X$} \vspace*{1mm}\\   \hline
  Galileon cosmology & $\frac{\beta\phi_0}{2\mpl}$ & $\frac{\dot{\phi}_0}{H_0\phi}$ & \cite{Chow:2009fm} \phantom{ \Huge $X$} \vspace*{1mm} \\   \hline
  %
  %BDK & $\frac{1}{2}\left ( \kappa(0)-\kappa(z\to \infty) \right)$ & $\frac{\partial_z \kappa(0)}{\kappa(z\to \infty) - \kappa(0)}$ & \cite{Bellini:2014fua,Sawicki2012}\\
  %
  %$f$(Gauss-Bonnet) & & & [] \\
  %
  $\alpha_M=\alpha_{M0} a^{\tilde{n}}$ & $\frac{\alpha_{M0}}{2{\tilde{n}}}$ & ${\tilde{n}}$ & \cite{Bellini:2014fua}  \phantom{ \Huge $X$} \vspace*{1mm}\\   \hline
  $\alpha_M=\alpha_{M0} \frac{\Omega_{\Lambda}(a)}{\Omega_{\Lambda}}$ & $-\frac{\alpha_{M0}}{6\Omega_{\Lambda}} \ln \Omega_m$  & $-\frac{3\Omega_{\Lambda}}{\ln \Omega_m}$ & \cite{Simpson:2012ra,Bellini:2014fua} \phantom{ \Huge $X$} \vspace*{1mm} \\   \hline
  $\Omega=1+\Omega_+a^{\tilde{n}}$ & $\frac{1}{2}\Omega_+$ & $\tilde{n}$ & \cite{Lombriser:2015sxa} \phantom{ \Huge $X$} \vspace*{1mm} \\  \hline
  %
  %Minimal self-acceleration & $ a_{acc}\exp \left ( \frac{C \chi_{acc}}{2}\right )$  & $\frac{\alpha_{M0}}{2(\Xi_0 -1)}$ & [] \\
  Minimal self-acceleration & $\lambda \left( \ln a_{acc} + \frac{C}{2} \chi_{acc} \right)$ & $\frac{C/H_0 - 2}{\ln a_{acc}^2 - C \chi_{acc}}$ & \cite{Lombriser:2016yzn}  \phantom{ \Huge $X$} \vspace*{1mm}\\   \hline  \hline
  %DHOST &? & ? &\cite{}
 % \\
 % Vector-Tensor theories &? & ? &\cite{}
%  \\
 % Bigravity &? & ? & \cite{}
 \end{tabular}
\caption{
 Mapping of the parametrisation in Eq.~(\ref{eq:param}) to a number of frequently studied,
 representative modified gravity models embedded in the Horndeski action~(\ref{eq:horndeski}) with luminal speed of gravitational waves. 
%Mappings of a number of frequently studied modified gravity models embedded in the Horndeski action~(\ref{eq:horndeski}) with luminal speed of gravitational waves to the parametrisation in Eq.~(\ref{eq:param}).
For simplicity, we have employed the approximations $\alpha_{M0}\ll1$ (and $n\sim1$).
}
\label{tab:mapping}
\end{table}
%%% TABLE %%%

%%% FIG %%%
%%% FIG %%%

%----%
%\subsection{Derivations}

\smallskip

The first model we shall inspect is $f(R)$ gravity~\cite{Buchdahl:1983zz}, where the Einstein-Hilbert action is generalised by $R\rightarrow R+f(R)$.
It can be mapped onto the action~(\ref{eq:horndeski}) by defining the scalar field $2G_4(\phi)\equiv\phi\equiv 1+f_R$ with $f_R\equiv \dd f/\dd R$ and $G_2\equiv -U(\phi)\equiv \frac{1}{2}[f(R) - f_R R]$.
Hence, one finds 
\be
\Xi_0=M_0=(1+f_{R0})^{1/2}\simeq 1 + \frac{1}{2} f_{R0}\, ,
\ee 
[where $f_{R0}\equiv (f_R)_0$ is the present value of $f_R$] as well as 
\be
n \simeq \(\frac{f_R'}{f_{R}}\)_0\, .
\ee
Adopting the particular functional form for $f$ proposed by Hu and  Sawicki (HS)~\cite{Hu:2007nk} 
\be
f(R) \simeq -2\Lambda - \frac{f_{R0} \bar{R}_0^{\tilde{n}+1}}{\tilde{n} R^{\tilde{n}}}\, ,
\ee 
one obtains 
\be
n \simeq \frac{3(\tilde{n}+1)\Omega_m}{(4-3\Omega_m)}\, .
\ee
These relations remain unchanged in a straightforward generalisation of the HS $f(R)$ model to Jordan-Brans--Dicke gravity~\cite{Brans:1961} with appropriate scalar field potential $U(\phi)$ (see, e.g., Refs.~\cite{Upadhye:2013nfa,Lombriser:2013eza}).
In that case $G_4\equiv\phi/2$ and $G_2=-U+X\omega(\phi)/\phi$, where  $\omega(\phi)$ is the Brans--Dicke function, and $\omega=0$ reduces to the HS $f(R)$ case.
We furthermore define $\phi\equiv1+\delta\phi$.
Alternatively to the HS $f(R)$ function, one may, for instance, also consider the designer $f(R)$ model~\cite{Song:2006ej} that exactly reproduces the standard cosmological expansion history.
It can be parametrised by the Compton wavelength parameter $B_0 \equiv [(H/H')f_R'/(1+f_{R})]|_0$.
For simplicity, we make use of the approximation $B_0\simeq-2.1\Omega_m^{-0.76}f_{R0}$~\cite{Ferraro:2010} to derive the expressions in Table~\ref{tab:mapping} (also see Fig.~\ref{fig:horndeskiparametrisation}).

\begin{figure}
\begin{center}
 \includegraphics[width=0.5\textwidth]{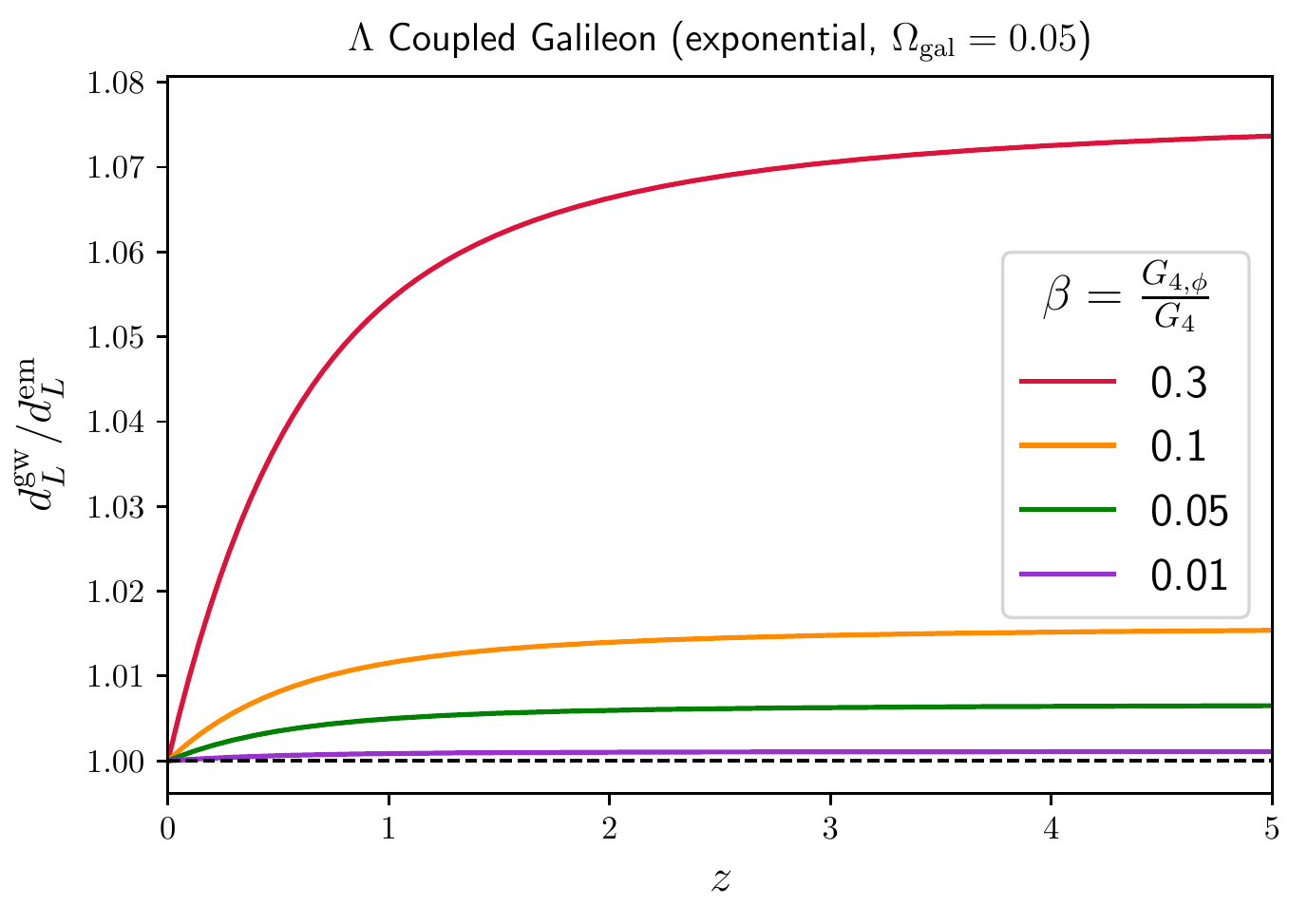}
\end{center}
\caption{Plot of
the luminosity distance ratio for
a  coupled Galileon gravity model, which saturates at low redshift; obtained with  \texttt{hi\_class} \cite{Zumalacarregui:2016pph}. 
 {\gt For sufficiently large values of $\beta$, the deviations of the parameter $\Xi_0$ from the GR value can be of order 5 percent.}
 }
\label{fig:brans_dicke_and_coupled_gals}
\end{figure}

\smallskip

Next we consider a cubic Galileon model conformally coupled to the Ricci scalar. The model is defined by the following choice of the Horndeski functions
\begin{equation}
G_4 = \frac{1}{2}\mpl^2\exp\left(\beta\phi/\mpl\right)\,,
\quad G_3 = -\frac{c_3}{\mpl H_0^2} X\,,
\quad G_2 = -\frac{c_2}{2} X\,,
\end{equation}
where $\beta$ regulates the strength of the coupling, with the uncoupled case in the limit $\beta=0$.
The coefficients $c_2,c_3$ are dimensionless, and fixed by the scalar field normalization and the value of 
 {\gt Galileon energy fraction} 
$\Omega_{\rm gal}$ (see refs.~\cite{Barreira:2014jha,Renk:2017rzu} for details). For consistency with expansion probes we consider models where the Galileon is a sub-dominant contribution to the energy density $\Omega_{\rm gal} = 0.05$, with $\Omega_\Lambda = 1- \Omega_M - \Omega_{\rm gal}$.
The initial condition for the scalar field is $\phi(t_{\rm ini}) = 0$ to ensure that $M_{\rm eff}^2\approx 1$ at early times, and the initial value of the field derivative $\dot \phi(t_{\rm ini})$ is fixed by the tracker solution.
This leads to a unique solution for $\phi_0$. Together with given $\Omega_{\rm gal}$, it fixes $\dot{\phi}_0$. It is then straightforward to express $\Xi_0$ and $n$ in terms of $\phi$ and its derivative today. We have $\Xi_0=\exp\left(\frac{\beta\phi_0}{2\mpl}\right)\simeq 1 + \frac{\beta\phi_0}{2\mpl}$ from which $n=H_0^{-1}\frac{\dot{\phi}_0}{\phi_0}$ using Eq.~\eqref{eq:nparam}.
%
%discuss the 
The coupling $\beta\neq0$ introduces a non-zero value of $\alpha_M$ and modifies the ratio of the luminosity distances, as shown in Figure \ref{fig:brans_dicke_and_coupled_gals}.
This is a minimal Galileon model that modifies the luminosity distance while remaining compatible with the GW speed.

Due to the wealth of models that can be realised with action~(\ref{eq:horndeski}),  instead of studying specific models one may adopt a phenomenological parametrisation for the time evolution of $G_4$.
Common options include a power law for $\alpha_{\rm M}$ in terms of the scale factor $a$ or, motivated by a connection to dark energy, a proportionality to the fractional dark energy density $\Omega_{\Lambda}(a)\equiv 8\pi G \Lambda/(3H^2)$ as well as a generalisation of that to an evolving effective dark energy contribution~\cite{Simpson:2012ra}.
For the power law $\alpha_M =\alpha_{M0} a^{\tilde{n}}$, one finds $M(z) = \exp[\alpha_{M0}a^{\tilde{n}}/(2{\tilde{n}})]$ and $\Xi_0 = e^{\alpha_{M0}/(2{\tilde{n}})} \simeq  1 + \alpha_{M0}/(2{\tilde{n}})$ as well as $n \simeq {\tilde{n}}$ (see Fig.~\ref{fig:horndeskiparametrisation}).
Similarly one may also consider a power law for the evolution of the Planck mass instead.
For instance, for the EFT function $\Omega=1+\Omega_+ a^{\tilde{n}}$, corresponding to $\Omega=M^2$ for $c_T=1$, one immediately finds that $\Xi_0 = M_0 \simeq 1 + \frac{1}{2} \Omega_+$ and $n\simeq \tilde{n}$.
Adopting $\tilde{n}=4$, a constraint of $|\Omega_+|<1.4\times10^{-2}$ was inferred from LISA forecasts in Ref.~\cite{Lombriser:2015sxa}, where for cosmic acceleration to be attributed to a running $\Omega$, one requires $\Omega_+\lesssim-0.1$.
For a scaling of the time evolution as $\alpha_M=\alpha_{M0} \Omega_{\Lambda}(a)/\Omega_{\Lambda} = \alpha_{M0} H_0^2/H^2$, one obtains $M^2=M_0^2 (a^3 H^2/H_0^2)^{\alpha_{M0}/(3\Omega_{\Lambda})}$ with $\Xi_0=M_0 = \Omega_m^{-\alpha_{M0}/(6\Omega_{\Lambda})} \simeq 1 - \alpha_{M0}/(6\Omega_{\Lambda}) \ln \Omega_m$ such that $M^2 = \Omega_m^{-\alpha_{M0}/(3\Omega_{\Lambda})}$ and $n\simeq-3\Omega_{\Lambda}/\ln \Omega_m$.

\smallskip

Finally, to more directly address the question whether a time evolution of $G_4$ can be made responsible for cosmic acceleration at late times, one can formulate the minimal evolution of the Planck mass required to produce a positive accelerated expansion without the contribution of dark energy or a cosmological constant~\cite{Lombriser:2016yzn}.
The approach defines a threshold $\alpha_{M,min} = C/(a H)
 - 2$, where $C=2H_0a_{acc}\sqrt{3(1-\Omega_m)}$ and $a_{acc} = (\Omega_m/(1-\Omega_m)/2]^{1/3}$.
This threshold is, however, in $3\sigma$ tension with cosmological observations, particularly of cross correlations of the {\gt CMB temperature}
%integrated Sachs-Wolfe effect
 with foreground galaxies~\cite{Lombriser:2016yzn}.
A simple rescaling of $\alpha_M=\lambda \alpha_{M,min}$ assesses the fraction of cosmic acceleration $\lambda$ that can be attributed to an evolving $G_4$.
For $\lambda\ll1$, one finds $\Xi_0 \simeq (a_{acc})^\lambda e^{\lambda C \chi_{acc}/2} \simeq 1 + \lambda \left( \ln a_{acc} + C \chi_{acc}/2 \right)$ and $n \simeq (C/H_0 - 2)/(\ln a_{acc}^2 - C \chi_{acc})$, where $\chi_{acc}$ denotes the comoving distance at the acceleration scale $a_{acc}$.

This list is far from being exhaustive of the range of Horndeski models that will be tested with Standard Sirens, and does in fact provide only a small sample. Rather than expanding Table~\ref{tab:mapping}, however, we shall address next the constraints that can be inferred on theories that generalise action~(\ref{eq:horndeski}) to include higher  derivatives in their equations of motion.

%%%%%%%%%%%%%%%%%%%%%%%%%%%%%%%%%%%%%%%%
  \subsubsection{DHOST theories of gravity}\label{sub_dhost}
%%%%%%%%%%%%%%%%%%%%%%%%%%%%%%%%%%%%%%%

The Horndeski action is built on the requirement of second order field equations.  However, although sufficient, this requirement is not necessary in order to avoid dangerous Ostrogradsky instabilities \cite{Ostrogradsky:1850fid,Woodard:2006nt}. This aspect,  noticed in \cite{Zumalacarregui:2013pma} using transformations of the metric, led initially to the so called beyond Horndeski theories \cite{Gleyzes:2014dya}. Subsequently, it was realised that higher order field equations are actually harmless provided that the Lagrangian is degenerate \cite{Langlois:2015cwa}, i.e. there exists an extra primary constraint that removes the Ostrogradsky mode. Exploiting this property, Degenerate Higher Order Scalar-Tensor (DHOST) theories (see \cite{Langlois:2018dxi} for a recent review) have been constructed \cite{Crisostomi:2016czh, Achour:2016rkg, BenAchour:2016fzp} and they
represent (so far) the most general scalar-tensor theories that propagate a single scalar degree of
freedom in addition to the helicity-2 mode of a massless graviton.
 
DHOST theories have been classified up to cubic order in second derivatives of the scalar field, and are divided in 41 classes whose general Lagrangian reads \cite{BenAchour:2016fzp}
\bea
S[g,\phi] = \int d^4 x \, \sqrt{- g }
\left(f_2 \, R + C_{(2)}^{\mu\nu\rho\sigma} \,  \phi_{\mu\nu} \, \phi_{\rho\sigma}
+ f_3 \, G_{\mu\nu} \phi^{\mu\nu}  +  
C_{(3)}^{\mu\nu\rho\sigma\alpha\beta} \, \phi_{\mu\nu} \, \phi_{\rho\sigma} \, \phi_{\alpha \beta} \right) \label{general action} \,,
\eea
where the functions $f_2$  and $f_3$ depend only on $\phi$ and $X \equiv \nabla_\mu \phi \nabla^\mu \phi$
 and the tensors 
$C_{(2)}$ and $C_{(3)}$ are the most
general tensors constructed with the metric $g_{\mu\nu}$ and the first derivative of the scalar field 
$\phi_\mu \equiv \nabla_\mu \phi$. Not all the 41 classes are healthy and many of them feature pathologies such as the absence of tensor modes or gradient instabilities at linear order \cite{deRham:2016wji, Langlois:2017mxy}.

Among the healthy classes a further restriction can be imposed if we require that tensor modes propagate with velocity $c_T = 1$ on cosmological background. 
The resulting DHOST theory is 
\bea
S[g,\phi]_{c_T = 1} = \int d^4 x \, \sqrt{- g }
\left[f_2 \, R + A_3 (\Box \phi) \phi^{\mu} \phi_{\mu \nu} \phi^{\nu} + A_4 \phi^{\mu} \phi_{\mu \rho} \phi^{\rho \nu} \phi_{\nu} + A_5 (\phi^{\mu} \phi_{\mu \nu} \phi^{\nu})^2
 \right] \label{DHOSTcT=1} \,, \nonumber \\
\eea
where $A_{3,4,5}$ are functions of $\phi$ and $X$, and the following relations have to be imposed to satisfy the degeneracy conditions that remove the Ostrogradsky ghost 
\bea
\label{degcond}
A_4= -\frac{1}{8 f_2}\left[
8 A_3 f_2 - 48 \left({f_{2_{X}}}\right)^2 - 8 A_3 f_{2_{X}} X + A_3^2 X^2
\right] \,, \quad A_5 = 
\frac{A_3}{2 f_2}(4 f_{2_X} + A_3 X) \,.
\eea
Notice that we can add a $K$-essence and a cubic Horndeski term to the above action~(\ref{DHOSTcT=1}), i.e. $K(\phi, X) + G_3(\phi, X)\Box \phi$, without altering the degeneracy conditions, to obtain the most general scalar-tensor theory with $c_T = 1$.
Horndeski theories discussed in the previous subsection are a special subclass of (\ref{DHOSTcT=1}) given by $A_{3,4,5}=f_{2_{X}}=0$.

\medskip

In order to concretely study the time dependence  of the effective tensor Planck mass $M_{\rm eff}$ within
DHOST theories, let us focus on a specific model that admits a tracker solution evolving toward a de Sitter fixed point responsible for the present time cosmic acceleration~\cite{Crisostomi:2017pjs} (see also \cite{Crisostomi:2018bsp}):
\begin{align}
K = c_2 X, \qquad G_3 = \frac{c_3}{\Lambda_3^3} X, \qquad f_2= \frac{M_{\rm Pl}^2}{2} + \frac{c_4}{\Lambda_3^6} X^2, \qquad A_3=-\frac{8 c_4}{\Lambda_3^6} - \frac{\beta}{\Lambda_3^6} \,,\label{DHOSTmodel}
\end{align}
where $M_{\rm Pl}$ is the Planck mass, $c_2, c_3, c_4, \beta$ are free constants and $\Lambda_3$ is a strong coupling scale. 
The tracker solution is characterised by the property that $\dot\phi = \xi/H $ ($\xi=$ const.) during both matter domination and the dS phase, each era with its own value of $\xi$.
If $\beta=0$ in (\ref{DHOSTmodel}), which corresponds to the case of beyond Horndeski theories \cite{Gleyzes:2014dya}, one recovers a unique value for $\xi$ throughout the evolution of the universe \cite{DeFelice:2010pv}\footnote{We do not consider this case here since it is excluded from the constraint derived in \cite{Creminelli:2018xsv} to avoid decay of GWs into Dark Energy. This constraint gives $A_3 = 0$ and therefore $\beta=-8 c_4$. As a consequence, $\beta=0$ would imply also $c_4 = 0$ and the model would simply reduce  to GR + K-essence + cubic Galileon, falling in the class of scenarios we discussed in the previous subsection.}.
Tracker solutions are important because they allow to constrain the model parameters independently of initial conditions (see \cite{Frusciante:2018tvu} for other tracker and scaling solutions in DHOST theories).
%Such comparison with data however is still missing, hence we still do not know what is a realistic region of parameter space for the quantities involved.

Working with dimensionless quantities, for this model we have
\begin{align}
M_{\rm eff}^2= 2 f_2 = 1 + 2 c_4 X^2 \,,
\end{align}
and therefore the ratio between the luminosity distances, see eq \eqref{dLgwdLemGeff}, reads
\begin{align}
\Xi(z)\,=\,\frac{d_L^{\rm gw}(z)}{d_L^{\rm em}(z)} = \sqrt{\frac{1+2 c_4 X^2(z=0)}{1 +2 c_4 X^2(z)}} \,,
\label{eq:Relation_luminosity_distance}
\end{align}
and the function $\delta(z)$ 
in equation \eqref{gen-ev-eq1}
is given by
\begin{align}
\label{eq:nu_parameters}
\delta(z) = - \frac{1}{ H} \frac{4 c_4 \dot{\phi}^3 \ddot{\phi}}{1 +2 c_4 \dot{\phi}^4} \,.
\end{align}
Thus, the GW luminosity distance is controlled by the dynamics of the scalar field and of the scale factor over the tracker solution.

\begin{figure}
	\includegraphics[width=7.5cm]{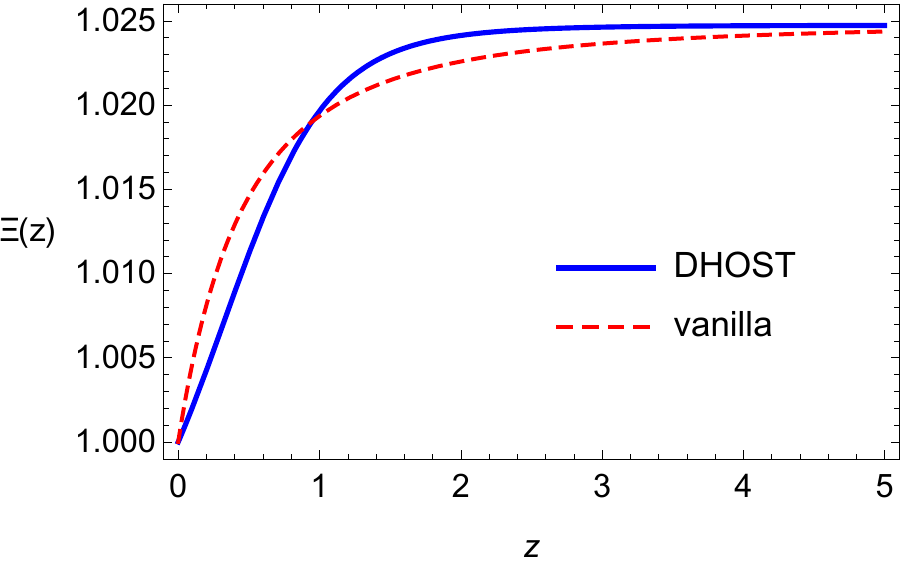}
	\hspace{5pt}
	\includegraphics[width=7.5cm]{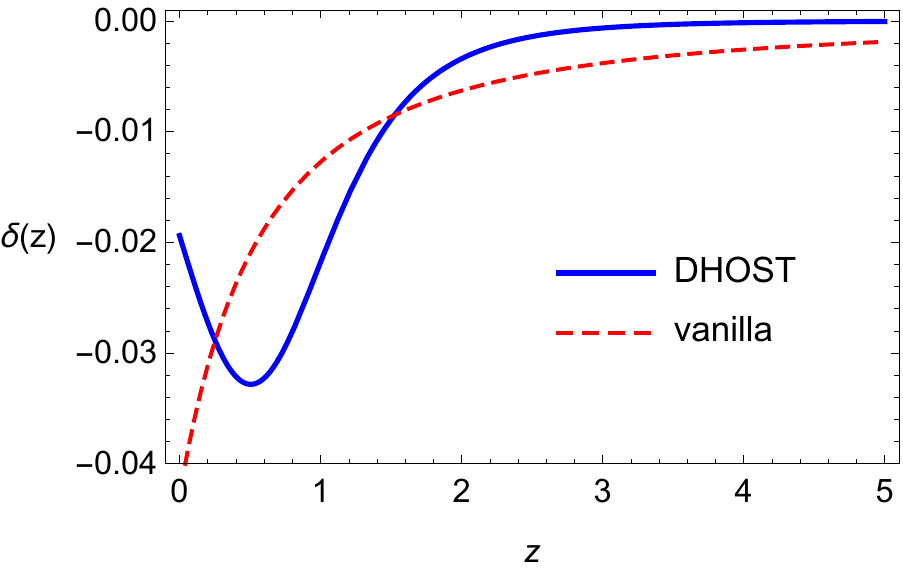}
	\caption{The ratio $\Xi(z)$ of the GW and EM luminosity distance (left panel), and the function $\delta(z)$ (right panel). The red (dashed) line is the best fit up to $z=5$ for  the vanilla parametrization \eqref{eq:param} with the best-fit values $\Xi_0\approx1.025$ and $n\approx2.174$, obtained 
	through a nonlinear fit of the theoretical points with the parametrization curve(s) with the least squares method.
	In representing   the theoretical (blue thick) DHOST curve, we have chosen the values $c_2=3,\;c_3=5,\;c_4=1,\;\beta=-5.3$ and rescaled units so that $M_{\rm Pl}=1=\Lambda_3$, as in \cite{Crisostomi:2017pjs}. {\gt Hence, in this
	set-up, the deviations of $\Xi_0$ from the GR value can be of order of percent level.}}
	\label{fig:Luminosity_distance}
\end{figure}

In Fig. \ref{fig:Luminosity_distance} we show the numerical solution for the  two quantities above, eqs \eqref{eq:Relation_luminosity_distance} and \eqref{eq:nu_parameters}, and the corresponding fit with the phenomenological parametrizations (\ref{eq:param}) and (\ref{nuprof}). 
In agreement with the discussion in Section~\ref{sect:phenopar} we see that, even if in this case the fit 
to $\delta(z)$ obtained from \eq{nuprof} is quite poor, and misses the presence of a local minimum, still the 
fit to the directly observed quantity $d_L^{\,\rm gw}(z)/d_L^{\,\rm em}(z)$ provided by \eq{eq:param}
is  qualitatively acceptable, and 
catches  well the overall  properties of the redshift dependent  profile for $d_L^{\,\rm gw}(z)/d_L^{\,\rm em}(z)$.
 As a general feature, the ratio of the luminosity distances reaches a plateau in the matter dominated era where the deviation with respect to GR vanishes; instead the function $\delta(z)$ features a local minimum and then goes to zero at large redshifts and in the future time where the universe reaches the de Sitter phase. %
In order to perform a proper fit with the parametrisation (\ref{eq:param}), a precise determination of the time today $t_0$ is required by fitting cosmological parameters. 
Indeed, it is easy to realise that a change in $t_0$ can shift the local minimum at smaller or negative redshifts in order to get a better agreement with~(\ref{eq:param}). A similar determination is beyond the scope of this work and the results presented here should only be seen as indicative of possible behaviors of $d_L^{\rm gw}(z) / d_L^{\rm em}(z)$ and $\delta(z)$ for self-accelerating cosmologies in DHOST theories. 

\subsubsection*{Additional parameterisations}

\begin{figure}
	\includegraphics[width=7.5cm]{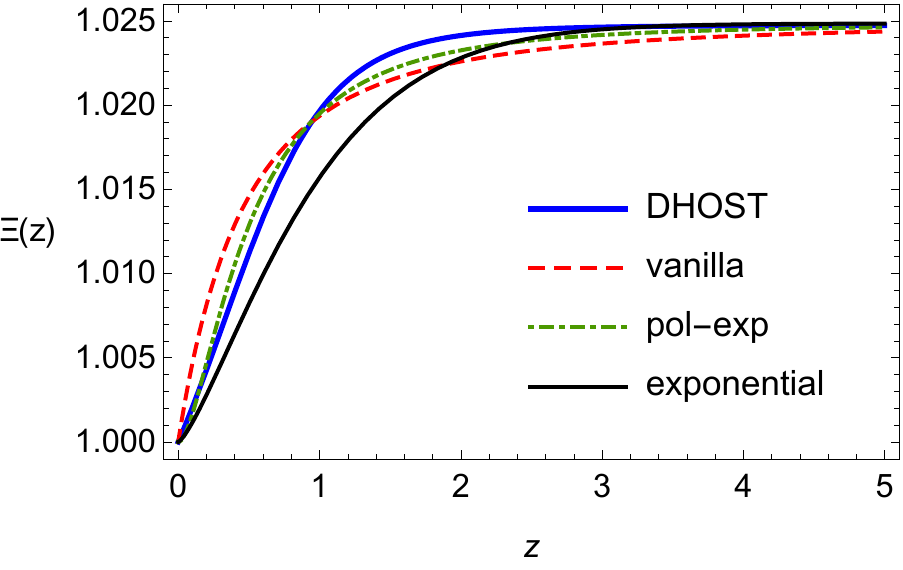}
	\hspace{5pt}
	\includegraphics[width=7.5cm]{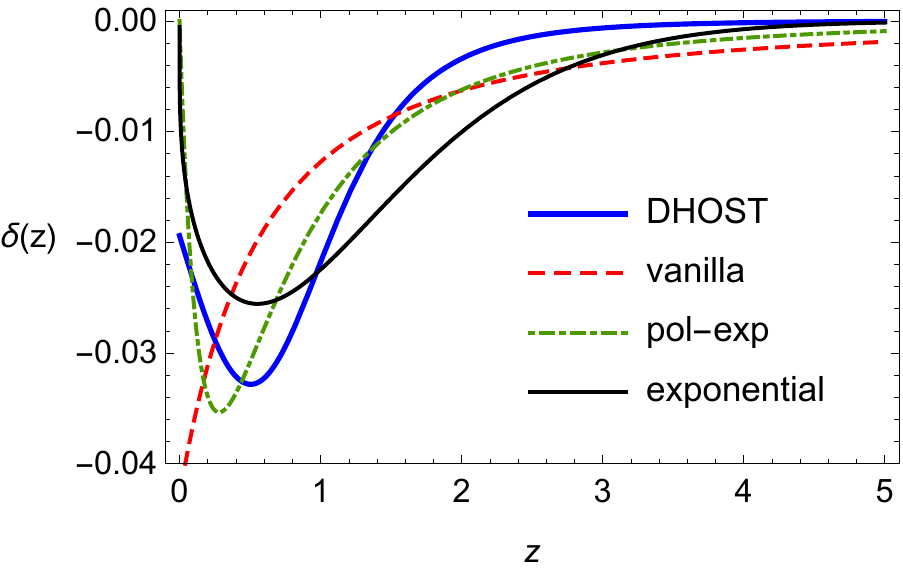}
	\caption{The ratio $\Xi(z)$ of the GW and EM luminosity distance (left panel), and the function $\delta(z)$ (right panel). The blue thick curve is the theoretical profile of the DHOST model with parameters chosen as in Fig. \ref{fig:Luminosity_distance}. The other curves are the vanilla parametrization \eqref{eq:param} (dashed red, $\Xi_0\approx1.025$, $n\approx2.174$), the polynomial-exponential parametrization	\eqref{eq:param2} (dot-dashed green, $\Xi_0\approx1.025$, $n\approx3.156$) and the exponential parametrization \eqref{param3} (thin black, $\Xi_0\approx1.025$, $n\approx1.322$). The parameter values are those of the best fits of $\Xi(z)$ up to $z=5$.}
	\label{fig:Luminosity_distance_alternative_parametrization}
\end{figure}

For these cosmologies, at the de Sitter fixed point we have $\delta = 0$ and therefore, in order to improve the fit, one can  modify the parametrization \eqref{eq:param} to get $\delta(z\ll1)\sim0$ 
%\be
%\delta(z\ll1)\sim0 \,, \label{nuzero2}
%\,
%\ee 
 via the addition of some extra terms (with no new parameters)
to eq \eqref{nuprof}.
A possible choice is 
\be
\delta(z)\,=\,\frac{n\left(1-\Xi_0\right)}{1-\Xi_0+\Xi_0 (1+z)^n}+
\frac{n\left(1-\Xi_0\right)}{(1+z)^n}
- \frac{2n\left(1-\Xi_0\right)}{(1+z)^{2n}} \,,
\ee
which has the desired asymptotic values at large and small redshift. 
{\gt By reverse engineering integrating eq. (2.37), we can analytically reconstruct}
%This choice allows  us [by reverse engineering integrating  eq.~\eqref{nuprof}] to analytically
  %reconstruct 
   a  profile for $\Xi(z)$ as
\be
\Xi(z)= e^{\frac{- (1-\Xi_0)\left[ 1-(1+z)^n\right]}{(1+z)^{2n}}}\left[ 
\Xi_0+  (1-\Xi_0) (1+z)^{-n} 
\right] \,, \label{eq:param2}
\ee
which introduces an overall exponential factor to the parametrization \eqref{eq:param}. This profile maintains the same number of free parameters as \eqref{eq:param} but with the disadvantage of being more complicated. Another alternative parametrization is
\begin{eqnarray}
\Xi(z) &=& \Xi_0+(1-\Xi_0)\,\rme^{-z^n}\,,\label{param3}\\
\delta(z) &=& \frac{n(1-\Xi_0) (1 + z)\, z^{n-1}}{1+\Xi_0(\rme^{z^n}-1)}\,.
\end{eqnarray}
Both the polynomial-exponential parametrization (\ref{eq:param2}) and the exponential parametrization (\ref{param3}) obey the asymptotic relations \eqref{Chias1} and \eqref{Chias2}. The three parametrizations are shown in Fig.~\ref{fig:Luminosity_distance_alternative_parametrization}.

To compare the three proposals \Eq{eq:param}, \Eq{eq:param2} and \Eq{param3} and to quantify which fits the theoretical curve $\Xi(z)$
``better'', we perform a simple Bayesian model selection, calculating the Bayes Information Criterion (BIC) \cite{Sch78} or the Akaike Information Criterion (AIC) \cite{Aka74} for the best fits and treating the theoretical curve as data.
 Our best fits are obtained 
	through a nonlinear fit of the theoretical points with the parametrization curve(s) with the least squares method. 
 Although differences in parametrizations are not especially relevant for LISA standard sirens at high redshifts (e.g., MBH mergers) due to the large error bars, the issue of model reconstruction and the use of Information Criteria may be useful when considering low-$z$ standard sirens (for instance, stellar-mass black holes), or if one uses parametrizations with a different number of free parameters.

The parametrization with smaller BIC or AIC is to be preferred. In the present case, the difference $\Delta:=|\textrm{(IC model 1)}-\textrm{(IC model 2)}|$ is the same for the BIC and the AIC, since the number of data points and the number of free parameters are exactly the same for all parametrizations. According to the classification of \cite{Jef61,KR95}, one finds weak evidence if $\Delta<2$, positive evidence if $2\leq\Delta<6$, strong evidence if $6\leq\Delta<10$, and very strong evidence if $\Delta
\geqslant 10$. A different number of data points can give rise to different orderings in this triple comparison, a fact to keep in mind when using real data. We make it clear that we are comparing parametrizations with a theoretical curve, which may give different results with respect to when one would use a sparse set of data points, as it will be the case for standard sirens.

\begin{figure}
	\includegraphics[width=7.5cm]{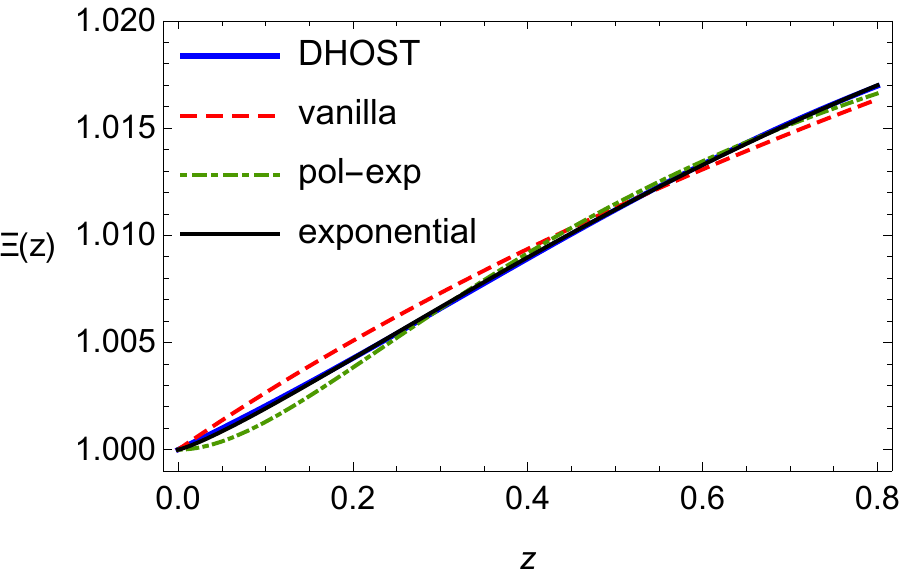}
	\hspace{5pt}
	\includegraphics[width=7.5cm]{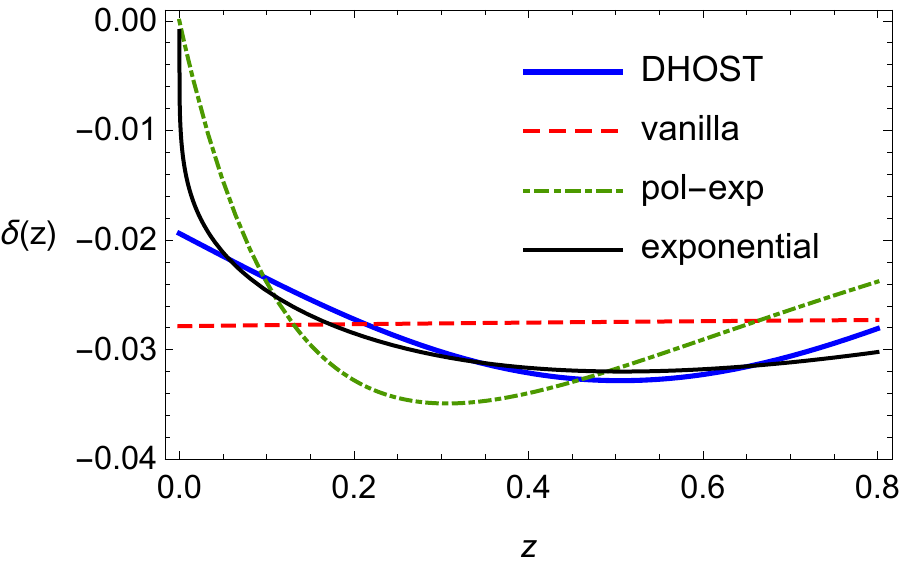}
	\caption{The ratio $\Xi(z)$ of the GW and EM luminosity distance (left panel), and the function $\delta(z)$ (right panel). The blue thick curve is the theoretical profile of the DHOST model with parameters chosen as in Fig. \ref{fig:Luminosity_distance}. The other curves are the vanilla parametrization \eqref{eq:param} (dashed red, $\Xi_0\approx4.872$, $n\approx0.007$), the polynomial-exponential parametrization	\eqref{eq:param2} (dot-dashed green, $\Xi_0\approx1.027$, $n\approx2.592$) and the exponential parametrization \eqref{param3} (thin black, $\Xi_0\approx1.032$, $n\approx1.206$). The parameter values are those of the best fits of $\Xi(z)$ up to $z=0.8$.}
	\label{fig:altparsmallz}
\end{figure}

Having said that, with very strong evidence ($\Delta\gg 10$) we find that fitting the $\Xi(z)$ data up to $z=5$, the polynomial-exponential parametrization	\eqref{eq:param2} is preferred over the vanilla parametrization \eqref{eq:param}, which in turn is preferred over the exponential parametrization \eqref{param3}. We represent this situation as POL$>$VAN$>$EXP. Fitting $\Xi(z)$ data up to $z=0.8$, the ordering changes as EXP$>$POL$>$VAN. As one can see from Fig.~\ref{fig:altparsmallz}, the exponential parametrization \eqref{param3} is considerably more accurate.

Notice that one could contemplate to use different parametrizations depending on whether one wants to fit what we measure directly (the function $\Xi(z)$) or to reconstruct the function $\delta(z)$ from data. In the second case, if we fit the reconstructed $\delta(z)$ up to $z=5$, then EXP$>$POL$>$VAN and the exponential parametrization \eqref{param3} is the most economic way between the three to achieve that goal. Our conclusion is that:
\begin{itemize}
\item For $\Xi(z)$, the polynomial-exponential parametrization \eqref{eq:param2} is the second best at small redshift and the best as a global fit, while for $\delta(z)$ it is the second best both at small redshift and as a global fit.
\item For $\Xi(z)$, the exponential parametrization \eqref{param3} is the best at small redshift but the worst as a global fit, while for $\delta(z)$ it is the best both at small redshift and as a global fit.
\end{itemize}
While the polynomial-exponential parametrization \eqref{eq:param2} is made ad hoc for DHOST and is the best global fit for $\Xi$, the exponential parametrization \eqref{param3} works as a small-$z$ fit for $\Xi$ not only for DHOST models, but also for models with extra dimensions and models of quantum gravity, as we will see in section~\ref{sec:QG}. Therefore, both parametrizations have their own advantages and drawbacks.

\subsection{GW propagation in nonlocal infrared modifications of gravity}\label{Sec:nonlocal}

``Nonlocal infrared modifications of gravity" is a generic denomination for models in which the fundamental theory is local, according to the standard lore of quantum field theory, but non-local terms, relevant in the   infrared (IR), emerge at some effective level. A situation where  this  can happen is when one takes into account quantum fluctuations. In that case, when in the spectrum there are  massless  particles (such as the graviton in GR) or, more generally, particles that are light with respect to the relevant energy scales,
the  quantum effective action, which is the quantity whose variation gives the equation of motion of the vacuum expectation value of the fields,   develops nonlocal terms. Terms that involve  inverse powers of the d'Alembertian are particularly interesting for cosmological applications, since they become
important  in the IR.  The idea that  quantum gravity  at large distances could induce cosmological effects is quite old, see e.g. \cite{Taylor:1989ua}, and several hints on non-trivial physics at large distances
emerge  for instance from the study of  IR effects in de~Sitter space~\cite{Tsamis:1994ca,Antoniadis:1986sb, Antoniadis:1991fa,Polyakov:2007mm,Polyakov:2012uc,Anderson:2013ila,Rajaraman:2016nvv,Dvali:2017eba}. However, a first-principle understanding of the IR limit of quantum gravity is a highly non-trivial task, and it makes sense to begin with a phenomenological approach, where one studies what sort of nonlocal terms could give a viable and interesting cosmology. In this spirit, a  nonlocal gravity model was  proposed 
by Deser and Woodard (DW) \cite{Deser:2007jk,Deser:2013uya} (generalizing earlier work in \cite{Wetterich:1997bz}), based on  the quantum effective action  
\be\label{GammaDW}
\Gamma_{\rm DW}=\frac{\mplr^2}{2}\int d^4x \sqrt{-g}\,\[ R- Rf(\iBox R)\]\, ,
\ee
with $f(X)$ a  function chosen  so to obtain the desired background evolution \cite{Deffayet:2009ca} (see \cite{Woodard:2014iga} for review). 
In the DW model  the nonlocal term involves a dimensionless function $f(X)$. 
Another  possibility that has been much investigated recently is that the nonlocal term rather involves a  mass scale. An explicit example of a model of this class is provided by the Dvali-Gabadadze-Porrati (DGP) model~\cite{Dvali:2000hr} that, when projected onto the four-dimensional brane and linearized around Minkowski space, is equivalent to a nonlocal theory for the fluctuations $\hmn$ whose covariantization, at least to linear order in $\hmn$, reads~\cite{Dvali:2002fz,Dvali:2002pe}
\be
\(1+\frac{m}{\sqrt{-\Box}}\)\Gmn=8\pi G\,\Tmn\, ,
\ee
where the mass scale $m=2M_5^3/M_4^2$ emerges from a combination of the 5-dimensional and 4-dimensional Planck masses. While DGP is not cosmologically viable on its self-accelerating branch, several other models with nonlocalities associated to a mass term have been studied  phenomenologically. In particular, 
a model of this type was proposed in~\cite{ArkaniHamed:2002fu} to introduce the 
degravitation idea,  and is based on the modified  Einstein equations  
\be\label{degrav}
\(1-\frac{m^2}{\Box}\)\Gmn=8\pi G\,\Tmn\, ,
\ee
where $m$ is a mass scale associated to the nonlocal operator $\iBox$.
However, a drawback of \eq{degrav} the energy-momentum tensor is not automatically conserved, since in curved space   $[\nabla^{\mu},\iBox]\neq 0$ and therefore  the Bianchi identity $\nabla^{\mu}\Gmn=0$ no longer ensures $\nabla^{\mu}\Tmn=0$. On the other hand, a symmetric tensor $\Smn$ can always be decomposed as 
\be\label{splitSmn}
S_{\mu\nu}=S_{\mu\nu}^{\rm T}+\frac{1}{2}(\nabla_{\mu}S_{\nu}+\nabla_{\nu}S_{\mu})\, , 
\ee
where $S_{\mu\nu}^{\rm T}$ is the transverse part of $\Smn$, 
$\nabla^{\mu}S_{\mu\nu}^{\rm T}=0$. This decomposition was used
in \cite{Jaccard:2013gla} to modify \eq{degrav}  into
\be\label{GmnT}
\Gmn -m^2\(\iBox\Gmn\)^{\rm T}=8\pi G\,\Tmn\, ,
\ee
where the superscript ``T" denotes the extraction of the transverse part from of the tensor $\(\iBox\Gmn\)$, so that energy--momentum conservation  becomes automatic. The  cosmological evolution of this model  turned out to be unstable, already at the background level, so this model  is not phenomenologically viable. Several variants of this idea have then been explored, and it has been found that it is quite difficult to build a model that is cosmologically viable. A rather extensive study of the various possibilities  (see   \cite{Maggiore:2016gpx,Belgacem:2017cqo}  for reviews) has finally pinned down two particularly interesting models. 
The first, proposed  in \cite{Maggiore:2013mea}, is a variant of \eq{GmnT} where the $\iBox$ operator acts on the Ricci scalar rather than on the Einstein tensor, and is defined by  the nonlocal equation of motion
\be\label{RT}
\Gmn -\frac{m^2}{3}\(\gmn\iBox R\)^{\rm T}=8\pi G\,\Tmn\, ,
\ee
(where the factor $1/3$ is a useful normalization for the mass scale $m$).
We refer to  it as
the ``RT" model, where R stands for the Ricci scalar and T for the extraction of the transverse part.
The second, proposed in \cite{Maggiore:2014sia}, is instead defined in terms of  a quantum effective action
\be\label{RR}
\Gamma_{\rm RR}=\frac{\mplr^2}{2}\int d^{4}x \sqrt{-g}\, 
\[R-\frac{m^2}{6} R\frac{1}{\Box^2} R\]\, .
\ee
We refer to  it as  the ``RR" model, after the two occurrences of the Ricci scalar in the nonlocal term.
It turns out that the RT and RR models are related by the fact that the equations of motion derived from \eq{RR}, when linearized over Minkowski space, are the same as the linearization of \eq{RT}.
However, if one goes  beyond linear order, or if one  linearizes over a different background such as FRW, the two models  are different. 

\begin{figure}[t]
\includegraphics[width=0.44\textwidth]{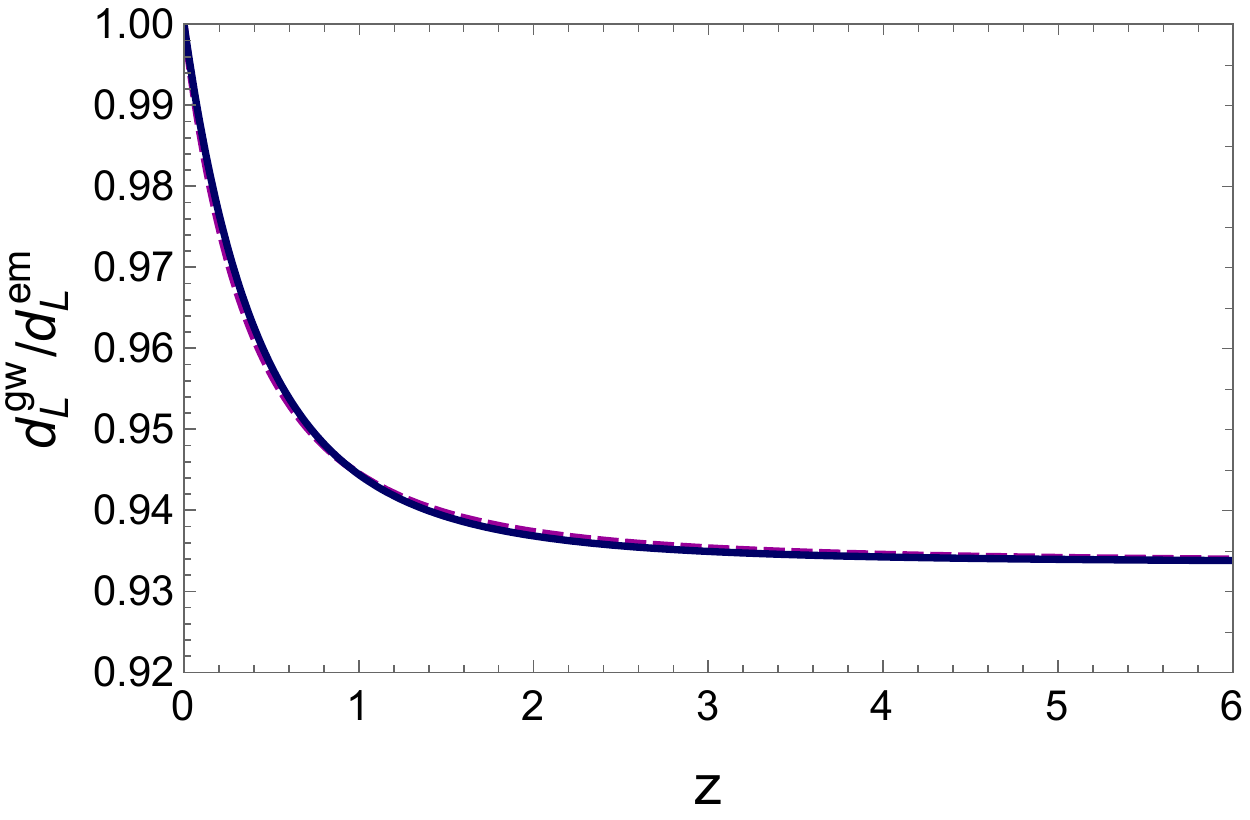}
\includegraphics[width=0.44\textwidth]{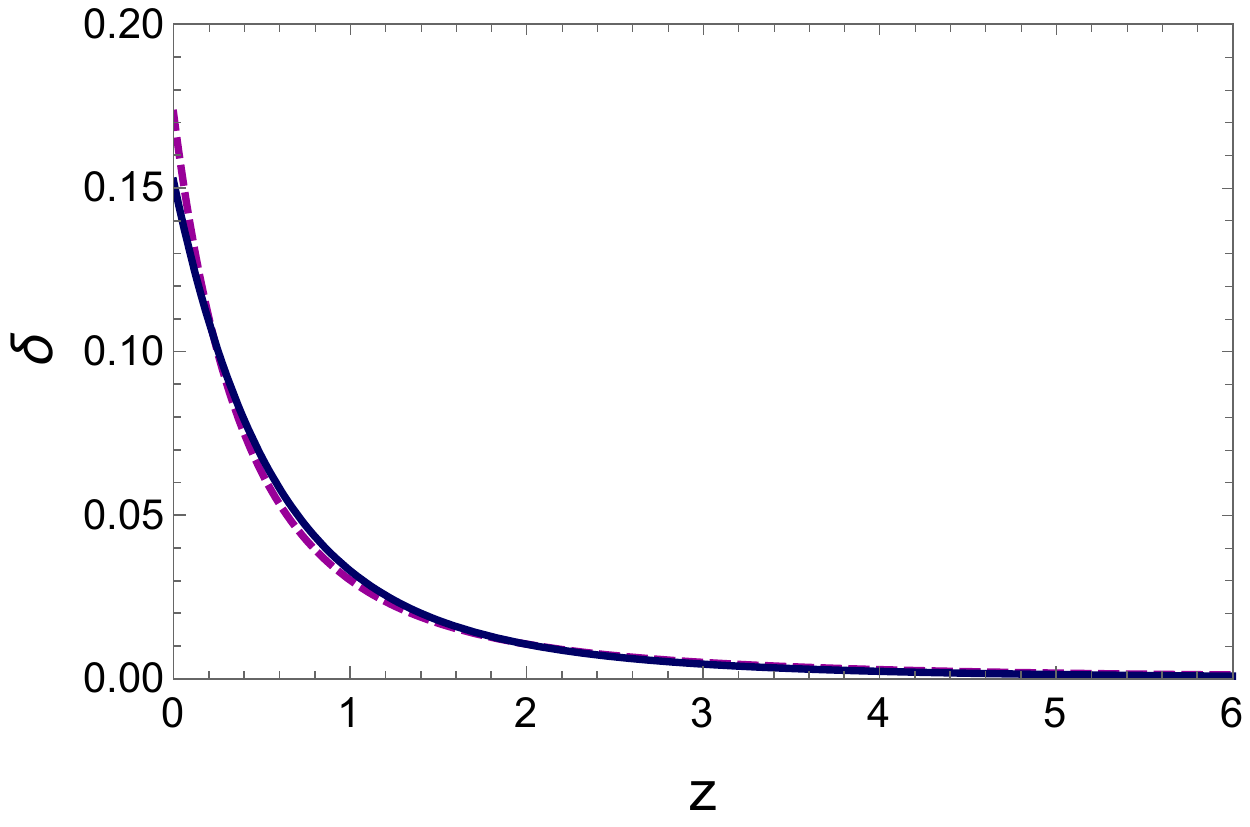}
\caption{ Left panel: the ratio $d_L^{\,\rm gw}(z)/d_L^{\,\rm em}(z)$ in  the RT nonlocal model, as a function of redshift (blue solid line) and the fit given by \eq{eq:param} (magenta dashed line). Right panel: the function $\delta(z)$  (blue solid line) and the fit given by \eq{nuprof}}
\label{fig:RT_dLGW}
\end{figure}

Detailed studies  \cite{Maggiore:2013mea,Maggiore:2014sia,Foffa:2013vma,Dirian:2014bma,Dirian:2016puz,Dirian:2017pwp,Belgacem:2017cqo} have shown that the RR and the RT models have a viable cosmological background evolution, with the nonlocal term driving accelerated expansion; they have stable cosmological perturbations both in the scalar and tensor sector;  tensor perturbations propagate at the speed of light, thus complying with 
the limit from GW170817/GRB~170817A;
both models fit CMB, BAO, SNe and structure formation data at a level statistically equivalent to $\Lambda$CDM (with the same number of parameters, since $m$ replaces $\Lambda$).
Very recently, however, it has been found in ref.~\cite{Belgacem:2018wtb} (elaborating on a discussion in \cite{Barreira:2014kra}) that the RR model is ruled out by the limit on the time variation of Newton's constant derived from Lunar Laser Ranging (LLR). This is due to the fact that the RR model predicts a time-varying effective Newton's constant $G_{\rm eff}(z)$ on cosmological scales, and does not have a screening mechanism on short scales. The analysis in~\cite{Belgacem:2018wtb}  then shows that the same time dependence would appear at the scale of solar system and of the Earth-Moon system, and would then violate the LLR bound. The RT model, in contrast, predicts a constant $G_{\rm eff}=G$ for modes well inside the horizon, and  therefore passes the LLR bound. In  \cite{Belgacem:2018wtb} it is also shown  
 that a screening mechanism that was proposed for the DW model is actually incorrect, and that the DW model is also ruled out by the LLR bound. Thus, it appears that, in this class of nonlocal models, the RT model is the only viable one (see also \cite{Tian:2019bla} for an analysis of LLR constraints that rules out other nonlocal gravity models studied in the literature). It is quite remarkable that the whole ensemble of conditions that have been imposed is so strong to single out a single model out of a rather broad class.\footnote{Recently, stimulated by the observation in \cite{Belgacem:2018wtb}, a new variant of the DW model has been proposed in \cite{Deser:2019lmm}. This new model solves a part of the problem raised, in the sense that a nonlocal quantity  $Y$ used in its construction  indeed changes sign between the static and the cosmological solution, in contrast to the original variable $X=\iBox R$, allowing for a screening mechanism based on setting to zero the function $f(Y)$ that defines the nonlocal model,  when $Y<0$. However, the rather general analysis in
 \cite{Barreira:2014kra,Belgacem:2018wtb} shows that, for both the RR model and the ``old" DW model,
 once one matches the static solution in the near region to the cosmological solution at large distances, one finds $X_{\rm tot}(t,r)\simeq X_{\rm static}(r)+X_{\rm cosmo}(t)$. Since $X_{\rm cosmo}(t)<0$ and numerically dominates over $X_{\rm static}(r)$, it follows that $X_{\rm tot}(t,r)<0$ independently of the sign of 
 $X_{\rm static}(r)$  (which, furthermore, in this case is also negative), and the screening mechanism proposed for the ``old'' DW model, based on setting $f(X)=0$ for $X>0$, cannot be realized. In general,  the same could happen  in the new model for the  field $Y=\iBox(\gMN\pam X\pan X)$, so that
$Y_{\rm tot}(t,r)\simeq Y_{\rm static}(r)+Y_{\rm cosmo}(t)$ with $Y_{\rm cosmo}(t)$ positive and numerically dominating over $Y_{\rm static}(r)$. If this is the case, again the model would never be in the regime $Y<0$ where the screening mechanism acts, and would then be ruled out. Further work is then needed to assess whether this model is indeed viable.}

These nonlocal gravity models predict  modified GW propagation, showing again that this phenomenon is completely general in modified gravity. The RR model prediction is fitted extremely well by the parametrization (\ref{eq:param}), with $\Xi_0\simeq 0.970$ and $n\simeq 2.5$; indeed, the prediction of the RR model was the original motivation for introducing this parametrization~\cite{Belgacem:2018lbp}.
The RT model also predicts a modified GW luminosity distance. The  equation for the tensor perturbations in this model was computed in \cite{Dirian:2016puz} and has again the form (\ref{prophmodgrav}), so GWs propagate at the speed of light, with a function $\delta(z)$  determined by the background evolution of some auxiliary field that are introduced to rewrite the model in a local form. Inserting the expression for the background evolution of these auxiliary fields found from the numerical integration of the modified Einstein equation, we get the results shown in Fig.~\ref{fig:RT_dLGW} for   $d_L^{\,\rm gw}(z)/d_L^{\,\rm em}(z)$ and $\delta(z)$. In these  plots the blue solid line are the predictions of the model, and the magenta dashed lines are the fit to the parametrization ({\ref{eq:param}) for $d_L^{\,\rm gw}(z)/d_L^{\,\rm em}(z)$ [and the corresponding 
parametrization (\ref{nuprof}) for $\delta(z)$], with the best-fit parameters
\be
\Xi_0\simeq 0.934\, ,\qquad n\simeq 2.6\, .
\ee
We see that the parametrization works extremely well, and the model predicts a relatively large value of $\Xi_0$, which differs from the GR value $\Xi_0=1$ by about $6.6\%$.

It should be observed that, in the RT model, for modes well inside the horizon
$G_{\rm eff}=G$~\cite{Nesseris:2014mea,Dirian:2014ara}. Thus, this is an example of a model where there is a modified GW propagation even if the effective Newton constant, or the effective Planck mass, does not evolve with time, and hence the relation (\ref{dLgwdLemGeff}) does not hold.\\

%BIGRAVITY SECTION
\subsection{GW propagation  in  bigravity}
\label{Sec:bigravity}

The linearised equations controlling the propagation of gravitational waves over cosmological distances can contain couplings with additional fields,  for example in scenarios with broken spatial diffeomorphisms or with anisotropic stress.
The extra dynamics is typically the result of a  dark energy or a dark matter component. 
In such cases, the evolution of tensor modes can be characterised by new phenomena, such as transitions and oscillations  between different states, analogous  to neutrino oscillations.  One ought to expect that in these scenarios a minimal parametrization for the GW luminosity distance, as the one discussed so far, is not sufficient to fully describe {the interplay between} different fields. A model-specific analysis is then necessary, and we shall consider one in this subsection. 
One of the best-developed  frameworks that exhibits these extra effects  is {\it bigravity}, where
a massless spin-2 field couples to a massive spin-2 particle in a specifically devised fashion so as to avoid exciting ghostly (i.e. unstable) degrees of freedom. 
We henceforth concentrate on this theory but stress here that
alternative scenarios exist (e.g. dark energy models with vector fields breaking space diffeomorphisms  \cite{Caldwell:2016sut,Caldwell:2018feo}) with  similar phenomenological consequences for the propagation of gravitational waves.

 A consistent theory of bigravity, free of Ostrogradsky instabilities at the fully non-linear level, has been proposed by Hassan and Rosen (HR) \cite{Hassan:2011zd}, adding an Einstein-Hilbert term for the reference metric of the so-called dRGT theory of massive gravity \cite{deRham:2010kj}
(see \cite{Hinterbichler:2011tt,deRham:2014zqa,Schmidt-May:2015vnx} for reviews).~Bigravity admits FRW cosmological solutions describing late-time acceleration that differ from $\Lambda$CDM  at the level of the background as well as for the dynamics of cosmological fluctuations \cite{Volkov:2011an,vonStrauss:2011mq,Comelli:2011zm,Comelli:2012db,Akrami:2012vf,Fasiello:2013woa,DeFelice:2014nja,Lagos:2014lca,Amendola:2015tua,Akrami:2015qga,Kenna-Allison:2018izo,Luben:2018ekw}.  
In this section  we study the evolution of tensor modes around homogeneous FRW configurations, developing techniques that will allow us to go beyond existing literature. We will be specifically concerned with GWs propagation (as opposed to generation) and will therefore be allowed to neglect the non-linearities  crucial in the strong-gravity GWs generation regime (see e.g. \cite{Dar:2018dra} for a recent analysis). Our analysis is focussed on the traceless symmetric part of the two tensor sectors: it will not include the scalar and the (typically decaying) vector degrees of freedom. For other works 
on GW propagation in bigravity, see for example \cite{Cusin:2014psa,Cusin:2015pya,Brax:2017pzt,Brax:2017hxh,Akrami:2018yjz}. 
 
We are going to address the question of how the coupling between different modes can affect the graviton propagation in a regime outside a late time de Sitter era, which is the one usually considered when studying oscillation effects in bigravity \cite{Narikawa:2014fua,Max:2017flc}. This
 question is particularly relevant for the purposes of this work, since GWs emitted from LISA standard sirens at large redshift  can probe phases of cosmological expansion  that are not captured 
  by a  pure de Sitter space approximation. We provide in Appendix \ref{app-bigravity} a brief review of bigravity theory, as well as the equations governing the homogeneous background evolution.
Tensor fluctuations  $h_{ij}^{(1,\,2)}$ are defined around two FRW line elements, associated to each one of the metrics involved:
\begin{eqnarray}
d s^2&=&a^2(\tau) \left(-d \tau^2+d \vec x^2 \right)\,,
\\
d \tilde s^2&=&\omega^2(\tau) \left(-c^2(\tau)\,d \tau^2+d \vec x^2 \right) \label{sec-hom-met}\,,
\end{eqnarray}
where $c(\tau)$ is the speed of the second tensor fluctuation. In what follows the ratio of scale factors is denoted by
\be 
\xi(\tau)\,=\,\frac{\omega(\tau)}{a(\tau)}\,,
\ee
while $\Hc=a'/a$ is the Hubble parameter corresponding to the first metric. 
The evolution of the tensor perturbations in bigravity 
 is described by a coupled system of linearised equations for the two tensor modes $h_1$ and $h_2$ \cite{Comelli:2012db}
\be \label{eq:bigravitymodes}
\lbra \frac{d^2}{d\tau^2} + \bpm 2\Hc & 0 \\ 0 & 2\left(\Hc + \frac{\xi'}{\xi} \right)-\frac{c'}{c}  \epm\frac{d}{d\tau} +\bpm 1 & 0 \\ 0 &  c^2\epm k^2 + m^2a^2f_1\bpm 1 & -1 \\ -\frac{c}{\kappa\xi^2} &  \frac{c}{\kappa\xi^2}\epm\rbra \bpm h_1 \\ h_2 \epm =0 \,, 
\ee
where we have dropped the tensorial indices since the propagation is the same for each transverse-traceless polarisation.
The constant $\kappa$ controls the relative size of the strength of gravitational interactions in the two sectors, while $m$ sets the scale of the bare graviton mass.
 The quantity $f_1(\tau)$ is a cubic function in $\xi(\tau)$ that depends on the bigravity parameters $a_i$:
  \be \label{def-f1}
 f_1(\tau)\,=\, 2\,\xi^2(\tau)\,\left[3\,a_3 \,c(\tau)\,\xi(\tau)+a_2\,(c(\tau)+1)\right]+a_1\,\xi(\tau)\,.
 \ee
The time-dependent coefficients in \eqref{eq:bigravitymodes} are controlled by the background solutions.
We will consider the branch of solutions where scalar and vector modes are not strongly coupled. In this branch, 
the propagation speed of the second tensor $h_2$ is determined by \cite{Comelli:2011zm}
\be \label{eq:bigravity_speed}
c(\tau)-1=\frac{1}{\Hc(\tau)}\frac{\xi'(\tau)}{\xi(\tau)}=\frac{d\ln\xi}{d\ln a}\,,
\ee
and in what follows we assume that matter only couples with the tensor perturbation $h_1$. 
Equation  \eqref{eq:bigravitymodes} implies that the cosmological propagation of GWs in bigravity is characterized by three distinctive effects:
  \begin{itemize}
\item[\emph{(i)}] the two tensor perturbations $h_{1,2}$ propagate at different speeds ($c\neq1$ if $\xi'\neq0$), 
\item[\emph{(ii)}] they have different friction terms, 
\item[\emph{(iii)}] they mix due to the non-diagonal mass matrix.
\end{itemize}
In general, given the time dependence of the parameters in \eqref{eq:bigravitymodes}, the propagation of GWs can not be solved analytically, and
the system of two tensor modes cannot be diagonalized.  On the other hand,  in our case
we can exploit the fact that the frequency of the GW  ($f\sim10^{-2}$Hz in the LISA band) is much larger than the universe expansion rate, $\mathcal{H}_0\sim 10^{-18}$Hz. 
Thus, the time variation of the parameters is  small compared to the frequency of the GW  and we can make use
of a WKB expansion to obtain approximate analytical solutions for the  tensor dynamics. That said, 
we nevertheless emphasize that for the range of redshifts probed with LISA standard sirens, $z\sim1-5$, there is {\it always}  some time dependence
left  in the parameters introduced by the scale factor $a(z)$. This is in contradistinction to LIGO sources at $z\ll1$, for which the
 approximation $a\simeq1$ can be consistently adopted, as  done in previous analysis \cite{Narikawa:2014fua,Max:2017flc}. 

The WKB solution for the system of equations \eqref{eq:bigravitymodes}  is described in Appendix \ref{app:wkb}, and 
is summarised by 
the following expression
\begin{align} \label{eq:h1sch}
a(\tau)\,h_1(\tau)&=\left[
 c_1\bphi_1(\tau)+c_2 \eM_{12}(\tau)\bphi_2(\tau)e^{i\int\delta\theta(\tau) d\tau}\right] e^{i\int\theta_1(\tau) d\tau}\,, \\
a(\tau)\,h_2(\tau)&=\left[
 c_1\eM_{21}(\tau)\bphi_1(\tau)e^{-i\int\delta\theta(\tau) d\tau}+c_2\bphi_2(\tau)\right] e^{i\int\theta_2 (\tau) d\tau}\,. \label{eq:h2sch}
\end{align}
The quantities
 $\theta_{1,2}$  
 and $\eM$ correspond respectively to the  eigenvalues and to the matrix of eigenvectors that would diagonalise
the system in the approximation in which the parameters in eq \eqref{eq:bigravitymodes} are constant. $\bphi_{1,2}$ denote  components of a vector controlling the  mode amplitudes. The quantities $c_{1,2}$ are constant fixed by initial conditions while $\delta\theta\equiv\theta_2-\theta_1$.\\ We relegate  technical details on these quantities and how to derive  \eqst{eq:h1sch}{eq:h2sch} to Appendix \ref{app:wkb}. Here we  emphasize that the possible mixing between $h_1$ and $h_2$ is controlled by the non-diagonal elements of the matrix of eigenvectors, i.e. $\eM_{12}$ and $\eM_{21}$. These entries are  non-vanishing whenever the mass matrix in \eqref{eq:bigravitymodes} is non-diagonal. On the other hand, we find that the size of the mixing is controlled by the relative difference of the velocities of each mode.
In particular, for the mixing not to be suppressed in the regime of large $k$, one needs to require
\be \label{eq:bigravityscales}
\left( c^2-1\right)\, k^2\lesssim m^2a^2\,,
\ee
which follows directly from the analytic expression of $\eM$ that can be found in Appendix~\ref{app:wkb}. If this inequality is satisfied one can have large mixings among modes with interesting phenomenological consequences, such as graviton oscillations. 
In Appendix \ref{app:wkb} we show that in the small mass regime, $m \sim\Hc_0$, this  inequality cannot be satisfied for viable cosmological scenarios. Hence, 
for the rest of this Section we focus on a large mass limit $m\gg\Hc_0$. In this regime $\xi$ is
approximately constant and, as a consequence, $c\simeq1$ (recall (\ref{eq:bigravity_speed})). In particular, we find  
\be
(c^2-1)\lesssim(\Hc^2-\mathcal{H}_0^2\Omega_{\Lambda})/(m^2a^2)\,,
\ee
where $\Omega_\Lambda$ is the density of DE. As a consistency check, we see that in the pure de Sitter limit, the speed is exactly luminal, $c_{dS}=1$. 
 The condition to have mode mixing, eq \eqref{eq:bigravityscales}, is satisfied for
\be \label{larmar}
m^4a^4\gtrsim k^2(\Hc^2-\mathcal{H}_0^2\Omega_{\Lambda})\,.
\ee
For LISA frequencies this leads to  a bound  $m \gtrsim10^8\Hc_0$, which improves by a couple of orders of magnitude the LIGO detectability range $m\gtrsim10^{11}\Hc_0$ \cite{Max:2017flc} (recall that $\Hc_0\sim10^{-33}$eV).  We stress here that in the large mass regime the evolution of the universe cannot reproduce the observed accelerated expansion unless we include an
 {effective} cosmological constant term compensating the large mass in the Friedmann equation (see Appendix \ref{app-bigravity}). This means that, by restricting the analysis to the large mass regime,  the present work does not fully probe bigravity: it does not capture the region of parameter space where $m\sim \Hc_0$, which is the most propitious for self-acceleration.
On the other hand, our analysis can also serve as a proxy for other scenarios supporting oscillations in the luminosity distance at late time. A case in point are models with vector gauge fields \cite{Caldwell:2016sut,Caldwell:2018feo}.

\smallskip

%-FIGURE-hS_oscillations_k_5.e-3_mg_2.e-24_tg_1-8
\begin{figure}[t]
\centering 
 \includegraphics[width=0.49\textwidth]{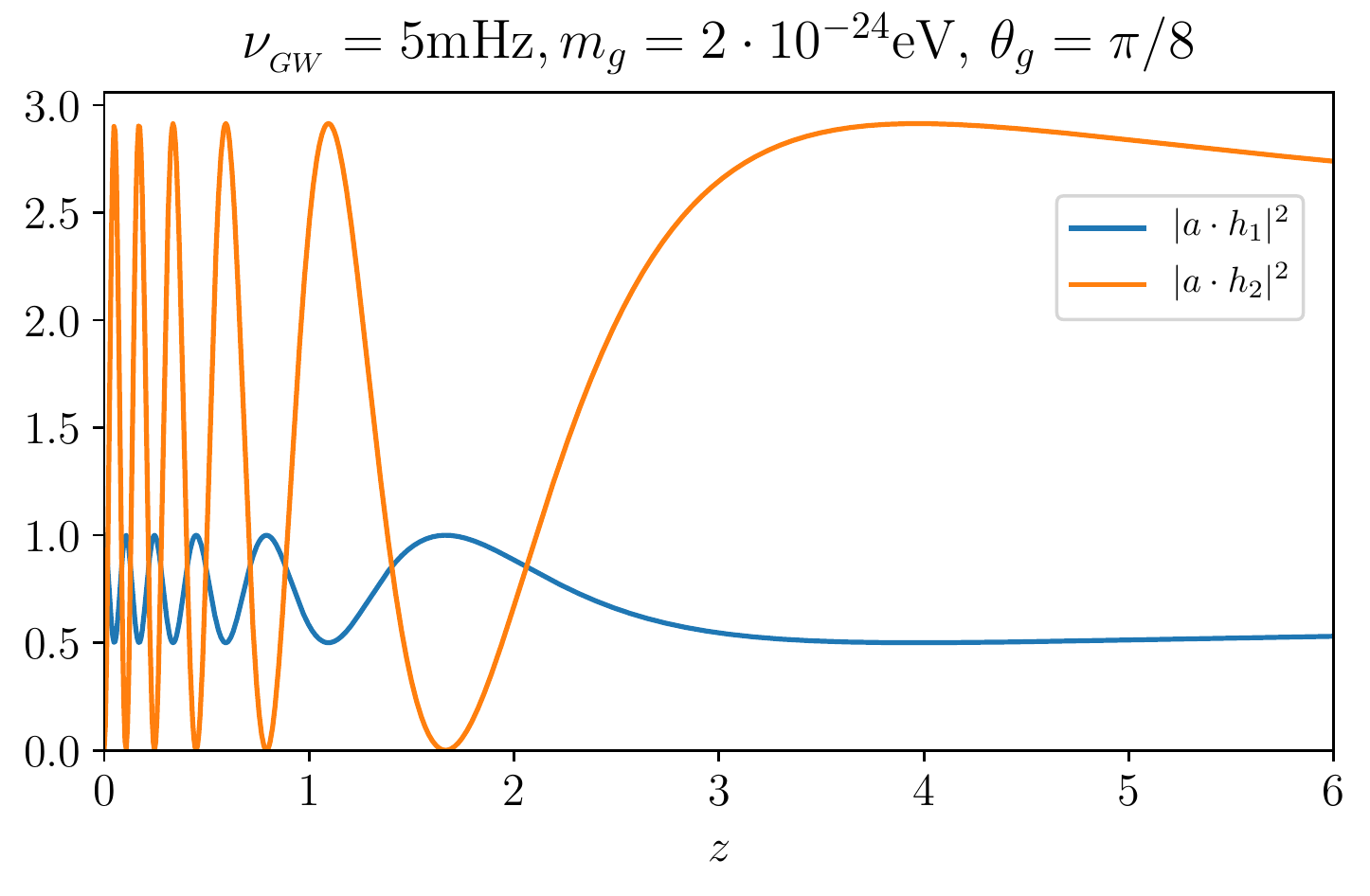}
  \includegraphics[width=0.49\textwidth]{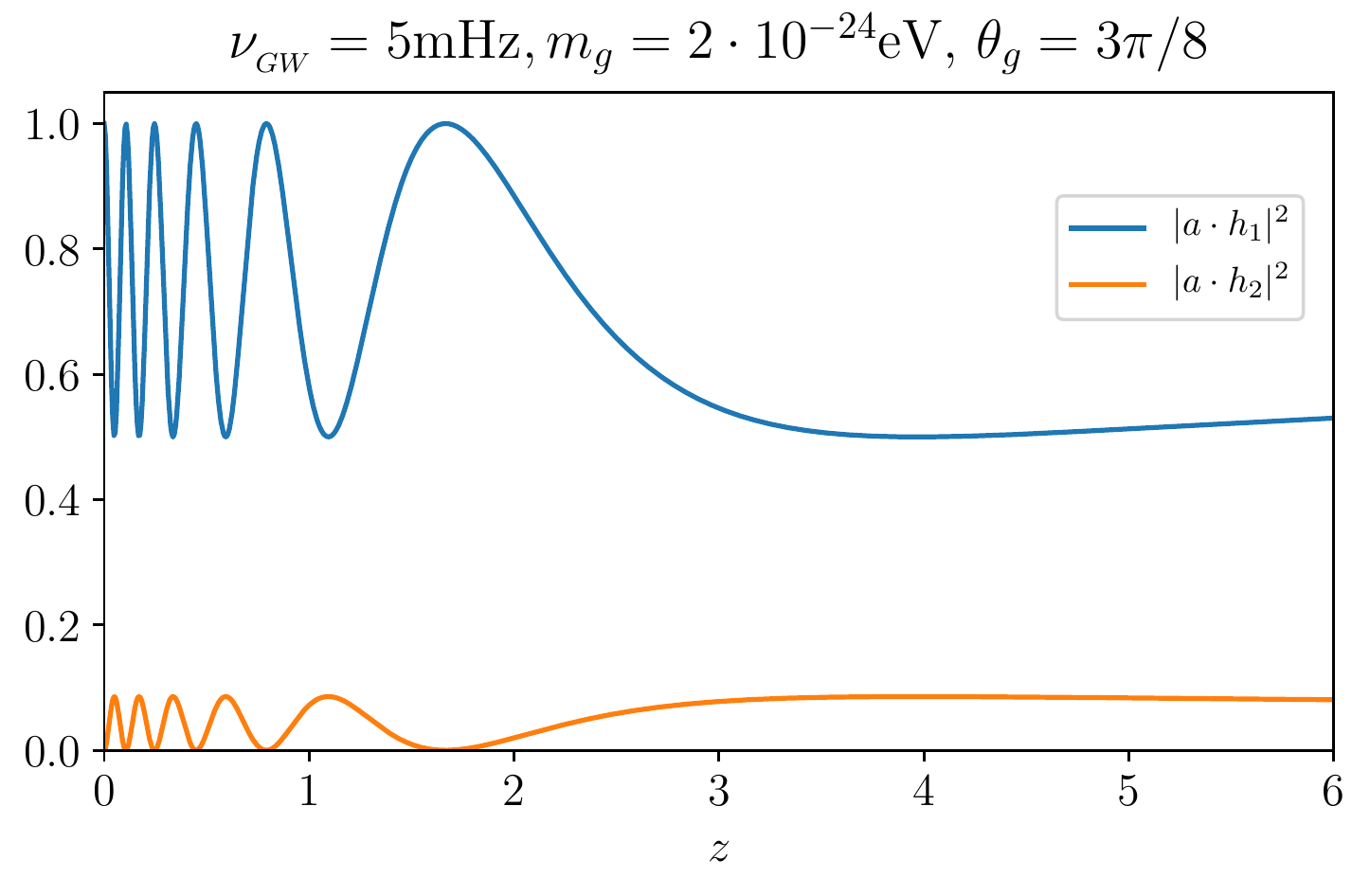}
 \caption{Amplitude of the tensor perturbations $h_1$ and $h_2$ for a GW emitted with frequency $\nu_{_{GW}}=5$mHz. The left and right panels correspond to different mixing angles for the same mass $m_g=2\cdot10^{-24}$eV and we have normalised the initial amplitude of $h_1$ to 1. See eq \eqref{defemma} for the definitions of $m_g$ and $\theta_g$. }
 \label{fig:bigravity_mixing}
\end{figure}
%------------

We henceforth restrict the analysis to the regime of \eqref{larmar} and investigate its phenomenological consequences. It is convenient to define an effective mass $m_g$ and a mixing angle~$\tg$:
\be \label{defemma}
m_g^2 \equiv m^2f_1\lpar \frac{1}{\kappa\,\xi^2} + 1 \rpar\,\quad \text{and} \quad \tg\equiv \tan^{-1}\lbra\sqrt{\kappa\,\xi^2}\rbra\,.
\ee 
Whenever $\left(c^2-1\right)\, k^2<m_g^2a^2$, one can expand the phases associated to each tensor mode as
\begin{align} 
\theta_1^2&=k^2\lpar1+\frac{\left(c^2-1\right)\kappa\,\xi^{2}}{1+\kappa\xi^2}\rpar-\Hc^2+\mathcal{O}\lpar\frac{\,\left(c^2-1\right)^2\,k^4}{m_g^4a^4}\rpar\,, \\
\theta_2^2&=k^2\lpar1+\frac{\,\left(c^2-1\right)\,}{1+\kappa\xi^2}\rpar+m_g^2a^2\lpar1+\frac{c^2-1}{2(1+\kappa\xi^2)}\rpar-\Hc^2+\mathcal{O}\lpar\frac{\,\left(c^2-1\right)^2\,k^4}{m_g^4a^4}\rpar\,.
\end{align}
Analogously, the eigenvectors simplify to
\be \label{eq:eigenvectors}
\eM=\bpm 1 & -\kappa\xi^2\lpar1-\lpar\frac{\left(c^2-1\right)k^2}{m_g^2a^2}+\frac{\left(c^2-1\right)}{2}\rpar\rpar \\ 1-\frac{\left(c^2-1\right) k^2}{m_g^2a^2} & 1\epm\,+\mathcal{O}\lpar\frac{\,\left(c^2-1\right)^2\,k^4}{m_g^4a^4}\rpar \,. 
\ee
In order to estimate how large is the correction to the GW amplitude w.r.t. the $c=1$ case, we can parametrize the effective mass with a dimensionless constant $\beta$ via $(m_g\,a)\sim\beta\cdot(k\,\Hc_0)^{1/2}$, which controls our complying with the large mass regime defined in (\ref{larmar}). It follows that the largest correction to the matrix of eigenvectors (\ref{eq:eigenvectors}) scales with $\beta^{-4}$. Therefore, provided that $\beta\gtrsim5$, we can neglect this correction in the amplitude. 

The mixing among modes affect the tensor speed and the luminosity distance.
The modification in the propagation speed of the lightest tensor $h_1$, defined as $\alpha_T\,\equiv\,c^2_{T}-1$, is given by
\be
\alpha_T=\frac{\left(c^2-1\right)\,\kappa\xi^{2}}{(1+\kappa\xi^2)}\,,
\ee 
and  scales as $\alpha_T\sim\beta^{-2}(k/\Hc_0)^{-2}$. 
For LISA, $\alpha_T$ is smaller than $10^{-16}$ in the large mass limit. However, it might still be observable given that this is a cumulative effect over long travel distances. A prompt EM counterpart can give constraints of the order of 

%-FIGURE-
\begin{figure}[t]
\centering 
 \includegraphics[width=0.49\textwidth]{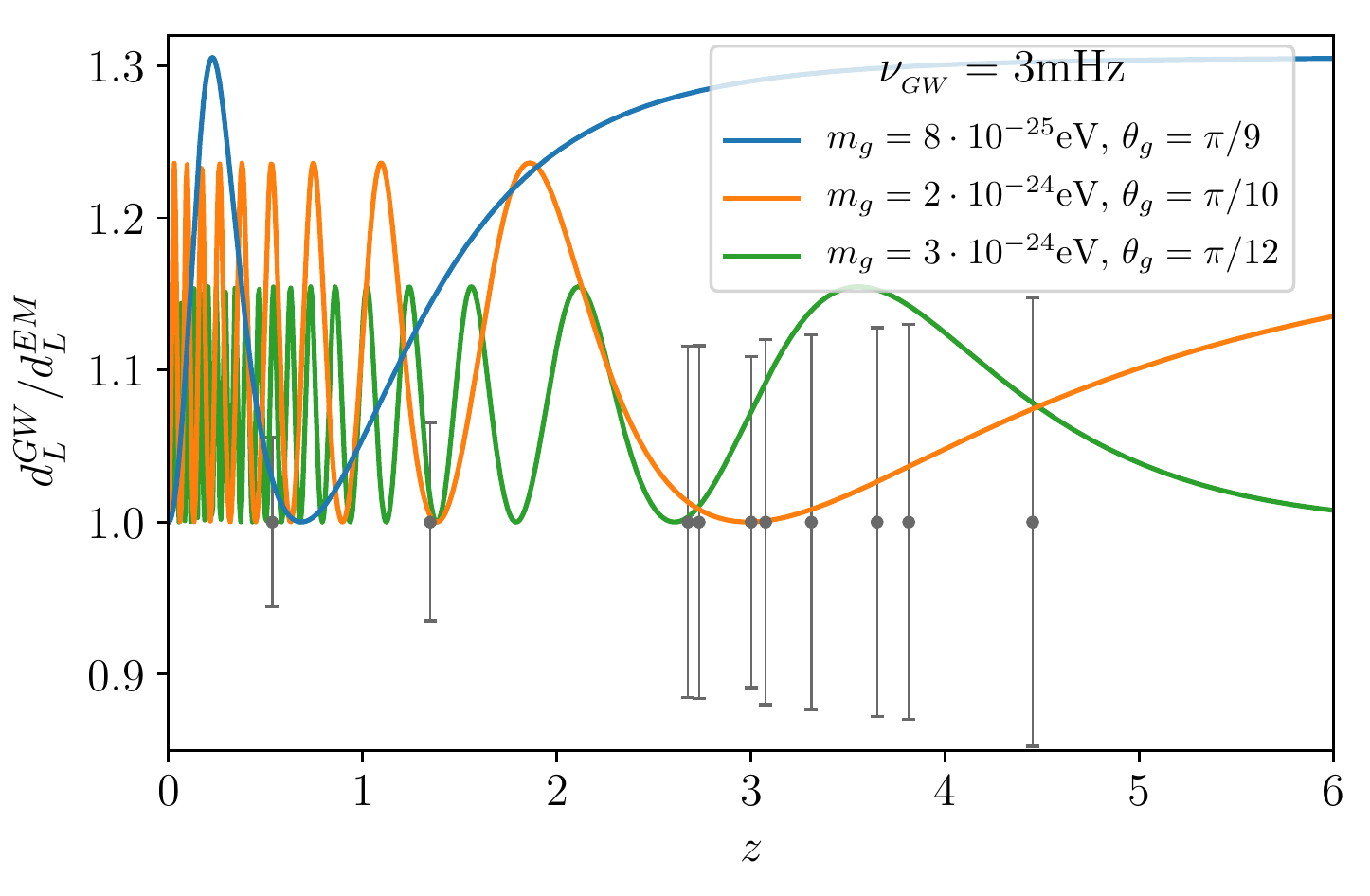}
  \includegraphics[width=0.49\textwidth]{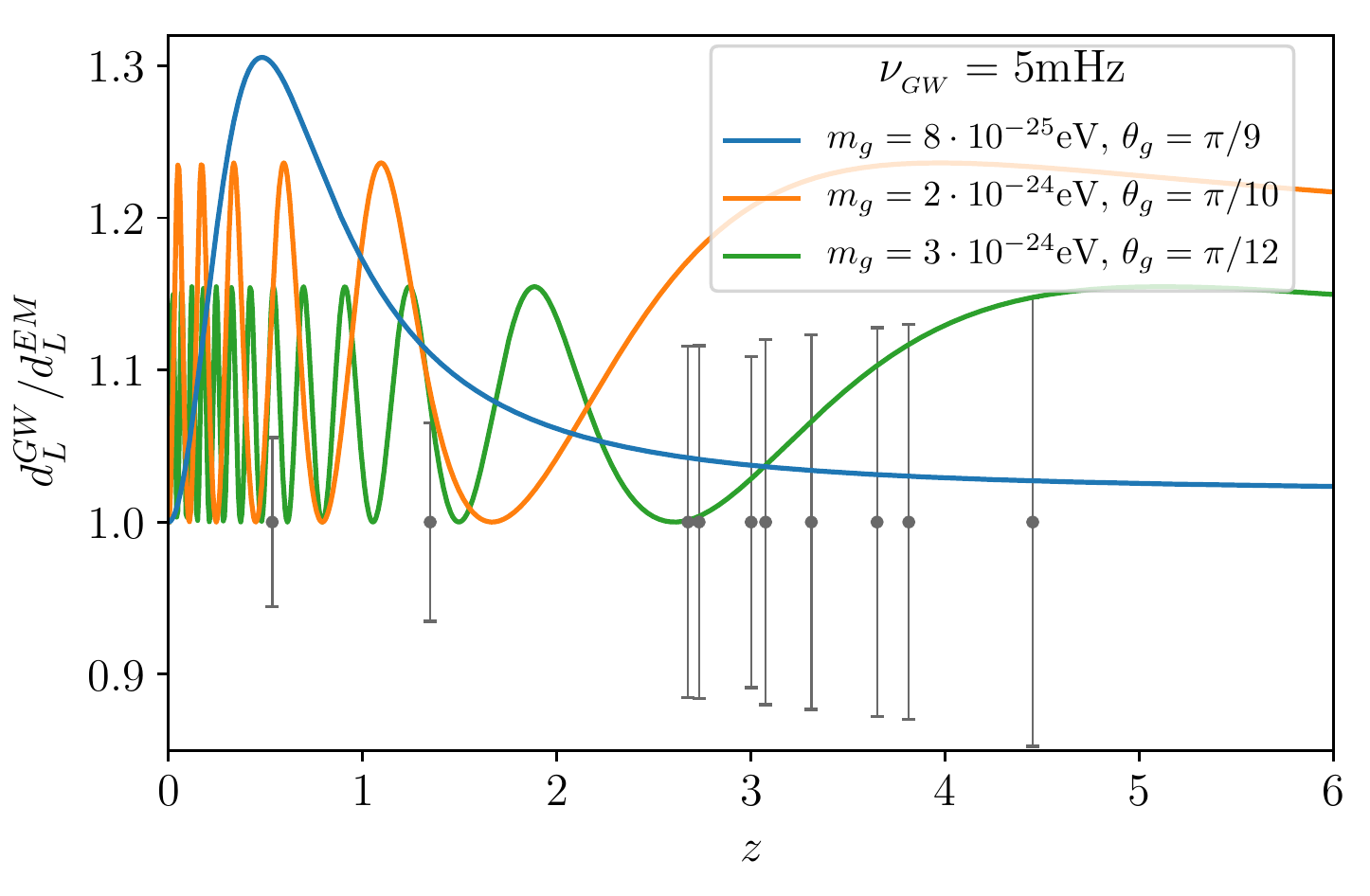}
 \caption{GW luminosity distance in the large mass limit of bigravity  for different values of $m_g$ and $\theta_g$. To visualize the range of masses and mixing angles that could be probed with LISA, we have included the errors of a representative catalog used in the analysis (without being scattered). In particular, we use
 the catalog  {\it hQ3} of the scenario 1 to be discussed in Section \ref{sect:catalogs}.}
 \label{fig:bigravity_dl}
\end{figure}
%---------

\be
\alpha_T\lesssim 2\times10^{-17}\lpar\frac{10\text{Gpc}}{D}\rpar\lpar\frac{\Delta t}{10\text{s}}\rpar\,,
\ee
where $\Delta t$ is the difference in the time of arrival and $D$ the distance to the source. 

Next, we study how  the mixing of the tensor modes $h_{1,2}$ can leave an imprint in the GW luminosity distance. Focussing
 on the regime $(m_g\,a)\gtrsim5\cdot(k\,\Hc_0)^{1/2}$, the only time dependence in the amplitude is introduced by the scale factor $a(\tau)$. 
Using the definitions \eqref{defemma} of the effective mass $m_g$ and the mixing angle $\theta_g$, the amplitude of the tensor component 
$h_1$ is
\be \label{eq:bigravityh1}
\vert a\,h_1\vert^2=h_0^2 \cos^4\tg \lpar1+\frac{\tan^4\tg}{\theta_2/\theta_2(\tau_e)}+\frac{2\tan^2\tg}{\sqrt{\theta_2/\theta_2(\tau_e)}}\cos\lbra\int_{\tau_e}^\tau\delta\theta d\tau'\rbra\rpar\,,
\ee
where $h_0$ is the amplitude in the $\theta_g=0$ case,
quantities depend on the time $\tau$, and $\tau_e$ is the time of emission. 
Also, $\theta_1^2=k^2-\Hc^2$, $\theta_2^2=k^2+m_g^2a^2-\Hc^2$,  $\delta\theta= m_g^2a^2/(2k)+\mathcal{O}(k^{-3})$. 
As a result, if $m_g^2a^2\sim k\Hc_0$, the oscillation frequency is of order $\Hc_0$. A similar expression can be obtained for $\vert a\,h_2\vert$. 
One should note that while $\vert a\,h_1\vert$ is never larger than the initial value $h_0$, for $\tg<\pi/4$, $h_2^2$ can exceed $h_0^2$ (up to $4h_0^2$). In the opposite limit, $\tg>\pi/4$, $h_2^2$ is always less than $h_0^2$. This could be observed by comparing the left and right panels of Fig. \ref{fig:bigravity_mixing}, where the mass is fixed but the mixing angles vary.
 
We can now compare the luminosity distance of GWs in bigravity with the one of EM radiation, $d_L^\text{gw}/d_L^\text{em}$. We  focus on the amplitude of the lightest tensor mode, $h_1$, which is the one we assume to be coupled to matter. 
%\be \label{eq:bigravitydLformal}
%\frac{d_L^\text{gw}}{d_L^\text{em}}=\frac{1}{\vert a\,h_1\vert}\,.
%\ee
From \eqref{eq:bigravityh1}, we obtain
\be \label{ratio-bigr}
\frac{d_L^\text{gw}}{d_L^\text{em}}\simeq\frac{1}{\cos^2\tg}\lpar1+\frac{\tan^4\tg}{\theta_2/\theta_2(\tau_e)}+\frac{2\tan^2\tg}{\sqrt{\theta_2/\theta_2(\tau_e)}}\cos\lbra\int_{\tau_e}^\tau \frac{m_g^2a^2}{2k}d\tau'\rbra\rpar^{-1/2}\,.
\ee
We see that this ratio can  become larger than one and display oscillatory patterns. Possible configurations range {in principle}\footnote{It is actually not possible to push the theory to the asymptotic regions for the following reasons: (i) $\kappa \rightarrow 0$ corresponds to a vanishing kinetic term for the second metric so that it becomes infinitely strongly coupled; (ii) taking the limit $\kappa \rightarrow \infty$, the second metric decouples and gets effectively frozen to a fixed background value.}
between $\tg=0$ (corresponding to $\kappa\xi^2=0$) and $\tg=\pi/2$ (corresponding to $\kappa\xi^2\rightarrow\infty$), and the maximum mixing occurs at $\tg=\pi/4$ ($\kappa\xi^2=1$). Moreover, in the limit in which $\theta_2/\theta_2(t_e)=1$ (which is a good approximation in the high $k$ limit), $d_L^\text{gw}/d_L^\text{em}$ is symmetric around $\pi/4$, i.e. $h_1^2(\pi/4-\varphi)=h_1^2(\pi/4+\varphi)$.  

%FIGURE
\begin{figure}[t]
\centering 
 \includegraphics[width=0.6\textwidth]{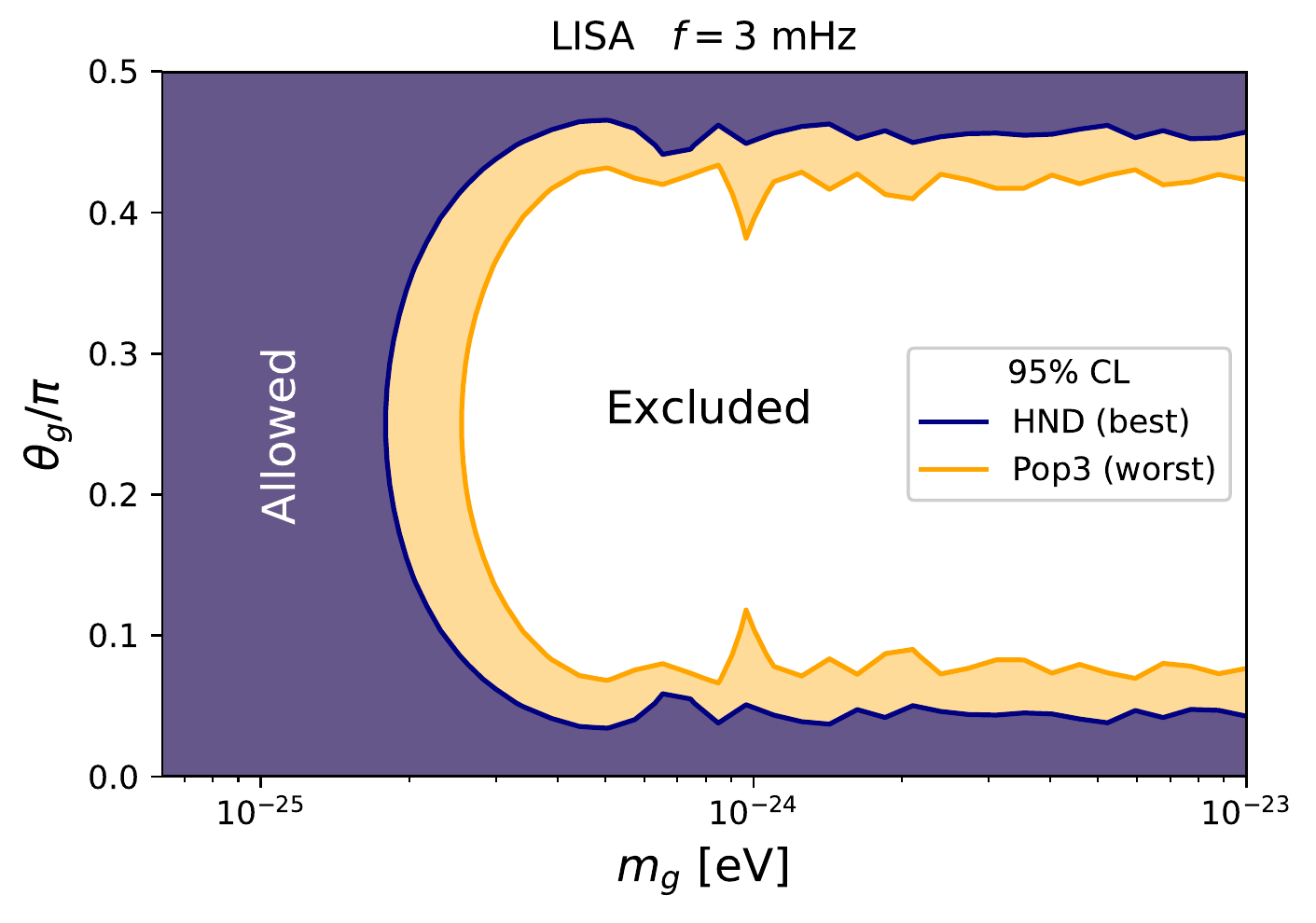}
 \caption{Expected constraints on GW oscillations {(mass and mixing angle)} in bigravity for the three different LISA catalogues in scenario 1 (see Section \ref{sect:catalogs}). The bounds assume the high-mass prediction (\ref{ratio-bigr}), fixed cosmology and a fiducial observation frequency. {The contours show the best case scenario (blue: heavy seed no delay, optimistic errors) and the worst case scenario (yellow: Pop3 seed, pessimistic errors), see Section \ref{sect:catalogs} for details.}}
 \label{fig:bigravity_bounds}
\end{figure}
%-------

Given that the ratio  \eqref{ratio-bigr}  oscillates as a function of redshift, 
  the two-parameter phenomenological parametrisation
of Section \ref{sec:general} is not expected to perform well in this case, and an  analysis specific
for this model is necessary.  
We plot the oscillatory behavior  of the GW luminosity distance for different masses and mixing angles in Fig. \ref{fig:bigravity_dl}. For a  qualitative analysis, we include  error bars in the measurements of luminosity distances  for a representative LISA catalogue (which we will discuss and use in Sections \ref{sect:catalogs} and \ref{sect:MCMC}).

Figure \ref{fig:bigravity_bounds} shows the projected constraints on GW oscillations in bigravity theories on the parameters $m_g$ and $\theta_g$.
The excluded regions have been obtained comparing the high-mass prediction for the luminosity distance ratio (\ref{ratio-bigr}) with several LISA catalogs, assuming fixed cosmology, and performing a $\chi^2$ analysis.
Oscillation effects can be observed for masses $m_g\gtrsim 2\cdot 10^{-25}\,$eV. The amplitude of the oscillation, and thus its detectability, increases as $\tg$ approaches to $\pi/4$ from above or below. 
The mixing angle range where oscillations can be detected is $0.05\pi \lesssim \theta_g \lesssim 0.45\pi$.
LISA will provide a $\sim 3$ order of magnitude improvement in mass sensitivity over the current LIGO limit, which probes $m_g \gtrsim 10^{-22}$eV \cite{Max:2017flc}, due to the larger oscillation baseline and the lower detection frequency. We conclude that the use of standard sirens will strengthen existing bounds towards smaller values for the mass. 

However, one should note that these bounds are based on standard sirens with a single fiducial frequency, i.e. a monochromatic GW. Since GW oscillation effects depend strongly on the frequency, a coallescing binary would experience a time-dependent modulation of the amplitude as the orbital frequency increases, leading to a distinct signal. 
Remarkably, no electromagnetic counterparts are necessary to study such effects. As a representative example, we plot in Fig. \ref{fig:gw_oscillations} the strain of a GW signal from a massive BH binary as a function of frequency and its modification in bigravity. The GW oscillations lead to a distinctive frequency profile, which would be interesting to further
characterise in order  to establish to what extent it can be probed with LISA. Moreover, the initial wave packet emitted might decohere while traveling. In that case the event would be followed by an ``echo'' signal, which may be detectable if the mixing is sufficiently large \cite{Max:2017kdc}.
The inclusion of decoherence effects and frequency dependence would provide further means to test GW oscillations. We leave this to future work.

%FIGURE
\begin{figure}[t]
\centering 
 \includegraphics[width=0.6\textwidth]{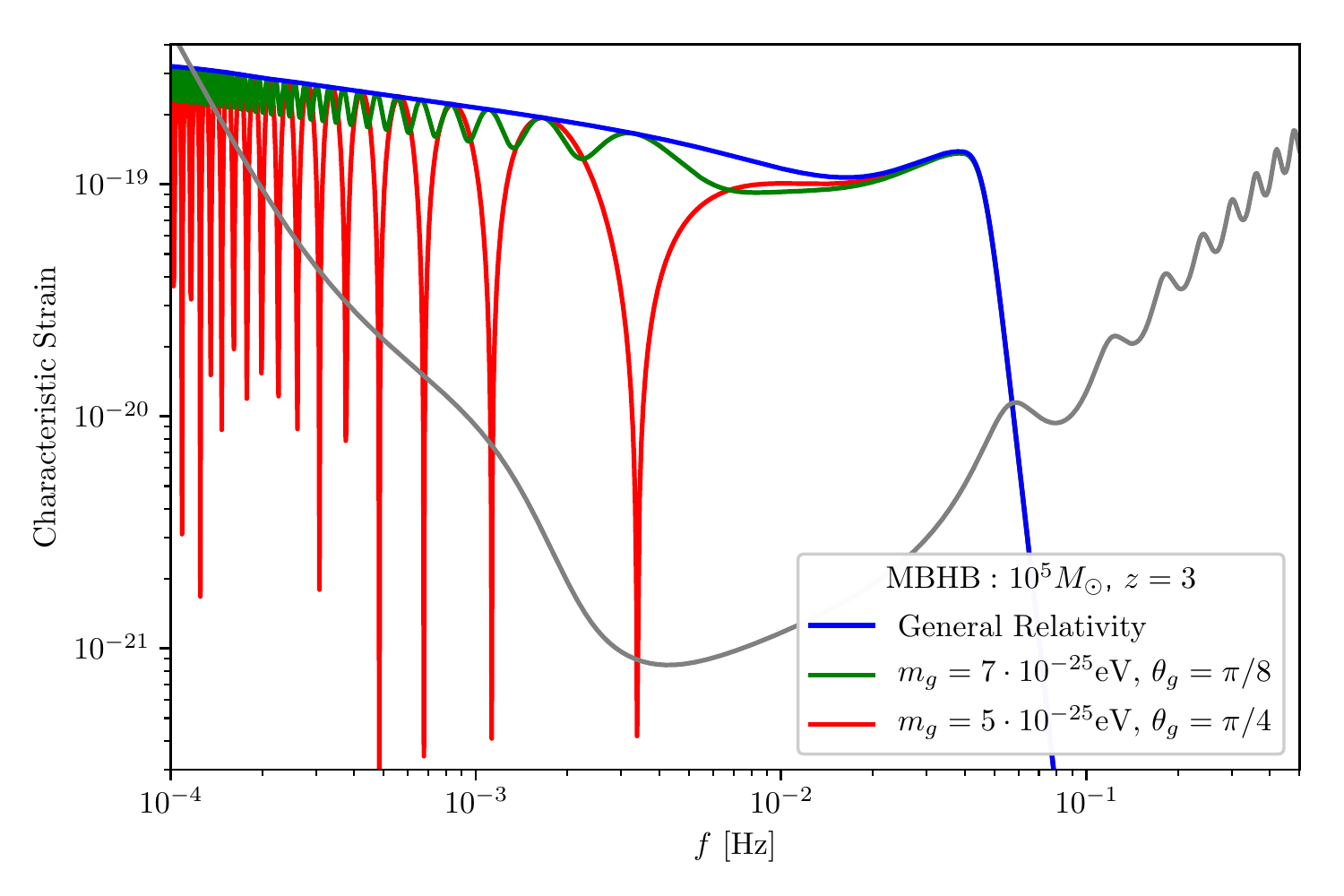}
 \caption{ 
GW strain versus frequency in bigravity for a signal from a massive BH binary (MBHB) merger event detectable with LISA (in grey the LISA sensitivity curve). We compare the signal strain for two   bigravity sets of parameters (fixed mass, changing mixing angle) with respect to GR. Notice the characteristic oscillations in frequency of the GW strain for this theory.}
 \label{fig:gw_oscillations}
\end{figure}
%------

\subsection{Models with extra and varying dimensions}\label{sec:QG}

Various models of modified gravity predict that gravitational waves
propagate in more than four dimensions, while photons and the other
Standard Model fields are localized on a four-dimensional
brane. Similar considerations apply to models of quantum gravity,
where spacetimes with dimensions different from four often enter in
theoretical attempts to find quantum theories of gravitational
interactions.

Examples of modified gravity scenarios are models in the DGP class
\cite{Dvali:2000hr}, where gravitational leakage into an extra fifth
dimension has been proposed as a way to explain dark energy, since
gravity becomes weaker at very large cosmological scales.  In theories
involving dimensions different from four, the GW propagation and the
luminosity distance relation can differ from GR, and it is interesting
to estimate whether GW observations can constrain such systems.

In \cite{Nishizawa:2017nef,Pardo:2018ipy,Abbott:2018lct}, constraints
have been imposed on the parameters entering the following
phenomenological expression for the luminosity distance
\cite{Deffayet:2007kf}, suitable for modified gravity models with
extra dimensions: \be\label{extradim}
%h \propto \frac{1}{\tilde d_L}\,,\qquad \tilde
\frac{d_L^{\rm gw}}{d_L^{\rm em}}\,=\,\left[1+\left(\frac{d_L^{\rm
      em}}{R_{\rm c}}\right)^{n_{\rm c}} \right]^{\frac{D-4}{2n_{\rm
      c}}}, \ee with $R_{\rm c}$ is the characteristic length scale
beyond which propagating GWs start to leak into the extra dimensions
and $n_{\rm c}>0$ is a steepness parameter denoting the transition
between GR and the onset of gravitational leakage. For sources
emitting GWs accompanied by an electromagnetic counterpart, one can
constrain large extra dimensions by comparing the distance inferred
from the GW signal with the one inferred from the electromagnetic one
\cite{Nishizawa:2017nef,Pardo:2018ipy,Abbott:2018lct}.  In particular,
using the GW170817 detection, the LIGO-Virgo collaboration
\cite{Abbott:2018lct} has recently set $90\%$ upper bounds on the
spacetime topological dimension $D$, assuming fixed transition
steepness and distance scale. For $R_{\rm c}$ between 1 and 20 Mpc,
$4\leq D<4.5$. Conversely, fixing $D=5,6,7$, a transition steepness
$n_{\rm c}\lesssim 10^{-1}$ is ruled out, while the $10\%$ lower limit
on $R_{\rm c}$ is about 20 Mpc for $n_{\rm c}>10$ and between 20 and
$10^4$ Mpc for $0.5<n_{\rm c}<10$.

Analogous constraints for quantum gravity (QG) can also be
investigated. Quantization of the gravitational force intimately
affects the dimensionality of spacetime. In fact, in QG approaches the
effective dimension of spacetime is not a fixed external parameter,
but a dynamical and scale-dependent one, which can change to a value
typically smaller (in some theories, larger) than four at very small
distances or very high energies \cite{tH93,Car17}. This
\emph{dimensional flow} \cite{Calcagni:2009kc} smoothly spans all
scales of the theory, from ultra-short to large scales where a
semi-classical, near-GR limit is reached. At an effective level, in a
regime where spacetime emerges as a continuum, the dimensional flow is
determined by the way volumes scale in position and momentum spaces
and by the dispersion relations associated with effective kinetic
operators. In turn, these features are encoded into the topological
dimension $D$ of spacetime ($D=4$ in physical models) and three
effective, scale-dependent ``dimensions,'' namely the Hausdorff
(called also fractal) dimension $\dh$ of position space, the Hausdorff
dimension $\dh^k$ of momentum space and the spectral dimension
$\ds$. In a classical spacetime, $\dh=\dh^k=\ds=D$. A general argument
based on the scaling of fields and their kinetic terms indicates that
correlation functions and distances depend on these dimensions through
a geometric parameter $\Gamma:=\dh/2-\dh^k/\ds$ that combines them
together \cite{Calcagni:2019kzo}. The relation between $\Gamma$ and the
parameter $\delta$ that expresses the modification of GW propagation
(and that, as we have seen in section~\ref{sect:timevarMpl}, in some
theories is related to the time variation of the effective Planck
mass) is \be\label{Gammapar}
\Gamma=-\delta+\frac{D-2}{2}\,, \ee
while the correction to the GW luminosity distance in a spacetime
characterized by dimensional flow is \be\label{dla}
%h \propto \frac{1}{\tilde d_L}\,,\qquad 
\frac{d_L^{\rm gw}}{d_L^{\rm em}}=1+\ve\left(\frac{d_L^{\rm
    em}}{\ell_*}\right)^{\g-1}, \ee where $\gamma\neq0$ is the value of the
parameter $\Gamma$ at small scales, $\ell_*$ denotes a length scale at
which quantum gravity effects can no longer be neglected  (hence it is of
order of the
Planck scale $\lp$) and, depending on the model, $\ve$ is
either a number $O(\g-1)$ or a random variable with zero average and
unit variance simulating an intrinsic spacetime uncertainty, i.e., a
geometry with microscopic fuzziness. Note that \Eq{dla} only relies
upon two assumptions. First, that there exists a continuum spacetime
limit where the position-space measure, the momentum-space measure and
the kinetic term of the tensor mode are deformed in a fairly generic
way. Second, that these deformations are such that
$\ds\neq0$. Theories where the spectral dimension vanishes in the UV,
such as quantum gravity with non-local UV modifications and the
low-energy limit of string field theory, must be treated separately
(and do not give rise to observable corrections)
\cite{Briscese:2019rii}.

The power-law correction in \Eq{dla} of quantum-gravity origin has a
strong resemblance with the similar expression \eqref{extradim} we
discussed above for modified gravity scenarios. Comparing \Eq{dla}
with \Eq{extradim}, we can match the parameters as follows. When the
scales $\ell_*$ and $R_{\rm c}$ are considered as free parameters,
they play the same role. However, in quantum gravity $\ell_*$ is
expected to be very small or even Planckian, which implies that the
power-law correction will be small when $\g-1<0$, large when $\g-1>0$
and $O(1)$ when $\g-1$ is a small positive number. Depending on the
theory and the scales at which \Eq{dla} holds, $\g$ can take very
different values, being either negative, zero, or positive.

In most scenarios, such as string theory in the low-energy limit,
group field theory, spin foams, loop quantum gravity, causal dynamical
triangulations, asymptotic safety, Stelle gravity,
Horava--Lifshitz gravity and various realizations of
$\kappa$-Minkowski spacetime, $\g<1$ in the deep UV and the correction
in \Eq{dla} is Planck-suppressed and negligible, as one would expect
from a generic reasoning based on perturbative effective field
theory. However, in some QG proposals based on discrete pre-geometries
(such as group field theory, spin foams and loop quantum gravity),
there is a possibility that $\g-1$ is a small positive number at
mesoscopic (far from the UV-regime) or near-IR scales for spacetimes
described by certain semi-classical states. Hence, in these cases the
correction in \Eq{dla} may be non-negligible, even if $\ell_*$ is
smaller by many orders of magnitude than any cosmological distance.

In analogy with the extra-dimensional case \Eq{extradim} studied in
\cite{Abbott:2018lct}, one can expect that LIGO-Virgo data, and later
LISA data, constrain the free parameters ($\ell_*, \g$) of \Eq{dla},
providing a test for QG proposals. This is indeed the case. The
model-independent analysis of \cite{Calcagni:2019kzo} shows that theories with
$\ell_*=O(\lp)$ and \be\label{gminus1} \g-1\gtrsim 0.02 \ee can fall
into the detectability range of interferometers. Such small non-zero
values of $\g-1$ can be obtained only in a near-IR regime. Most QG
models do not achieve this (because, in this regime, either $\g-1<0$
or $\g-1$  {\gt is positive but too close to zero}) and do not affect the
luminosity distance. However, QG proposals where spacetime emerges
from a discrete  structure (group field theory, spin
foams and loop quantum gravity), where $0<\g-1\ll 1$ ($\g$ close to
1), may leave an observable imprint depending on the considered quantum state of
geometry (which can be described in terms of geometric indicators,
such as the Hausdorff and spectral dimensions). In this
latter class of QG models, the spacetime dimension turns out to be
highly state-dependent, and hence the effective length scale
characterising IR nearly-classical geometries may be several orders of
magnitude above the Planck length.  {\gt
Whether such quantum states are realistic and viable within the corresponding theoretical framework remains an open question and a potential subject of future study.}

 It is worth noting that the quantum correction in
\Eq{dla} is not the same as one could find, for instance, in
particle-physics or atomic observables, where constraints on new
physics are tight to an extreme. In most theories of quantum gravity,
these corrections are actually not present, since only the
gravitational sector of the dynamics is affected by non-perturbative
phenomena such as the dimensional flow. This is indeed the case for
two of the three QG  proposals that support the parameter space
\Eq{gminus1}, namely spin foams and loop quantum gravity. In these two
models there is no modification in  the Standard Model of particle
physics, nor in any fundamental constant of nature. For
group field theory though, this is less obvious.
However, for theories where
particle and atomic physics are also affected, corrections are still of a
power-law type $(\ell/\ell_*)^\a$ but, typically, with a power $\a$
different from the special combination of dimensions \Eq{Gammapar}
\cite{Calcagni:2019kzo}. Therefore, GW and particle-physics bounds are (at least
partially) independent, leading to complementary constraints to the fundamental theories.
On the other hand, solar-system tests can yield stronger bounds on $\gamma$ in QG scenarios where also the scalar sector is affected \cite{Calcagni:2019kzo}. However, in many QGs models the correction to the Newton potential is not known, while the spin-2 sector is under better control and GWs can be a robust and, for the time being, unique way to constrain such models \cite{Calcagni:2019kzo}.

To conclude, we comment on the parametrization \Eq{eq:param} for the
systems considered in this subsection. For models that predict
relations such as (\ref{extradim}) and \Eq{dla}, the parametrization
(\ref{eq:param}) is no longer appropriate. Indeed, \Eq{eq:param}
yields $ d_L^{\rm gw}(z)/d_L^{\rm em}(z)\simeq 1+O(z)$ at small
redshift, while \Eq{extradim}, for $D>4$ and $d_L^{\rm em}\ll R_{\rm c}$,
gives \be\label{extradimlowz} \frac{d_L^{\rm
    gw}(z)}{d_L^{\rm em}(z)}\simeq 1+\frac{D-4}{2n_{\rm
    c}}\left(\frac{z}{H_0R_{\rm c} }\right)^{n_{\rm c}}, \ee which is
of the form $1+{\cal O}(z)$ only for $n_{\rm c}=1$. On the other hand,
assuming that the electromagnetic sector is not modified, \Eq{dla}
gives \be\label{qglowz} \frac{d_L^{\rm gw}(z)}{d_L^{\rm em}(z)}\simeq
1+\ve\left(\frac{z}{H_0\ell_*}\right)^{\g-1}, \ee which is of the form
$1+{\cal O}(z)$ only for $\g=2$, valid only in two specific models
which leave no imprint on the luminosity distance
\cite{Calcagni:2019kzo}. Moreover, quite importantly,  the ratio $d_L^{\rm gw}/d_L^{\rm em}$ in
\Eq{extradim} and \Eq{dla} does not tend to a constant at large
redshifts -- which is the interesting range
for the LISA Standard Sirens  that we consider in this work -- hence it can not be reproduced well with the parametrization \Eq{eq:param} in the high-$z$ regime.

A possible parametrization inspired by the low-$z$ behaviour
(\ref{extradimlowz}) and (\ref{qglowz}), but such that $d_L^{\rm
  gw}/d_L^{\rm em} \simeq {\rm const}$   at large redshift, is given by
\Eq{param3}. At large $z$, this tends to the constant value $\Xi_0$,
while in the low-$z$ limit it goes to \be \frac{d_L^{\rm
    gw}(z)}{d_L^{\rm em}(z)}\simeq 1+(\Xi_0-1)z^n\,, \ee reproducing
\Eq{extradimlowz} when $\Xi_0=1+(D-4)(H_0R_{\rm c})^{-n_{\rm
    c}}/(2n_{\rm c})$ and $n=n_{\rm c}$, and \Eq{dla} when
$\Xi_0=1+\ve(H_0\ell_*)^{1-\g}$ and $n=\g-1$. A similar behavior can be obtained with the parametrization (\ref{eq:paramzm}).

%%%%%%%%%%%%%%%%%%%%%%%%%%%%%%%%%%%%%%%%%%%%%%%%%%%%%%%%%%%%%%%%%%%%%%%%%%%%%%%%%%%%%%%%%%%%%%%%%%%%%%%%%%%%%%%%%%%%%%%%%%%%%%%%%%

\section{Construction of catalogs of LISA standard sirens}
\label{sect:catalogs}

LISA will detect three classes of binary black holes (BHBs) at cosmological distances \cite{Audley:2017drz}, namely, massive BHBs (MBHB), extreme mass ratio inspirals (EMRIs) and stellar mass BHBs.
These GW sources have different properties and they are expected to be observed in different redshift ranges.
In particular, stellar mass BHBs will mainly be detected at $0.01 \lesssim z \lesssim 0.1$~\cite{DelPozzo:2017kme}, EMRIs at $0.1 \lesssim z \lesssim 2-3$~\cite{Babak2017} and MBHBs at $z \gtrsim 1$~\cite{Klein16}, implying that LISA will constitute a cosmological probe able to map the expansion of the universe from local to very large distances \cite{Tamanini:2016uin}.
Nevertheless only MBHBs are expected to produce powerful EM counterparts, since they are believed to merge in a gas rich environment that may power EM emission through jets, disk winds or accretion: this fact will allow us 
to determine precisely the object positions in the sky. MBHBs will thus likely be the main standard siren sources for LISA, though the number of EM counterpart detections might not exceed few tens of events~\cite{Tamanini:2016zlh}.
They will also provide useful data to test modified theories of gravity through a modified cosmological propagation of GWs.
In this section we outline our strategy to build simulated catalogs of LISA MBHBs with EM counterparts, while in section~\ref{sect:MCMC} we will use these catalogs to constrain modified gravity theories.
To produce catalogs of MBHB standard sirens with EM counterpart, we employ a similar approach to the one adopted in \cite{Tamanini:2016zlh}.
In what follows, we summarize the procedure to construct such catalogs, highlighting the differences with respect to the analysis of \cite{Tamanini:2016zlh}.

To describe the cosmological evolution of massive black holes, and specifically MBHBs, we 
use the semi-analytic galaxy formation model of \cite{EB12} (with later incremental improvements described in \cite{Sesana14,Antonini_long}).
This model was extensively used to predict massive black hole merger rates/stochastic background levels for various LISA/pulsar timing array configurations, respectively 
in \cite{Klein16,Bonetti_triple} and \cite{Dvorkin2017,Bonetti2018b}; to study LISA standard sirens 
in \cite{Tamanini:2016zlh}; to investigate LISA event rates for extreme mass-ratio inspirals in \cite{Babak2017}; and to study projected LISA ringdown constraints 
on the no-hair theorem in \cite{ringdown}.

We adopt here the same scenario  used in \cite{Klein16,Antonini_long}. 
In more detail,
the model follows the evolution of baryonic structures on top of a Dark Matter merger tree produced
with an extended Press--Schechter formalism~\cite{1974ApJ...187..425P},
 modified to reproduce the results of numerical N-body simulations~\cite{Parkinson2008}. The baryonic structures
that are evolved include a chemically pristine diffuse inter-galactic medium, which is either shock heated to the virial temperature of
Dark Matter halos, or flows into the halos along cold filaments~\cite{cold_flows}; a cold, chemically enriched inter-stellar medium (in both a disk and
bulge component); stellar disks and bulges, forming from the inter-stellar medium (with supernova feedback included in the
star formation description); nuclear star clusters, forming from {\it in situ} star formation and from the migration of globular clusters to
the nuclear region~\cite{Antonini_long,Antonini_short}; and massive black holes, accreting from nuclear gas reservoirs (which, in turn, form after major galactic mergers or galactic
disk instabilities, which trigger bursts of star formation in the bulge and inflows of gas to the nuclear region). The black holes
affect the overall evolution via the feedback of 
their active galactic nucleus (AGN) phases, which ensures also
that the observed local correlations between the properties of massive black holes and those of their galactic hosts are correctly reproduced~\cite{EB12,Sesana14,Barausse2017}.

Most relevant for our purposes are two ingredients of the model, namely the prescriptions for the initial mass function
of the massive black hole seeds, and for the ``delays'' between galaxy and massive black hole mergers. While we
refer to \cite{Antonini_long} for a complete description of these two aspects of our model, let us here briefly recall
their main features.

Regarding the seeding model, we adopt two distinct scenarios: a light seed scenario in which
the massive black hole population evolves from the remnants of population III (popIII) stars (with remnant masses  $\sim 100 M_\odot$) at high redshifts $z\gtrsim 15$~\cite{2001ApJ...551L..27M};
and a heavy seed scenario in which the black hole seeds form, again at high redshifts $z\gtrsim 15$,
from bar instabilities of protogalactic disks, with relatively large seed masses $\sim 10^5 M_\odot$~\cite{Volonteri2008}.
As for the delay times, our model tracks the  evolution of Dark Matter halos driven by dynamical friction, from the moment when 
they first touch to the final halo/galaxy merger~\cite{Boylan-Kolchin2008}, including also environmental effects such as tidal disruption and evaporation~\cite{Taffoni2003}. Nevertheless,
the evolution of MBHBs that form in the aftermath of a galaxy merger is still poorly understood. Indeed, while
MBHBs are generally thought to sink quite efficiently down to separations of a few pc, under the effect
of dynamical friction from the background stars and gas, their evolution from pc-separation down to coalescence 
depends sensitively on the distribution and velocities of stars and gas in the nuclear region (see e.g. \cite{Colpi2014} for a review). 
As a result, MBHBs may even stall
at pc separations (the so-called final pc problem), at least in the absence of triple massive black hole systems (which, however, naturally form 
as a result of the hierarchical nature of galaxy evolution~\cite{Bonetti2018b,Ryu2017,Bonetti_triple}).
To account for these uncertainties, we adopt a model in which the final pc problem is
solved   efficiently, i.e.  we assume no delays between the merger of galaxies/halos and black hole coalescence; 
and a model in which we employ a more physical description of the evolution of MBHBs. In more detail, in the latter scenario we account for the interaction of the MBHB with the nuclear gas (if that is present), which
may trigger mergers on timescales as short as $10^7$ yr, and with the stellar background (``stellar hardening''), which typically induces black hole mergers
on timescales of a few Gyr (see  e.g. \cite{Colpi2014} for a review of the physics of these processes). We also account for the formation of triple massive black hole systems, which trigger 
coalescences on timescales again of a few Gyr, using the simplified model described in \cite{Antonini_long} (see \cite{Bonetti2016,Bonetti2017,Bonetti2018a,Bonetti2018b,Ryu2017,Bonetti_triple} for more work on triple massive black
hole interactions).

The results of these semi-analytical simulations are catalogs of MBHB mergers across the whole cosmic evolution, specifically from $z \simeq 20$ to $z=0$.
Using the data of these simulations, in particular the (luminosity) distance and masses of MBHB mergers, we are able to characterize the GW emission of all merger events within the catalogs, and thus to compute the expected GW signal when it reaches the detector. We follow the approach used in~\cite{Klein16,Tamanini:2016zlh} to assess the parameter estimation capabilities of LISA for such catalogs.

We use Fourier-domain inspiral-only precessing waveforms using 3.5 post-Newtonian (PN) phase evolution with 2PN spin-spin and 3.5PN spin-orbit couplings~\cite{klein-2014}, and spin precession equations at 2PN spin-spin and 3.5PN spin-orbit orders. Using a set of precessing inpiral-merger-ringdown waveforms, in \cite{Klein16} was shown that the error on the luminosity distance decreased by a factor of the SNR ratio when compared to an inspiral-only analysis, while the errors on the sky location area decreased by a factor of the SNR squared.

In this analysis, we compute Fisher matrices to assess the detector parameter estimation capabilities. 
We first define the noise-weighted inner product in the space of signals by (see e.g. \cite{Finn:1992wt,Babak:2017tow})
\begin{align}
 (h_1 | h_2) &= 4\, \mathrm{Re} \int \frac{\tilde{h}_1(f) \tilde{h}_2^*(f)}{S_n(f)} df,
\end{align}
where $\tilde{h}(f)$ is the Fourier transform of the waveform $h(t)$ which folds in the detector response, a star denotes complex conjugation, and $S_n(f)$ is the one-sided power spectral density of the detector noise. For the detector noise we use the LISA  noise curve~\footnote{In order to obtain our Eq
\eqref{exp_Sn},  we square Eq 
 (3) of \cite{Petiteau:2016} and we remove its $20/3$ factor (that comes from sky-averaging, which we do not perform). Eq. (1) of \cite{Petiteau:2016}  results then equivalent to our Eq. \eqref{exp_Sacc}.
  %noticing that $2.22\cdot 10^{-5}  \simeq (3\cdot 10^{-5})^{5/4} / (10^{-4})^{1/4}$.
 Notice also that the requirements reported in the public LISA proposal document  \cite{Audley:2017drz} are  simplified compared to the formulas in \cite{Petiteau:2016}. In Section 4.2 of \cite{Audley:2017drz} it is proposed to set requirements above $0.1$ mHz, for which    low-frequency contributions are not too relevant. There are extra factors that affect the high-frequency acceleration noise and the low-frequency displacement noise, but in general the acceleration noise dominates at low frequencies and the displacement noise at high frequencies. It would be interesting, in future works, to  analyse how
our results change when changing sensitivity curves, whose features will be  modified with a better definition of the instrument properties, and of  astrophysical
backgrounds.
 }given by \cite{Petiteau:2016}
\begin{align}\label{exp_Sn}
 S_n(f) &= \frac{4 S_{\text{acc}}(f) + S_{\text{other}}}{L^2} \left[ 1 + \left(\frac{2fL}{0.41c}\right)^2 \right] + S_{\text{conf}}(f),
\end{align}
where the acceleration noise is given by
\begin{align}\label{exp_Sacc}
 S_{\text{acc}}(f) &= \frac{9\times 10^{-30} \text{m}^2 \text{Hz}^{3}}{(2 \pi f)^4} \left[ 1 + \left(\frac{6 \times 10^{-4} \text{Hz}}{f}\right)^2 \left( 1 + \left(\frac{2.22 \times 10^{-5} \text{Hz}}{f}\right)^8 \right)  \right]\,.
\end{align}
The other noise sources combine to
\begin{align}
 S_{\text{other}} &= 8.899 \times 10^{-23} \text{m}^2 \text{Hz}^{-1},
\end{align}
and the %4-year
  confusion noise from unresolved binaries is approximated by
\begin{align}
 S_{\text{conf}} (f) &= \frac{A}{2} e^{- s_1 f^\alpha} f^{-7/3} \left\{ 1 - \tanh \left[ s_2 (f - \kappa ) \right]\right\},
\end{align}
with $A = (3/20) 3.2665 \times 10^{-44}$~Hz$^{4/3}$, $s_1 = 3014.3$~Hz$^{-\alpha}$, $\alpha = 1.183$, $s_2 = 2957.7$~Hz$^{-1}$, and $\kappa = 2.0928 \times 10^{-3}$~Hz.
 %\eb{we do not do any sky average right? For the confusion noise, where do these expressions come from? they seem slightly different than those of
%https://arxiv.org/pdf/1803.01944.pdf; have we checked our SNRs against their python script?}

We follow~\cite{Klein16,Tamanini:2016zlh} and use a low-frequency approximation of the signal in LISA~\cite{cutler-98}, in which the instrument is equivalent to two independent Michelson interferometers. We refer to~\cite{Klein16} for details of the implementation.

The SNR of a signal in one interfermometer can be estimated with
 $\rho = (h | h)^{1/2}$, and the signals from the two interferometers combine as $\rho^2 = \rho_I^2 + \rho_{II}^2$.
The Fisher matrix is defined as 
\begin{align}
 \Gamma_{ij} = \left( \frac{\partial h}{\partial \theta^i} \bigg| \frac{\partial h}{\partial \theta^j} \right),
\end{align}
where $\partial h/\partial \theta^i$ is the derivative of the wave signal $h$ with respect to parameter $\theta^i$. The diagonal elements of its inverse $\Sigma = \Gamma^{-1}$ provide estimates of statistical errors due to the detector noise, as $\Delta \theta^i \approx \sqrt{\Sigma^{ii}}$. The 
 Fisher matrices  
 in the two detectors combine as $\Gamma = \Gamma^I + \Gamma^{II}$.

%{\color{orange} [NT: Antoine please complete here with a description of the FM code (the one you used for the LISA call with inspiral+merger+ringdown which differs from the inspiral-only code used in \cite{Tamanini:2016zlh}, please mention the differences). Please also mention explicitly that we use the noise curve of the LISA call, and not the ones used in \cite{Tamanini:2016zlh}] }
%{\color{red} [AK: I didn't use an IMR code, I don't have that. I used the same rescaling trick as before.]}

For each system in our catalogs, we estimate the error on the luminosity distance using the Fisher matrix method described above using an inspiral-only signal, and rescale it by the ratio of inspiral-merger-ringdown to inspiral-only SNR computed estimated using a phenomC waveform~\cite{Santamaria10}, as $\Delta d_L = (\rho_{\text{insp}} / \rho_{\text{IMR}}) \sqrt{\Sigma^{d_L d_L}} $.
To this experimental error one must add the uncertainties due to weak lensing and peculiar velocities, and we follow the prescriptions outlined in \cite{Tamanini:2016zlh} for estimating and incorporating the error budget on $\Delta d_L$ of these effects.

Among all the LISA detections with SNR$>8$ we select the events with a sky localization $\Delta \Omega < 10\, {\rm deg}^2$.
As in \cite{Tamanini:2016zlh} we characterize the EM emission at merger by assuming the production of an optical accretion-powered luminosity flare and of radio flares and jets, based on results from general-relativistic simulations of merging MBHBs in an external magnetic field \cite{Palenzuela:2010nf,Giacomazzo:2012iv}.
The semi-analytical simulations of MBHB catalogs described above contain all the necessary data (mass in stars, nuclear gas, etc.) needed to compute the magnitude of the EM emission of each MBHB merger in both the optical and radio bands.
Following \cite{Tamanini:2016zlh} we can thus determine the number of counterparts detected by future EM facilities, specifically LSST \cite{LSSTweb}, SKA \cite{SKAweb} and ELT \cite{EELTweb}.
Note however that over the sky location region of the source there might appear several EM transients that could likely be identified with the EM counterpart of the MBHB merger.
As in \cite{Tamanini:2016zlh} we will however assume that the true EM counterpart can be efficiently identified (e.g.~through timing, EM spectrum characterization and evolution, or information on the host galaxy), and leave a more thorough investigation of this issue to future analyses.
Spotting the EM counterpart helps  localize the source in the sky, thus allowing one to determine the host galaxy and to improve the errors on the GW parameters by fixing the two sky localization angles.

The identification of the host galaxy in principle allows one to obtain the redshift of the MBHB event using either spectroscopic or photometric techniques.
In what follows we distinguish two scenarios depending on the error estimates for these redshift measurements:
\begin{itemize}

	\item \textbf{Scenario 1 (realistic)}: Photometric measurements have a relative 1$\sigma$ error on $z$ given by $\Delta z_{\rm photo} = 0.05 (1+z)$ at best, according to Euclid~\cite{Amendola:2016saw}, while spectroscopic measurements have a relative error given by $\Delta z_{\rm spect} = 0.01 (1+z)^2$, assuming we can define a volume-complete spectroscopic sample. However, the flux limit for that sample evolves with redshift. This may explain the redshift dependence, since luminosity distance evolves as approximately the square of $(1+z)$ at $z>1$~\cite{DeVicente:2015kyp}. % Part of the origin will also be in the width of the optical fiber of the spectrograph, which at high redshift induces confusion from multiple sources. 
	 Although extrapolating those expressions beyond $z\sim3$ might seem unrealistic~\cite{Kruhler:2010jw}, they are the only guesses we can use at the moment to estimate future redshift measurements at high redshift. 
	%\\
	%{\color{orange} [NT: Juan please add a couple of sentences and refs to explain these errors]}
%On the other hand, when a GRB redshift is determined from afterglows the best error is 5%, see arXiv:1011.1205  [2011A&A...526A.153K] 
	\item \textbf{Scenario 2 (optimistic)}: Photometric measurements have a relative 1$\sigma$ error on $z$ given by $\Delta z_{\rm photo} = 0.03 (1+z)$, while spectroscopic measurements have a fixed relative error given by $\Delta z_{\rm spect} = 0.01$. Moreover here we assume a delensing procedure able to reduce by 50\% the luminosity distance uncertainty due to weak lensing. This scenario is equivalent to the ``optimistic scenario'' considered in \cite{Tamanini:2016zlh}.

\end{itemize}
For both scenarios the redshift errors are propagated to the error budget of the luminosity distance using the fiducial cosmology, i.e.~$\Lambda$CDM with Planck values for the cosmological parameters.

The total result of the whole procedure described above are thus mock catalogs of MBHB standard sirens, specified by values of the luminosity distance, redshift and total error on the luminosity distance.
We simulate 22 4-year catalogs~\footnote{Notice that although the nominal mission
lifetime is 4 years (with possible extensions), the duty cycle is only $75\%$ of the total nominal time. It would be interesting to adjust our analysis to properly take into account how frequently data gaps occur: we leave a dedicated study of this point to a future work.} for each MBHB astrophysical model (popIII, heavy seeds with and without delays), for a total of 66 catalogs and 264 years of data.
For each catalog we then perform a quick Fisher matrix cosmological analysis assuming $\Lambda$CDM.
We ranked the 22 catalogs of each MBHB astrophysical model according to the area of the 1$\sigma$ contour ellipses in the $(\Omega_m, H_0)$ parameter space of $\Lambda$CDM.
This quantity provides us with a rough measure of the cosmological constraining power of each catalog.
Finally we select the median catalog among all ranked 22 catalogs as the representative catalog for each MBHB astrophysical model, similarly to the recipe employed in \cite{Cai:2017yww}.
We perform this procedure for Scenario~1 and then use the same selected catalogs for Scenario~2, but using different assumptions for the redshift errors as explained above.
This  allows us to directly compare results between the two different scenarios.
The final product are thus six catalogs (corresponding to the three different MBHB astrophysical models, each one combined with  the two scenarios for the redshift measurement accuracy) to use as mock data for the MCMC analysis presented in the following section.
The six selected MBHB catalogs are plotted in Fig.~\ref{fig:catalog_dl}. 

%-FIGURE-
\begin{figure}[t]
\centering 
 \includegraphics[width=0.49\textwidth]{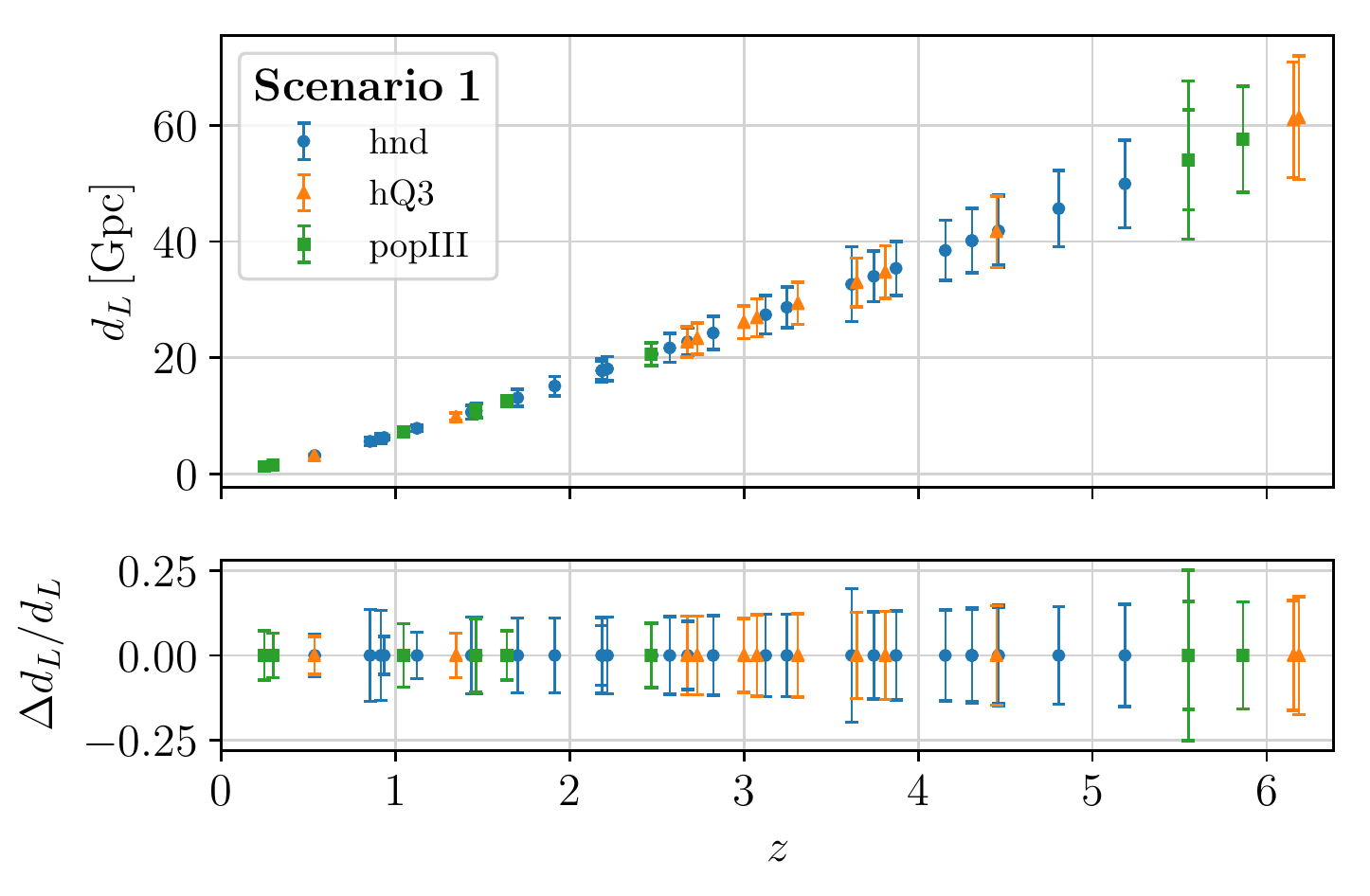}
  \includegraphics[width=0.49\textwidth]{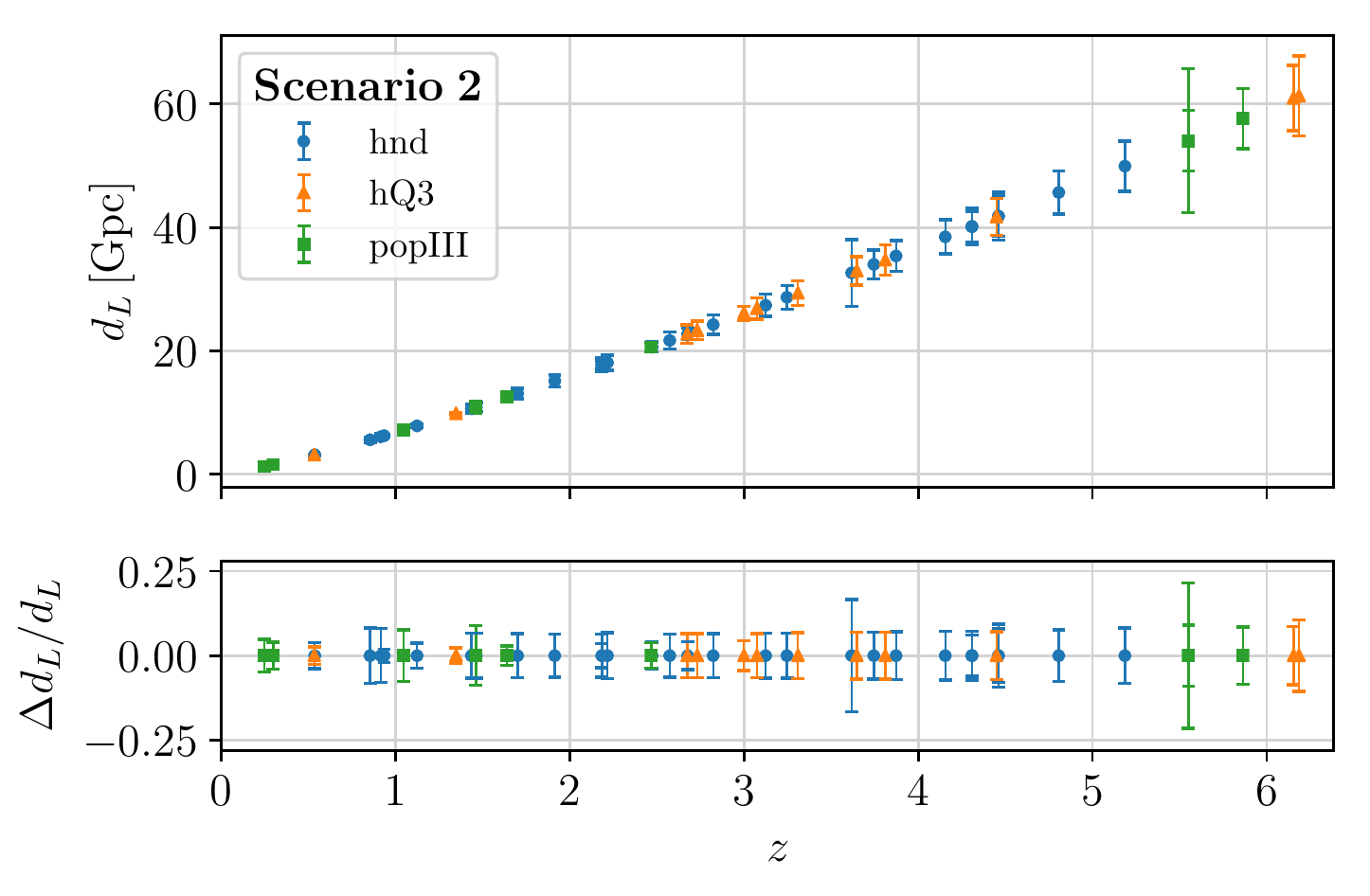}
 \caption{GW luminosity distance as a function of redshift for the different sources in the three catalogs:
heavy seeds and no delay (``hnd"), heavy seeds with a delay included and the Toomre parameter set to $Q=3$ (``hQ3") and light seeds due to pop III stars (``pop~III"), and the two scenarios for estimate of the error on the redshift:  scenario 1 (realistic,  left) and scenario 2 (optimistic, right). The bottom subpanels show the relative error in $d_L$.}
 \label{fig:catalog_dl}
\end{figure}
%---------

%%%%%%%%%%%%%%%%%%%%%%%%%%%%%%%%%%%%%%%%%%%%%%%%%%%%%%%%%%%%%%%%%%%%%%%%%%%%%%%%%%%%%%%%%%%%%%%%%%%%%%%%%%%%%%%%%%%%%%%%%%%%%%%%%%

\section{Cosmological parameters and dark energy with LISA}\label{sect:MCMC}

We now use the mock catalogs discussed in the previous section to study the effect of LISA standard sirens on the estimate of the cosmological parameters, with particular emphasis on  the DE sector, as described by the parameters $(w_0,w_a)$ for the DE equation of state, \eq{w0wa}, and by the parameters $(\Xi_0,n)$ that enter
\eq{eq:param} for modified GW propagation. We adapt to LISA  a similar  analysis performed in \cite{Belgacem:2018lbp} for the Einstein Telescope.

As discussed in section~\ref{sect:catalogs}, we provide our results for  three models  for the formation of MBH binaries: heavy seeds and no delays (``hnd'' for short), heavy seeds with a delay included and the Toomre parameter (see \cite{Tamanini:2016zlh}) set to $Q=3$ (``hQ3''),  
and light seeds due to pop III stars (``popIII'').  
%\eb{do we use the version with or without delays? in any case, we should mention
%adding delays or not makes little difference for popIII; also please use the same acronyms as in Tamanini et al 2016 and in Klein et al 2016 otherwise its confusing. I think we used popIII, Q3d and Q3nod}.  
 For each formation model, we use the two scenarios (1) and (2) of section~\ref{sect:catalogs} for estimating the error in the measurement of the redshift: thus, overall, we consider  six different possibilities. 

From the point of view of the cosmological model, we  consider different cases. We will start with $\Lambda$CDM, to investigate  how LISA standard sirens   help to constrain its parameters. We will then study the extension of the DE sector obtained introducing a non-trivial DE equation of state, parametrized by $w_0$ only  (without modified GW propagation, i.e. setting $\Xi_0=1$); finally, we will introduce modified GW propagation, extending the DE sector through the parameters $(w_0,\Xi_0)$ and fixing $w_a=0$ and $n$ to a reference value.\footnote{We do not introduce more than two extra parameters at the same time  in the DE sector, since the constraining power would decrease and the convergence time of the MCMC chains would become very long. Indeed, already with the two parameters $(w_0,w_a)$, we found that, because of the limited number of sources in the LISA catalogs, the MCMC chains do not resolve the degeneracy between these parameters and do not reach a good convergence.}
For each of these three cosmological models we run MCMCs for the six scenarios describing  the formation model and the estimate of the redshift error, as discussed above.

Furthermore, for each of these $3\times 6$ cases we run separate MCMCs to
compute the constraints that can be obtained with standard sirens only, and those obtained by combining them with CMB, BAO and SNe data. In that case we use 
the {\em Planck} temperature and polarization power spectra \cite{Ade:2015rim}, the JLA SNe dataset \cite{Betoule:2014frx} and a set of isotropic and anisotropic BAO data (see Section 3.3.1 of~\cite{Belgacem:2017cqo} for details). 
%{\bf \color{red}  Notice that since we 
%constrain possible deviations from  $\Lambda$CDM and GR, we use information on these dataset as based  on such fiducial framework.  }

To generate the catalogs discussed in section~\ref{sect:catalogs} we  assume a fiducial cosmological model, that we always take to be $\Lambda$CDM with $H_0=67.64$ and $\oma=0.3087$, which are the fiducial values obtained from the CMB+BAO+SNe dataset that we use. So, in particular, our fiducial values for the extra parameters in the DE sector are $w_0=-1$ and $\Xi_0=1$,  and we use $\Lambda$CDM  and standard perturbation theory within GR as fiducial theoretical framework for treating and analyzing our dataset. 
 Our aim is to evaluate to what  accuracy LISA, alone or in combination with other datasets, can reconstruct these  fiducial  values.
We then generate our simulated catalog of events by  assuming that, for a source at redshift $z_i$, the actual luminosity distance will be $d^{\Lambda\rm CDM}_L(z_i;H_0,\oma)$, with the above values of $H_0$ and $\oma$. The ``measured" value of the luminosity distance is then randomly extracted from a Gaussian distribution centered on this fiducial value, and  with a width 
obtained from the estimate of the error on the luminosity distance 
discussed in section~\ref{sect:catalogs}.

The relevant cosmological parameters for evaluating the theoretical values of the GW luminosity distance $d_L^{\,\rm gw}(z)$ are $\oma$ and $H_0$ in the $\Lambda$CDM case, with the addition of $w_0$ in $w$CDM and the further addition of $\Xi_0$ in the $(\Xi_0,w_0)$ case. Apart from a constant addend coming from normalization, the logarithm of the likelihood assigned to a given set of values $(H_0,\oma,w_0,\Xi_0)$ is given by
\begin{equation}\label{loglkl}
\ln(L(H_0,\oma,w_0,\Xi_0))=-\frac{1}{2}\sum_{i=1}^{N_s}\frac{\left[d_L^{\,\rm gw}(z_i;H_0,\oma,w_0,\Xi_0)-d_i\right]^2}{\sigma_i^2}\, ,
\end{equation}
where $N_s$ is the number of mock sources in the catalog, $d_L^{\,\rm gw}(z_i;H_0,\oma,w_0,\Xi_0)$ is the theoretical value of the GW luminosity distance for the $i$-th source and $d_i$ is the ``measured" value of its luminosity distance contained in the catalog. The quantity $\sigma_i$ is the error on luminosity distance and it also takes into account the error on redshift determination, which is simply propagated to the luminosity distance using the fiducial $\Lambda$CDM cosmology. The used MCMC code explores the cosmological parameters space, accepting or rejecting the points according to a Metropolis-Hastings algorithm based on the likelihood specified above. The priors assumed on cosmological parameters are Gaussian (or truncated Gaussian) distributions with mean and standard deviation, in this order, given by: $67.8$ and $1.2$ for $H_0$ (in $\rm{km} \, \rm{s}^{-1} \rm{Mpc}^{-1}$), $0.02225$ and $0.00028$ for $\omega_b$=$\Omega_b h^2$ (baryon density fraction), $0.1192$ and $0.0027$ for $\omega_c=\Omega_c h^2$ (cold dark matter density fraction), $-1.0$ and $1.0$ for $w_0$ (restricted from $-3.0$ to $-0.3$), 1.0 and 0.5 for $\Xi_0$ (restricted to positive values).

As we see from Fig.~\ref{fig:catalog_dl}, the number of total available standard sirens in our sample is quite limited. For instance, 
in the specific realization of the catalogs that we use   there are 32, 12 and 9  sources in the ``hnd'', ``hQ3'' and ``popIII'' scenarios, respectively. It is important to realize that, since these numbers are relatively small, the scatter of the mock data around their fiducial values unavoidably induces fluctuations in the reconstruction of the mean value of the cosmological parameters inferred from standard sirens alone, so that the reconstructed mean values in general will not coincide with those  obtained from other cosmological observations such as CMB, BAO and SNe. Thus, depending on the specific realization of the catalog, the 
contours  in parameter space of the  likelihood  obtained from standard sirens could show mild tensions with those obtained from other datasets. This is unavoidable, and will also happen in the actual experimental situation. Thus, in the plots shown below, the   overlap (or lack of it) between different contours has little meaning, as it depends on the specific random realization of the catalog. What carries the important information is the relative size of the contours, that will tell us to what extent the addition of standard sirens to other datasets can improve our knowledge of the cosmological parameters.
However, when combining contours from different datasets to obtain a combined estimate on the error on a parameter, one must be careful not to use a specific realization where the separate contours are in mild tension among each other.

\subsection{\texorpdfstring{$\Lambda$}{}CDM}\label{sect:LCDMresults}

We begin by studying the effect of  standard sirens on cosmological  parameter estimation 
in  $\Lambda$CDM, as a benchmark for our  subsequent generalizations, and we will then  introduce different extensions of the dark energy sector.
As  baseline  $\Lambda$CDM model we use the standard set of six independent
cosmological parameters:  the Hubble parameter today
$H_0 = 100 h \, \rm{km} \, \rm{s}^{-1} \rm{Mpc}^{-1}$, the baryon density fraction today $\omega_b = \Omega_b h^2$ , the cold dark matter density fraction today  $\omega_c = \Omega_c h^2$,  the amplitude  $A_s$ and tilt $n_s$ of the spectrum of  primordial scalar perturbations,   and  the reionization optical depth  $\tau_{\rm re}$. We keep the sum of neutrino masses fixed, at the value $\sum_{\nu}m_{\nu}=0.06$~eV, as in the {\em Planck} baseline analysis~\cite{Ade:2015xua}. We then run a series of MCMC, using 
the  CLASS Boltzmann code \cite{Blas:2011rf} (or our modification of it, in the case of non-trivial GW propagation). 

In $\Lambda$CDM, assuming flatness, $d_L(z)$ depends only on  $H_0$ and $\oma$, so these are the parameters for which the inclusion of standard sirens has the most significant impact. Fig.~\ref{fig:H0OmaLCDMreal} shows the two-dimensional likelihood in the $(\oma,H_0)$ plane in $\Lambda$CDM, comparing the 
contribution from  CMB+BAO+SNe (red) to the contribution from LISA standard sirens   (gray), and the overall combined contours, for the three formation scenarios and the scenario (1) for the determination of the error on the luminosity distance, while
Fig.~\ref{fig:H0OmaLCDMopt} shows the results for the scenario (2).

\begin{figure}[t]
\begin{center}
\includegraphics[width=0.4\textwidth]{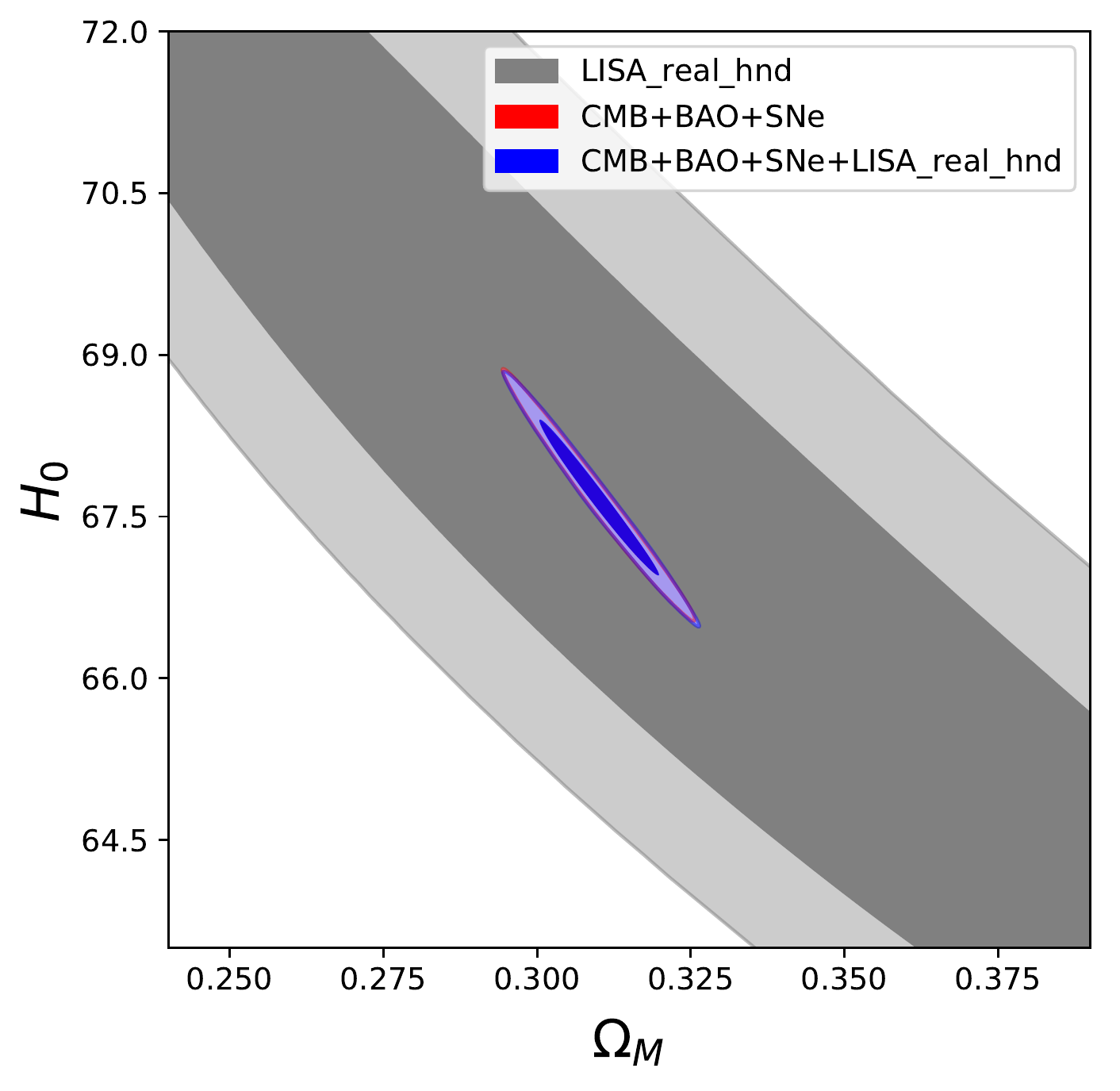}
\includegraphics[width=0.4\textwidth]{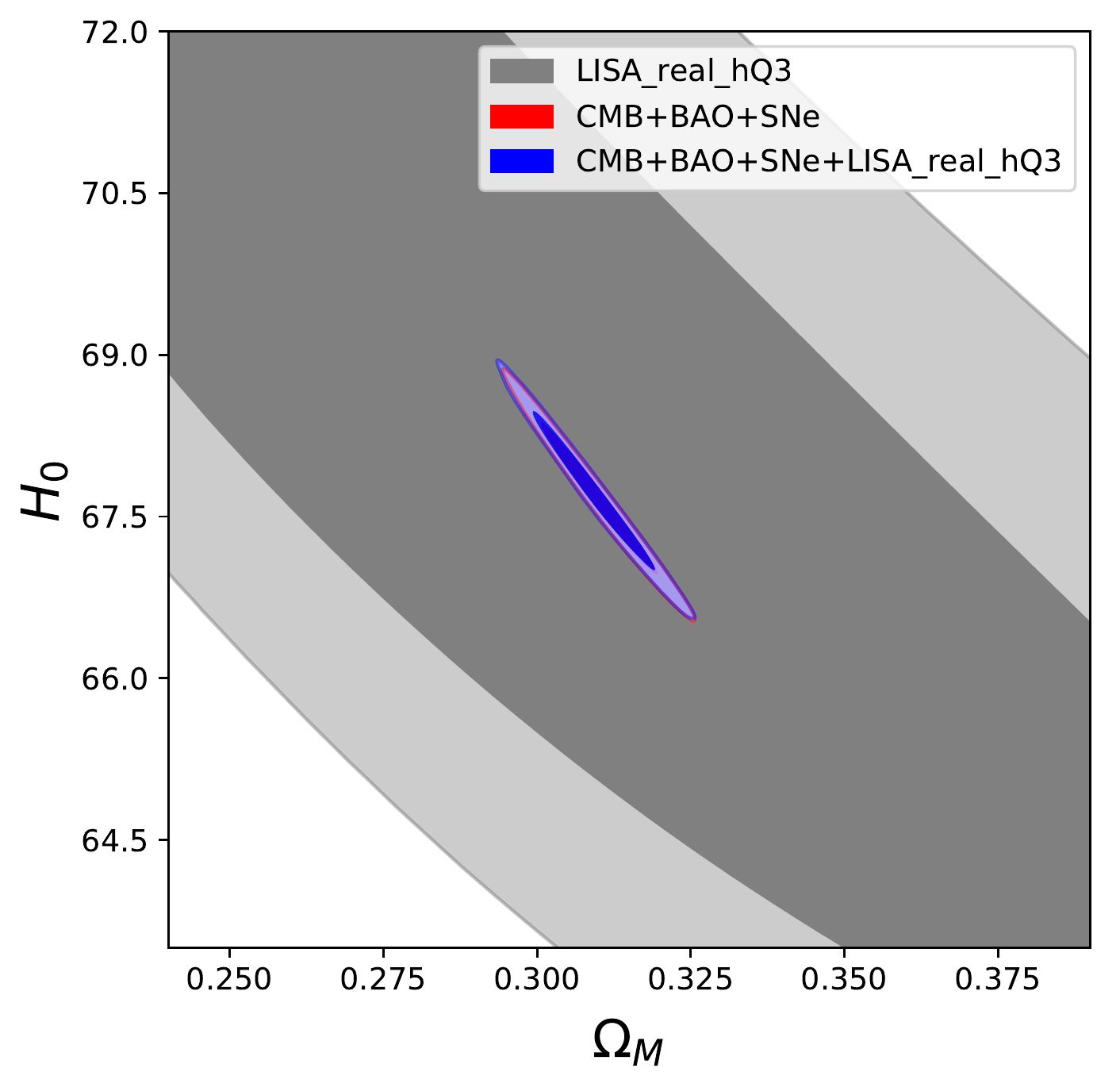}
\includegraphics[width=0.4\textwidth]{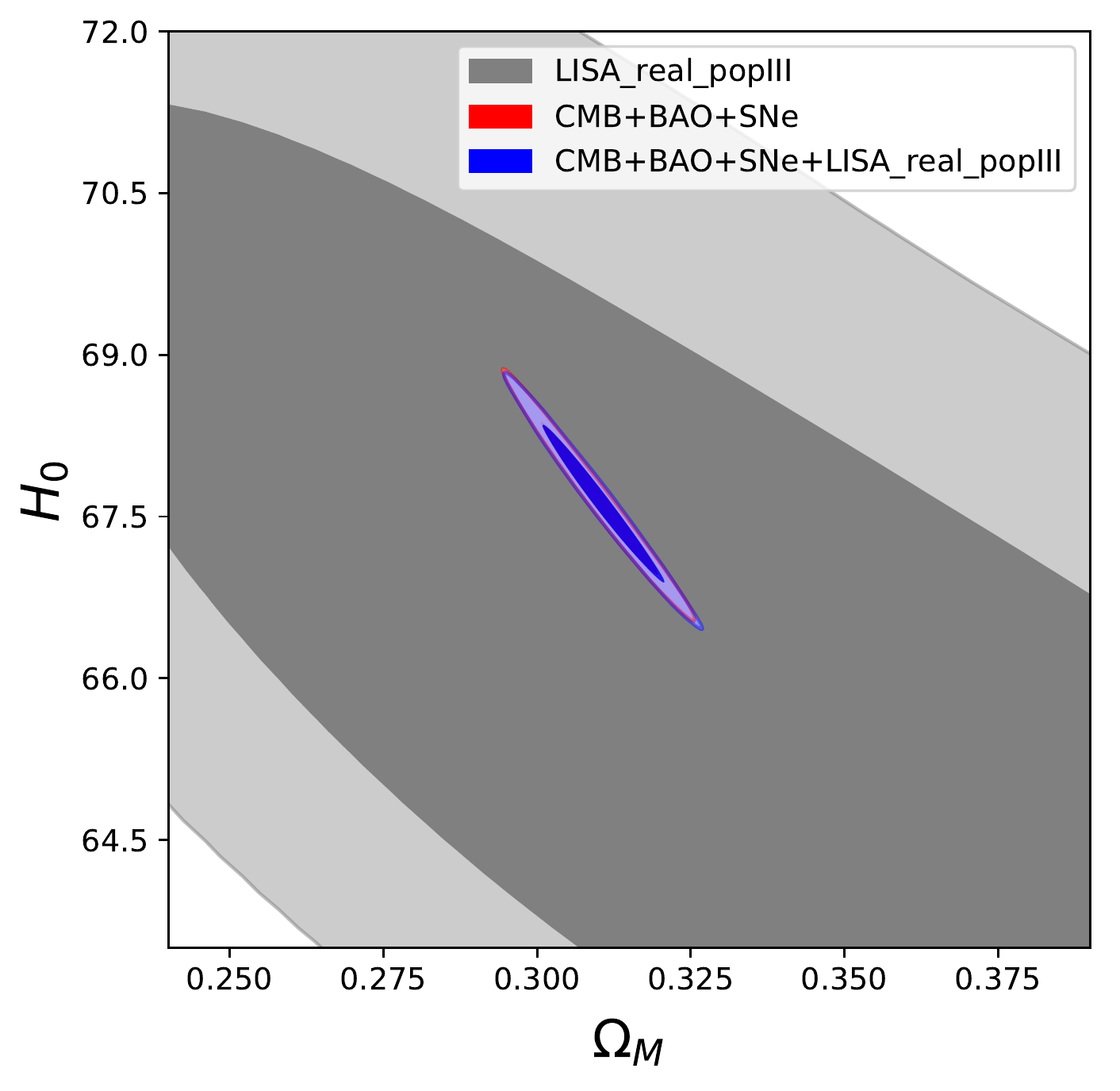}
\end{center}
\caption{The  $1\sigma$ and $2\sigma$
contours  of the two-dimensional likelihood in the $(\oma,H_0)$ plane in $\Lambda$CDM, with the 
contribution from  CMB+BAO+SNe (red), the contribution  from LISA standard sirens  (gray) and the overall combined contours (blue), in the scenario (1) (``realistic") for the  error on the luminosity distance. Upper left: heavy no-delay (``hnd") scenario; Upper right: ``hQ3" scenario; lower panel: ``pop~III" scenario. $H_0$ is given in the usual units $\rm{km} \, \rm{s}^{-1} \rm{Mpc}^{-1}$. Notice that the red contours are almost
superimposed to the blue ones.}
\label{fig:H0OmaLCDMreal}
\end{figure}

\begin{figure}[t]
\begin{center}
\includegraphics[width=0.4\textwidth]{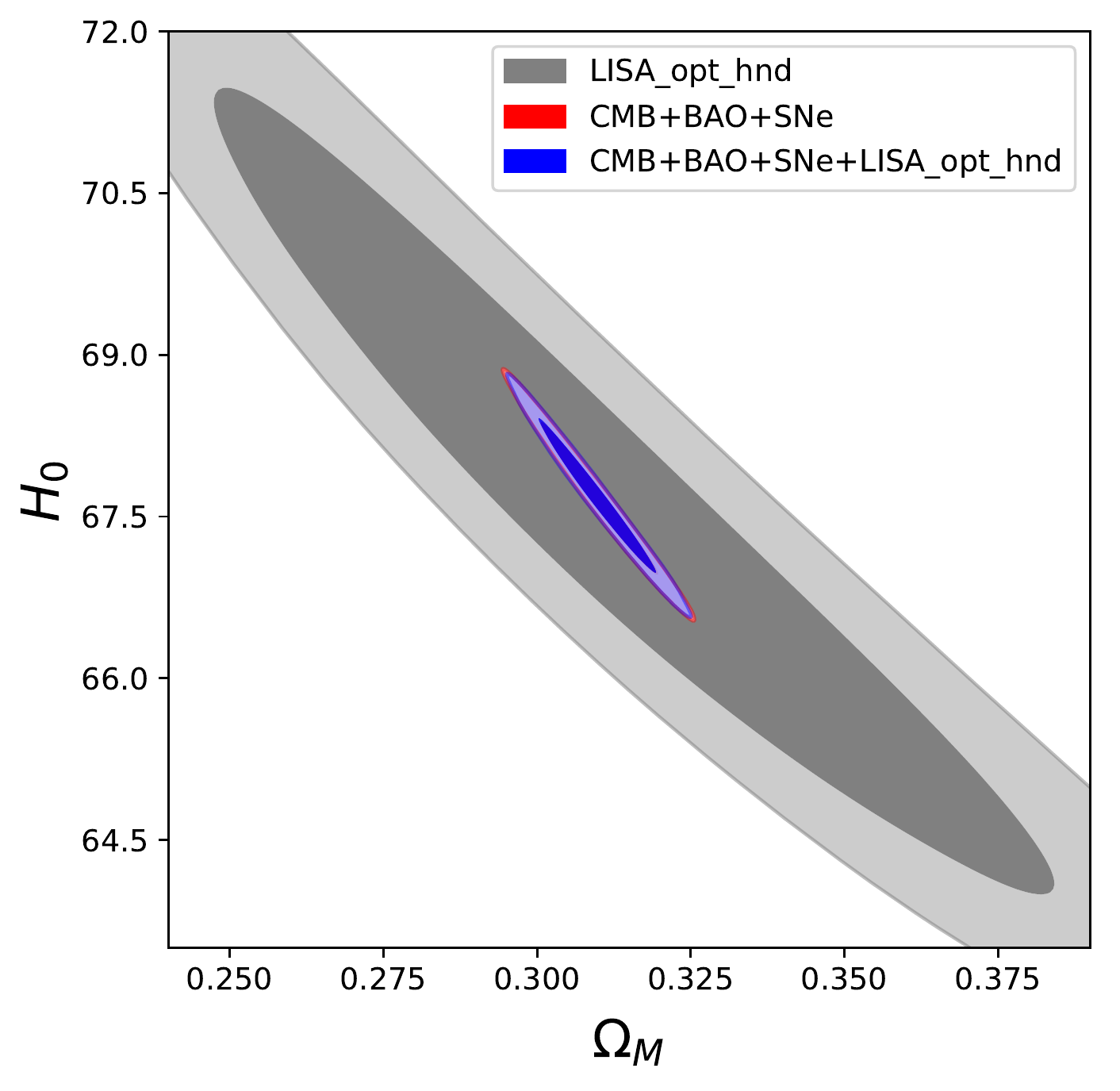}
\includegraphics[width=0.4\textwidth]{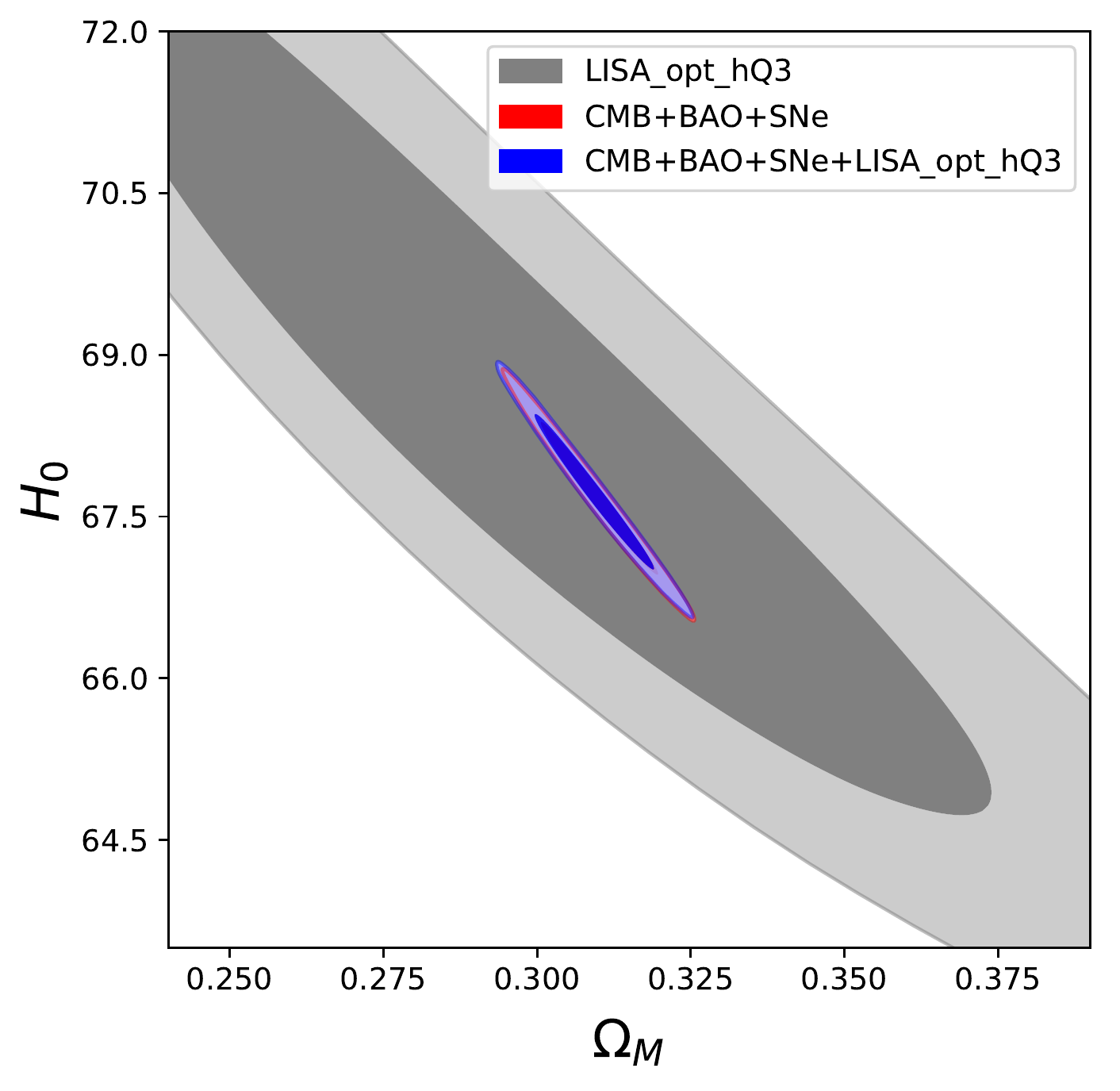}
\includegraphics[width=0.4\textwidth]{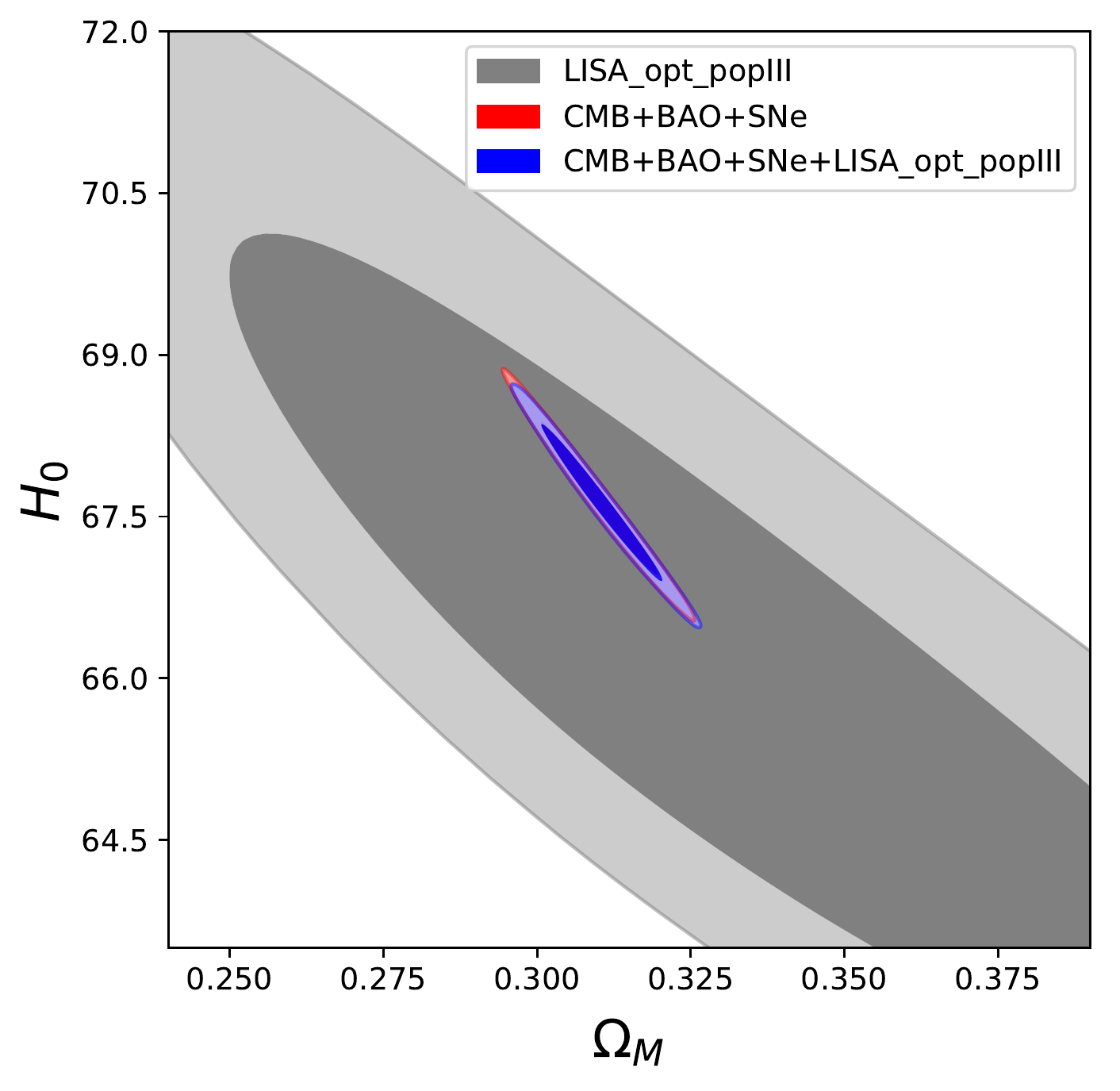}
\end{center}
\caption{As in Fig.~\ref{fig:H0OmaLCDMreal}, for the scenario (2) (``optimistic") for the error on the luminosity distance.}
\label{fig:H0OmaLCDMopt}
\end{figure}

In particular in the most favorable scenario (``hnd" seeds and optimistic errors on $d_L$), from the corresponding one-dimensional marginalized likelihood  we find that, with standard sirens only, the relative error on $H_0$ is 
\be
\frac{\Delta H_0}{H_0}=3.8\%\, ,
\ee 
(which raises  to $7.7\%$ in the ``realistic" scenario with ``hnd" seeds) and the one on $\oma$ is $\Delta\oma/\oma=14.7\%$; using our CMB+BAO+SNe dataset we get instead 
$\Delta H_0/H_0=0.7\%$ and $\Delta\oma/\oma=2.1\%$; combining CMB+BAO+SNe+standard sirens we get 
\be
\frac{\Delta H_0}{H_0}=0.7\%\, ,\qquad
\frac{\Delta \oma}{\oma}=2.0\%\, .
\ee
Therefore,  from MBH binaries at LISA, we do not  find a significant  improvement on the   accuracy on $H_0$, compared  to current results from  CMB+BAO+SNe. It should however be observed that 
the measurement from standard sirens is still useful because it has completely different systematic errors from that obtained with 
CMB, BAO and SNe. 

These results should be compared with those found in \cite{Tamanini:2016zlh}. To perform the comparison, we should however note that the best results quoted in \cite{Tamanini:2016zlh} referred to a LISA configuration with 5~Gm arms, and with all other design specifications to their most optimistic possible choices,
while our results use the  current LISA configuration with 2.5 Gm arms, given in \cite{Audley:2017drz}.
  The corresponding LISA  sensitivity curve used in our study 
is quite different and generally gives worse results. For example the number of standard sirens used in our analysis is roughly half the number used in the 2016 paper~\cite{Tamanini:2016zlh}.

Furthermore, at the methodological level there are some differences  between our analysis and that in \cite{Tamanini:2016zlh}. First, as discussed above, to generate our catalogs   we have scattered the values
$d^{\Lambda\rm CDM}_L(z_i;H_0,\oma)$, extracting the ``measured" values of the luminosity distance  from a Gaussian distribution centered on this fiducial value, and  with a width $\Delta d_L$
determined by  the projected error estimate.
Ref.~\cite{Tamanini:2016zlh} did not  consider scattered data, thus the statistical error due to scatter was not included in the analysis. The Fisher matrix analysis performed in  \cite{Tamanini:2016zlh} was performed on non-scattered data,
%works pretty much the same with scattered and non-scattered data,
% \eb{Nicola please rewrite this sentence}, 
  but the MCMC analysis of the present paper  takes also into account the statistical uncertainty due to the scattering and thus gives more realistic results.\footnote{On the other hand, the MCMC analysis is based only on one realization of the dataset, in contrast to the Fisher matrix  analysis which can be averaged over many realizations. However, as discussed in section~\ref{sect:catalogs}, the catalog used is chosen as a typical ``average'' catalog, between many different realizations.}
 Second, in the present analysis the degeneracies between the cosmological parameters are fully taken into account by our MCMC by  freely varying all cosmological parameters of $\Lambda$CDM (and, in the following subsections, of its extensions), while in the Fisher matrix analysis of   \cite{Tamanini:2016zlh} the most stringent results (that eventually, together with all other assumptions mentioned above, led to the estimate that $H_0$ could be measured to $0.5\%$ with standard sirens only) where obtained by varying only $H_0$ and assuming a fixed prior on $\oma$. The procedure of the present paper therefore gives a larger estimate of the error on the cosmological parameters that can be  obtained with LISA standard sirens.

On the other hand, one should also be aware of the fact that, before drawing final conclusions on the sensitivity of LISA to $H_0$, as well as to the DE parameters that will be discussed below, much more work is needed. For instance, the counterpart model used in \cite{Tamanini:2016zlh}
makes a number of assumptions, which will only be validated when (if) LISA electromagnetic counterparts are actually observed. 
Another caveat is that the error on the localization used in this paper is based on the use of the inspiral waveform only, with corrections based on phenomenological  waveforms, and it is still an open issue how this will change when including a full description of the merger and ringdown phases. The estimate of the redshift error also involves other factors, with respect to those that we have discussed, such as the exposure time, that  will depend on the availability of telescope time and the pointing accuracy,  which are currently difficult to estimate, as well as on the duration of the electromagnetic transient, whose estimation  in turn involves the modeling of the counterpart.  It should also be observed that the number of sources observed depends strongly on the sensitivity curve at low frequency, where there is potential room for improvement with respect to the sensitivity curve that we have adopted. 
Finally, in this paper we are only using MBH binaries with counterpart as standard sirens, and we are not considering stellar mass BH binaries and EMRIs, that are not expected to have a counterpart and must be treated with the statistical methods.
These issues are currently under active investigations in the LISA Consortium, and are well beyond the scope of this paper. 
The results of this paper should therefore be understood as the forecast that can be obtained under the specific set of assumptions that we have made.

\subsection{$w$CDM}

\begin{figure}[t]
\begin{center}
\includegraphics[width=0.4\textwidth]{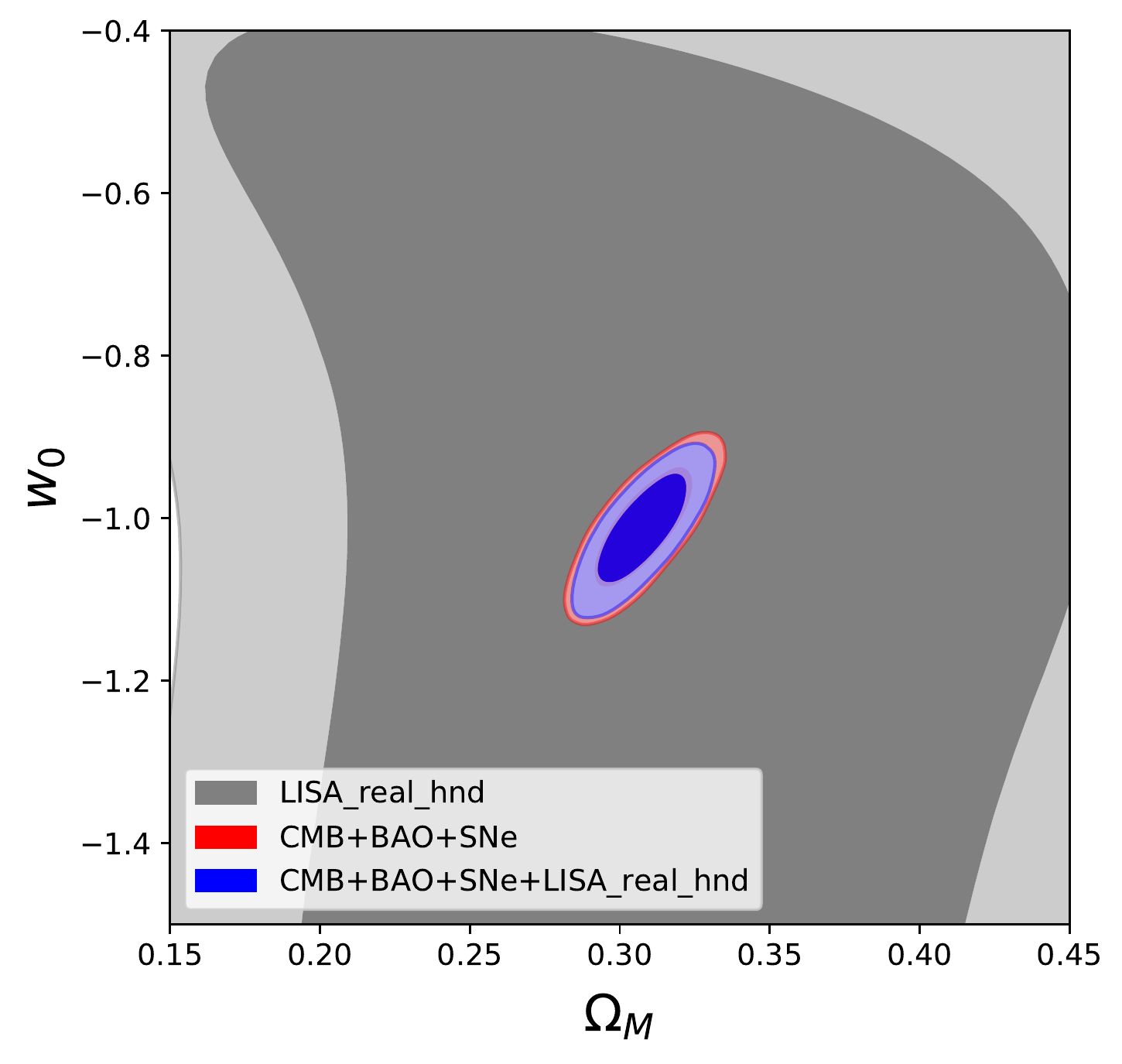}
\includegraphics[width=0.4\textwidth]{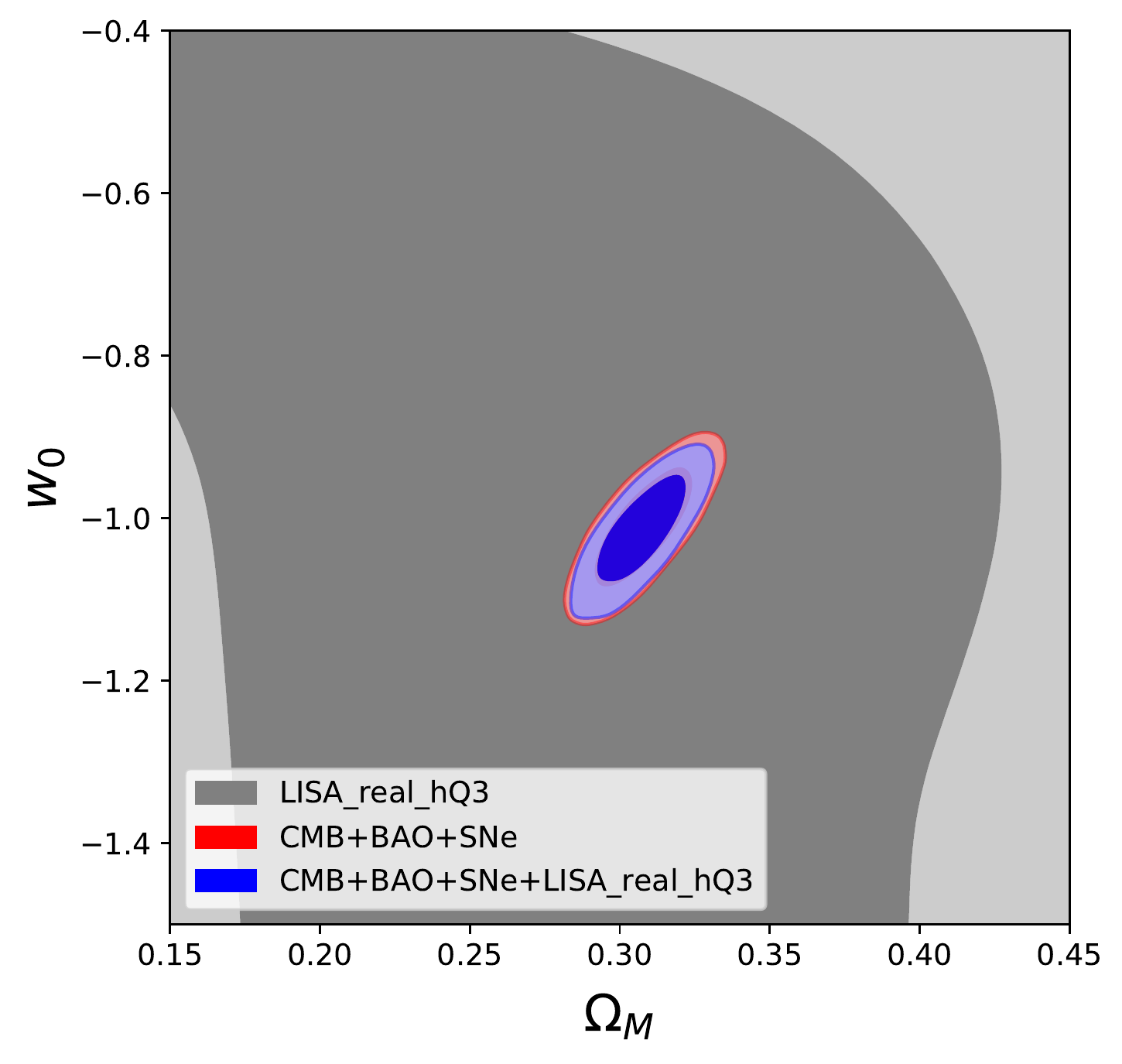}
\includegraphics[width=0.4\textwidth]{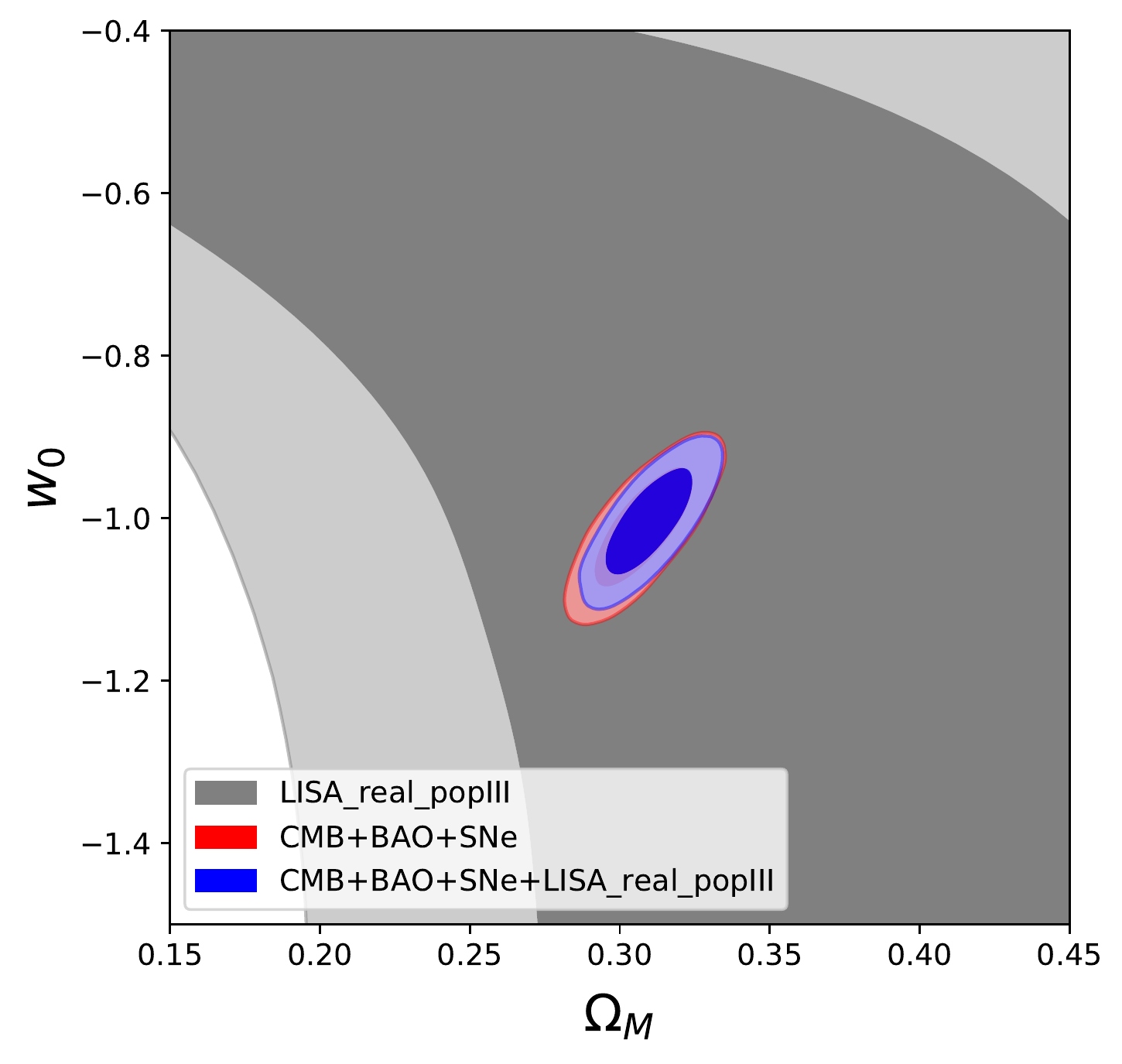}
\end{center}
\caption{The  $1\sigma$ and $2\sigma$
contours  of the two-dimensional likelihood in the $(\oma,w_0)$ plane in $w$CDM, with the 
contribution from  CMB+BAO+SNe (red), the contribution  from LISA standard sirens  (gray) and the overall combined contours (blue), in the scenario (1) (``realistic") for the  error on the luminosity distance. Upper left: heavy no-delay (``hnd") scenario; Upper right: ``hQ3" scenario; lower panel: ``pop~III" scenario.}
\label{fig:w0OmaLCDMreal}
\end{figure}

\begin{figure}[t]
\begin{center}
\includegraphics[width=0.4\textwidth]{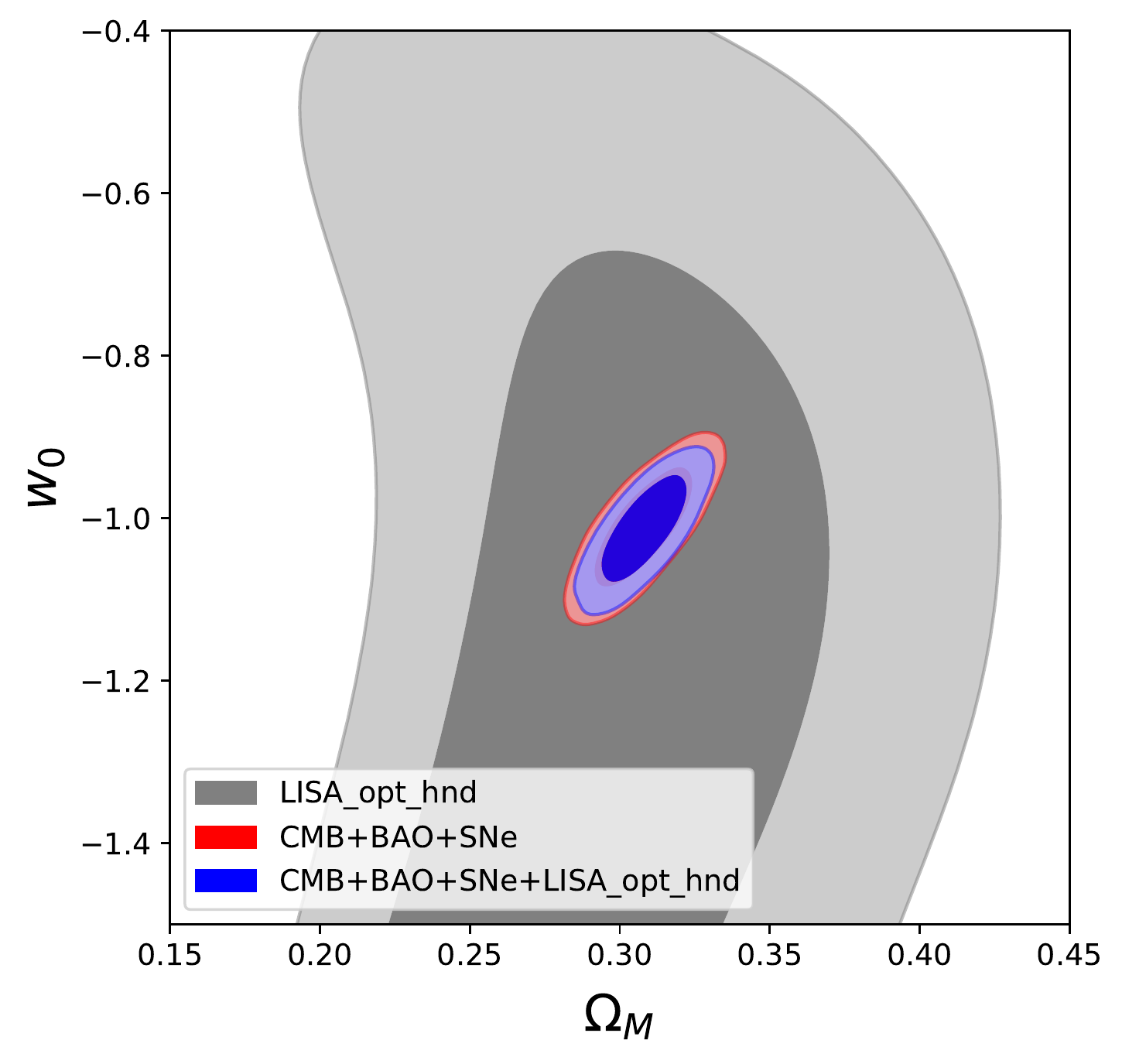}
\includegraphics[width=0.4\textwidth]{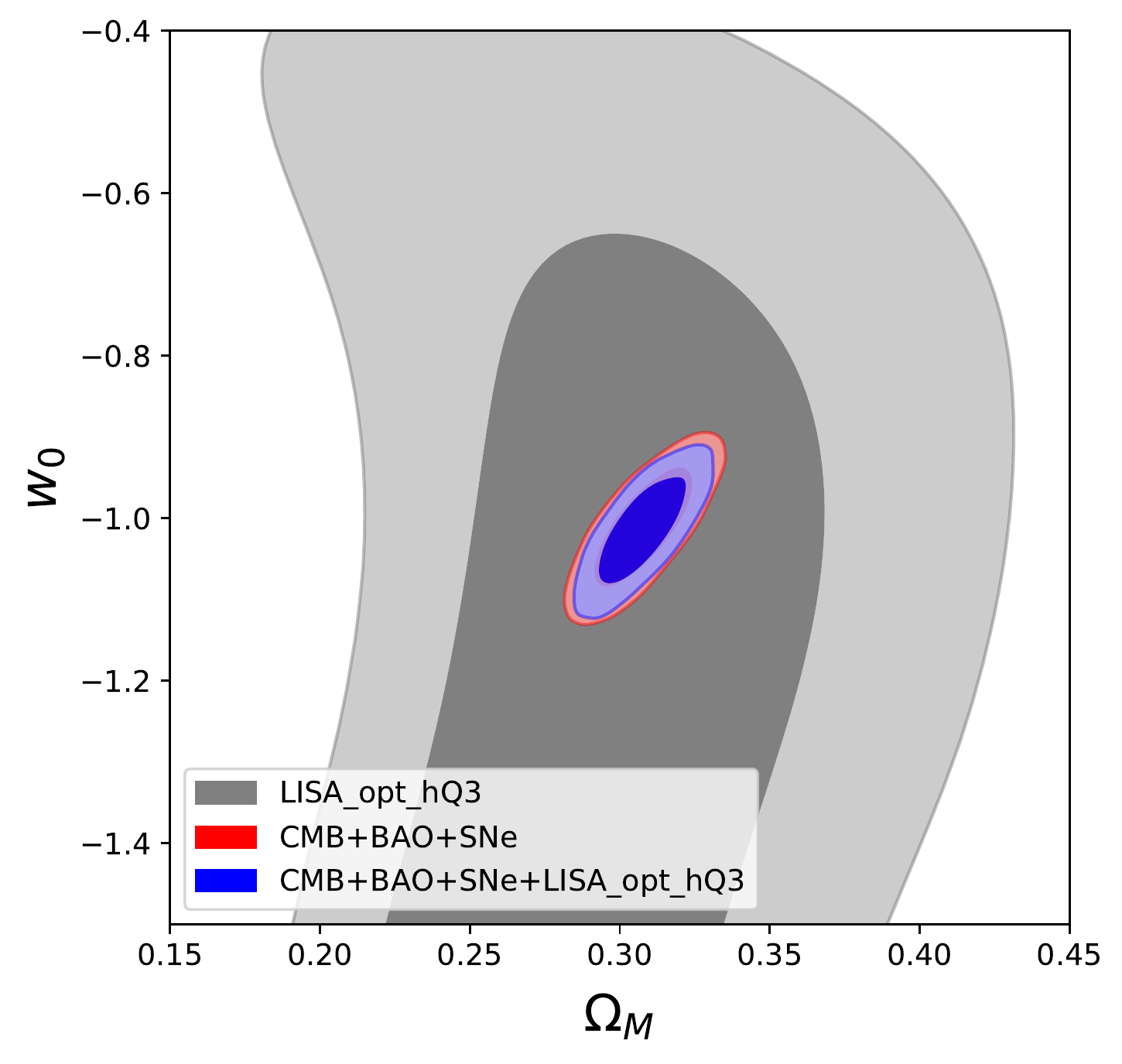}
\includegraphics[width=0.4\textwidth]{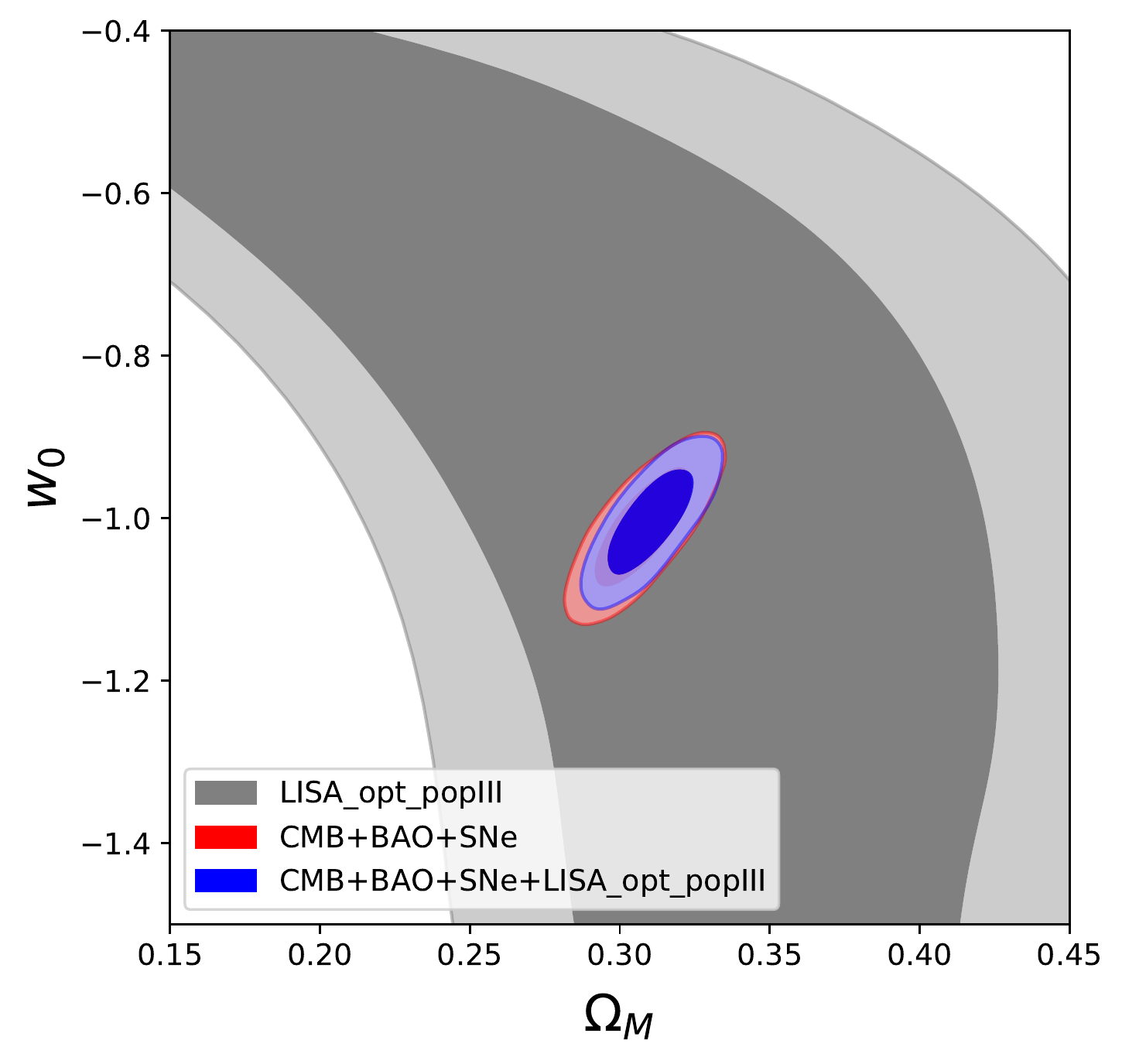}
\end{center}
\caption{As in Fig.~\ref{fig:w0OmaLCDMreal}, for the scenario (2) (``optimistic") for the error on the luminosity distance.}
\label{fig:w0OmaLCDMopt}
\end{figure}

We next add the parameter  $w_0$, corresponding to  the so-called $w$CDM model, where the  DE equation of state is taken to be constant, or $w_a=0$ in \eq{w0wa}.\footnote{A further natural extension would be to the $(w_0,w_a)$  set of parameters. However we have found that, with the limited number of sources in the LISA catalogs, it is not possible to disentangle the degeneracy between these two parameters, and the MCMC chains do not reach a good convergence.}} 
Since the model used to generate the catalog of sources is still $\Lambda$CDM, we are actually asking to what accuracy we can find back the fiducial  value $w_0=-1$.
Figs.~\ref{fig:w0OmaLCDMreal} and \ref{fig:w0OmaLCDMopt}  show the two-dimensional likelihood in the $(\oma,w_0)$ plane, displaying  the combined  contribution from 
CMB + BAO + SNe (red),  and the total combined result (blue), for the $3\times 2$ scenarios considered.

Even in the most optimistic scenario that we have considered,  
we learn from the plots that
LISA standard sirens  alone  do not give any significant constraint on $w_0$, and, when combined   to  CMB+BAO+SNe data, they only induce a very marginal improvement. 
From the corresponding one-dimensional likelihoods, from CMB+BAO+SNe only,  we find that $w_0$ can be reconstructed with the accuracy 
$\Delta w_0=0.045$. Combining  CMB+BAO+SNe with standard sirens, 
in the ``optimistic hnd" scenario  we get
\be\label{ourDeltaw0Deltawahnd}
\Delta w_0=0.044 \, ,
\ee
so the improvement due to SMBH standard sirens is quite negligible.
Basically the same results are obtained in  all other scenarios considered.

These results  significantly degrade the estimates presented in~\cite{Tamanini:2016zlh}. As discussed in section~\ref{sect:LCDMresults}, the difference is due to the use of the updated sensitivity curve of LISA, the more realistic assumptions in the construction of the catalogs and estimates of the errors, and the fact that the degeneracies between cosmological parameters are now fully taken into account through a full MCMC.
However,  as for $H_0$, one should be aware of all the assumptions and uncertainties that entered in the construction of the source catalogs, and that could significantly alter the results. 
In particular these estimates can either be improved by combining the cosmological data collected with SMBHBs with the ones collected from other LISA sources, such as EMRIs and SOBHBs, or by extending the observational period of LISA. In the first case the different redshift ranges where EMRIs and SOBHBs are expected to be observed, will help breaking some degeneracies in the cosmological parameters, such as for example the degeneracy between $H_0$ and $\Omega_M$ in $\Lambda$CDM. In the second case instead the estimated errors are expected to improve roughly as $\sim\sqrt{N}$ or better, where $N$ is the number of SMBHB merger with EM counterpart observed by LISA, which linearly depends on the observational period. Both these improvements should reduce the error associated with the statistical scatter of the SMBHB data, which was not considered in \cite{Tamanini:2016zlh} and in the new analysis performed here, which takes into account more realistic SBHB catalogs and an updated LISA noise sensitivity curve, appears to be the most relevant source of error. Other improvements can be achieved by a more realistic and detailed characterization of the emission and detection of the EM counterparts of SMBHB mergers, which here as in \cite{Tamanini:2016zlh} has been modeled only with optical and radio EM emissions, but the consideration of other EM signatures (X-rays, $\gamma$-rays, ...) could lead to a higher number of cosmologically useful events.

\subsection{\texorpdfstring{$(\Xi_0,w_0)$}{}}

We next extend the DE sector by introducing  the parameter $\Xi_0$. In order to keep under control the number of new parameters, which is necessary to ensure the convergence of the MCMC chains,
we only  take $(\Xi_0,w_0)$ as the parameters that describe the DE sector of the theory, fixing $w_a=0$ and $n=2.5$; the latter value is of the order of that suggested by the RR and RT nonlocal models, see section~\ref{Sec:nonlocal}. However, the precise value of $n$ is not very important for the forecasts that we present, 
{ since the uncertainty  on this quantity is  large for $|\Xi_0 - 1| \ll 1$ \cite{Belgacem:2018lbp}}. 
%(delta n/n) = delta(Xi0)/(|1- Xi0|)  (eq 97 of Ref 16).  
 Again, we assume $\Lambda$CDM as the fiducial model used to generate the catalog of sources, so  the  fiducial values for the  parameters $\Xi_0$ and $w_0$ are $\Xi_0=1$ and $w_0=-1$.

\begin{figure}[t]
\begin{center}
\includegraphics[width=0.4\textwidth]{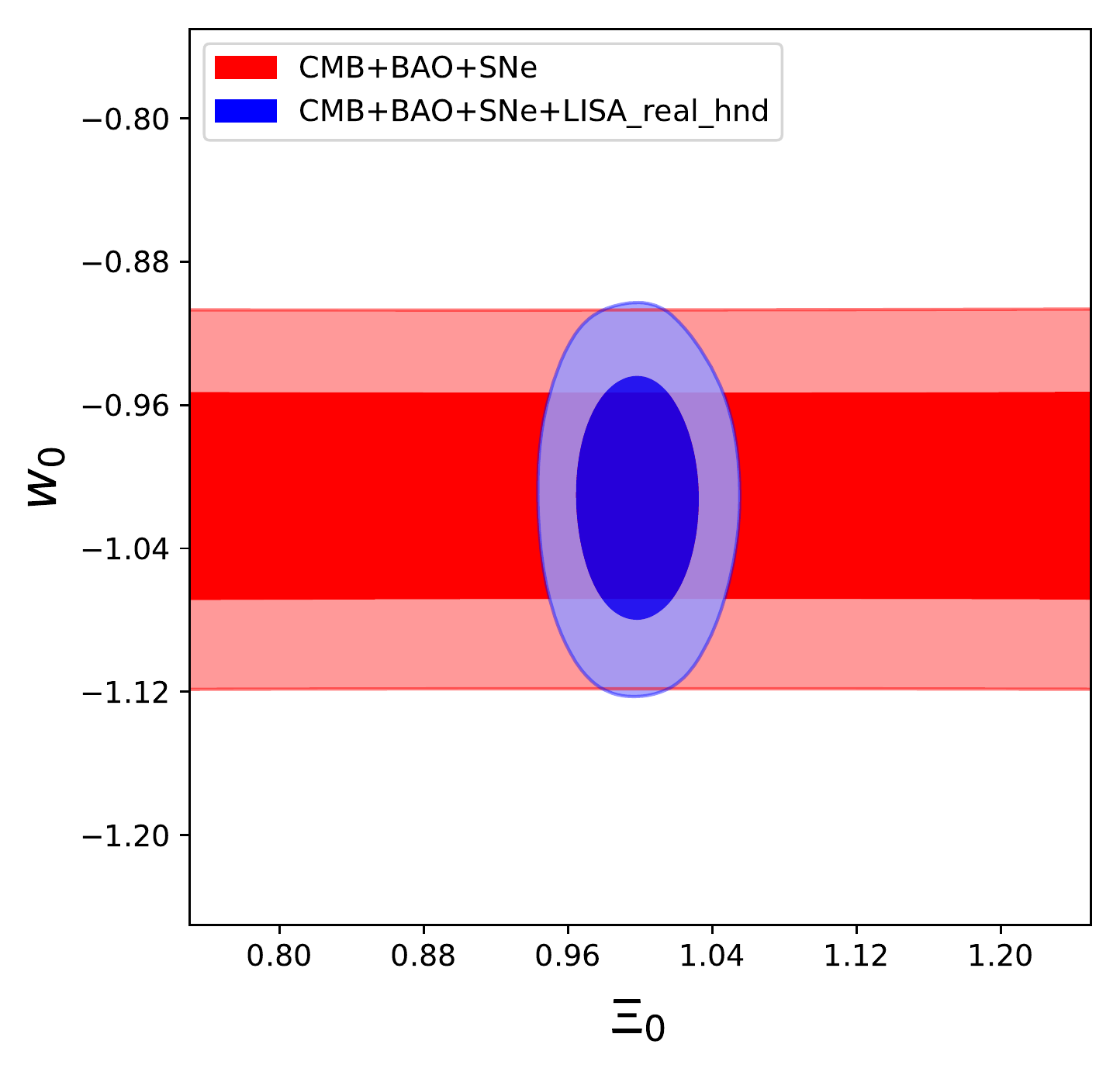}
\includegraphics[width=0.4\textwidth]{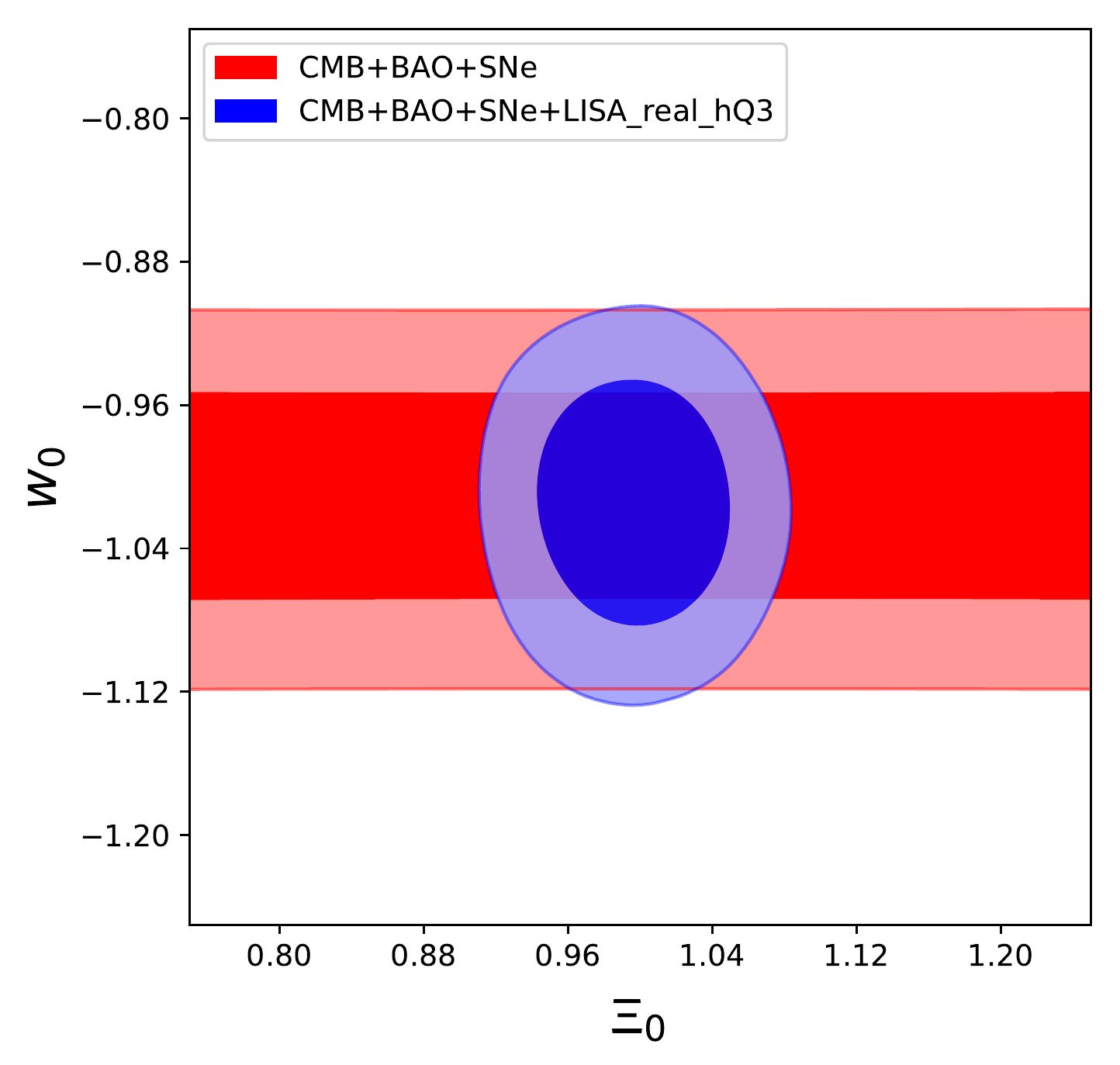}
\includegraphics[width=0.4\textwidth]{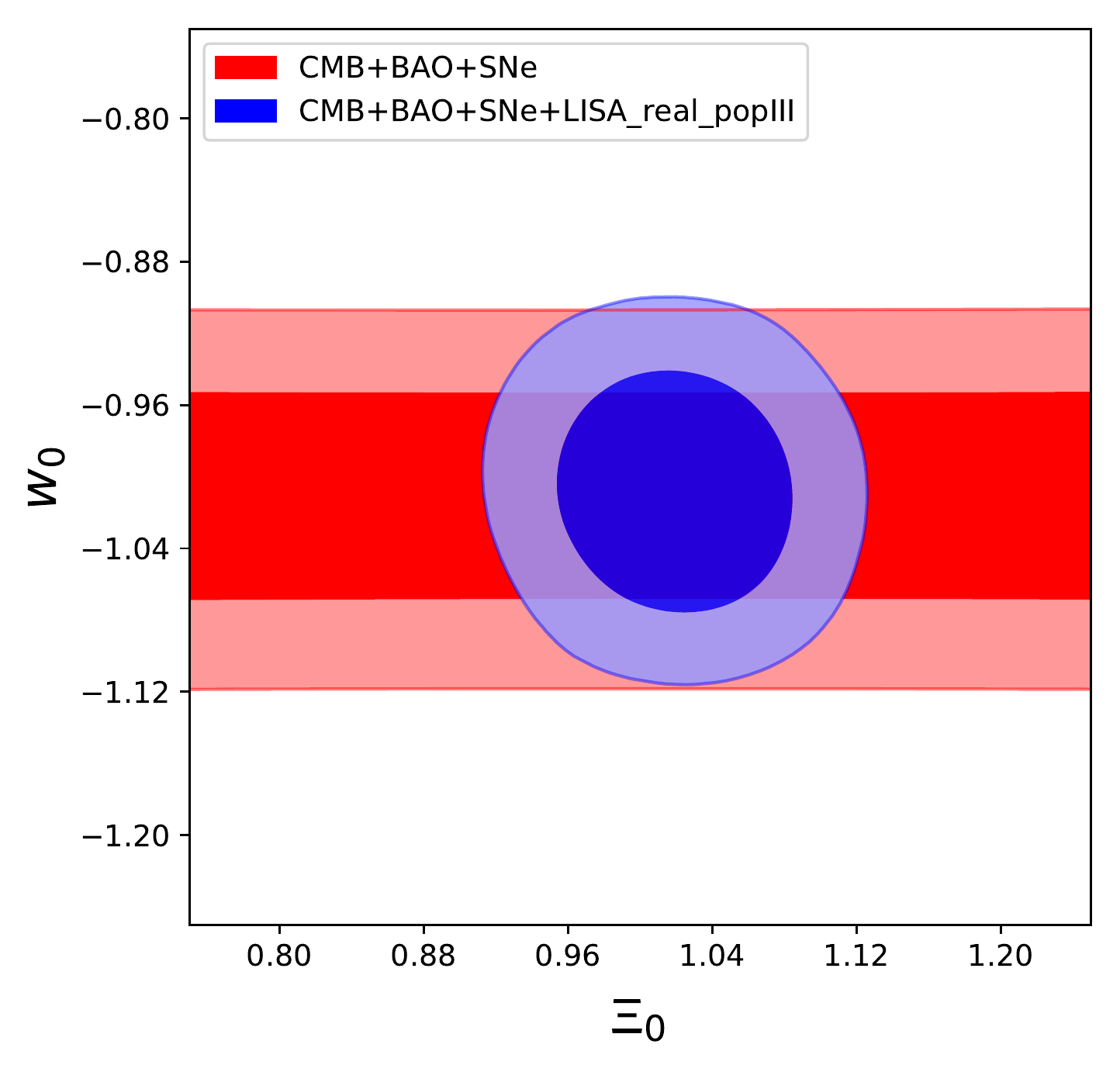}
\end{center}
\caption{The  $1\sigma$ and $2\sigma$
contours  of
 the  two-dimensional likelihood in the $(\Xi_0,w_0)$ plane, with the combined  contribution from 
CMB+BAO+SNe (red)  and the  combined contours
from  CMB+BAO+SNe+LISA standard sirens (blue), in the scenario (1) (``realistic") for the  error on the luminosity distance.  Upper left: heavy no-delay (``hnd") scenario; Upper right: ``hQ3" scenario ; lower panel: ``pop~III" scenario.}
\label{fig:xi0w0real}
\end{figure}

\begin{figure}[t]
\begin{center}
\includegraphics[width=0.4\textwidth]{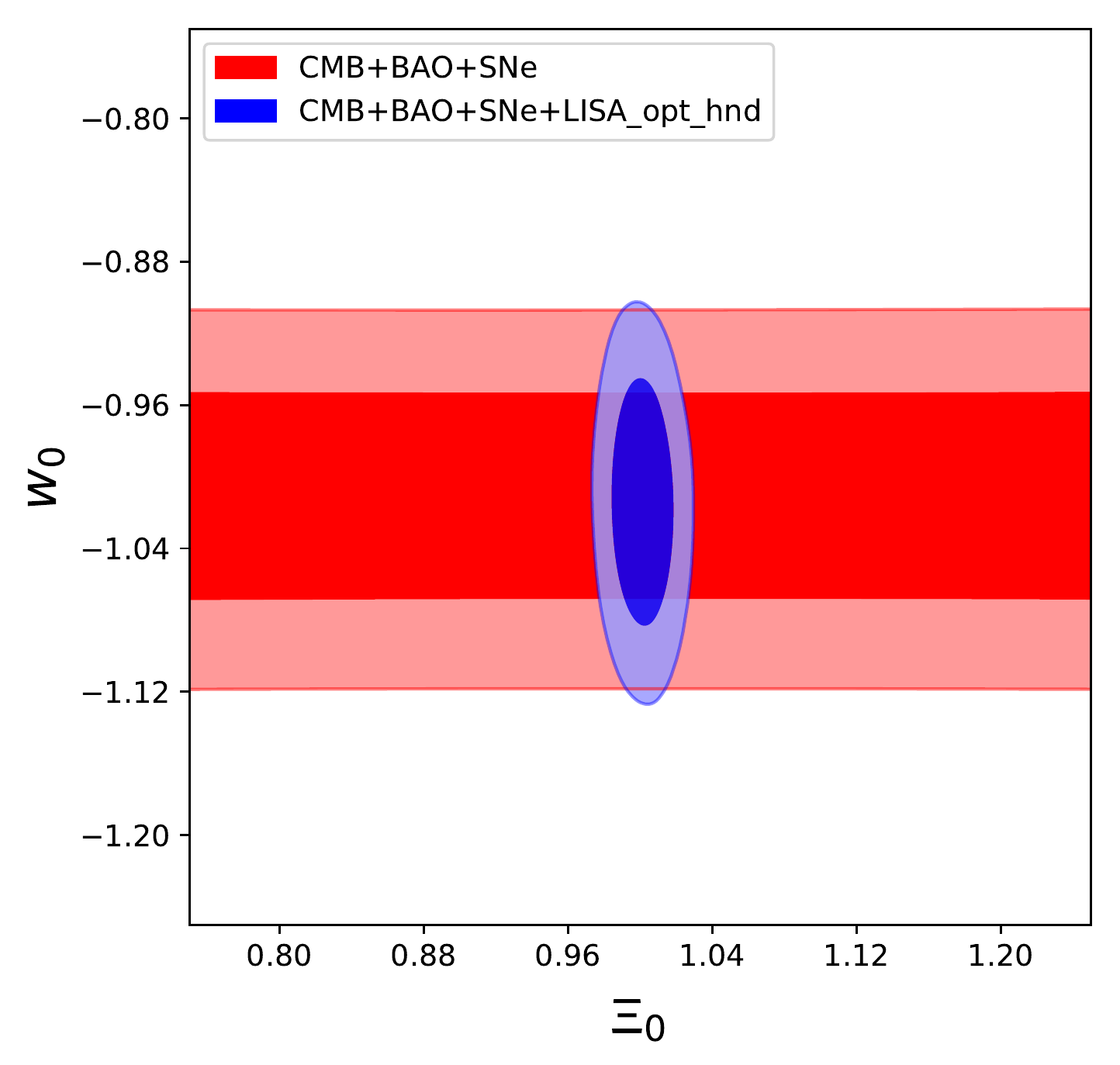}
\includegraphics[width=0.4\textwidth]{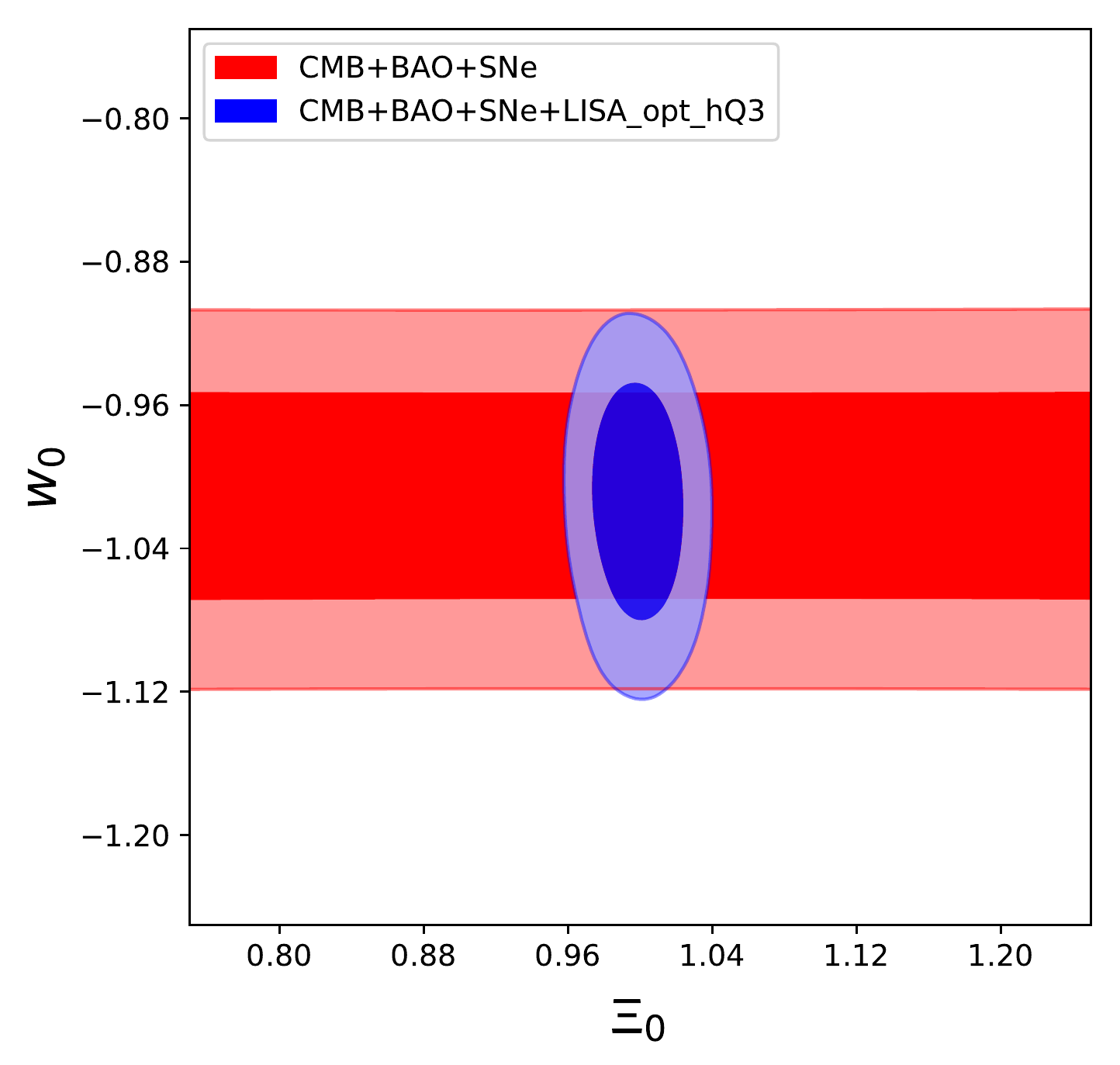}
\includegraphics[width=0.4\textwidth]{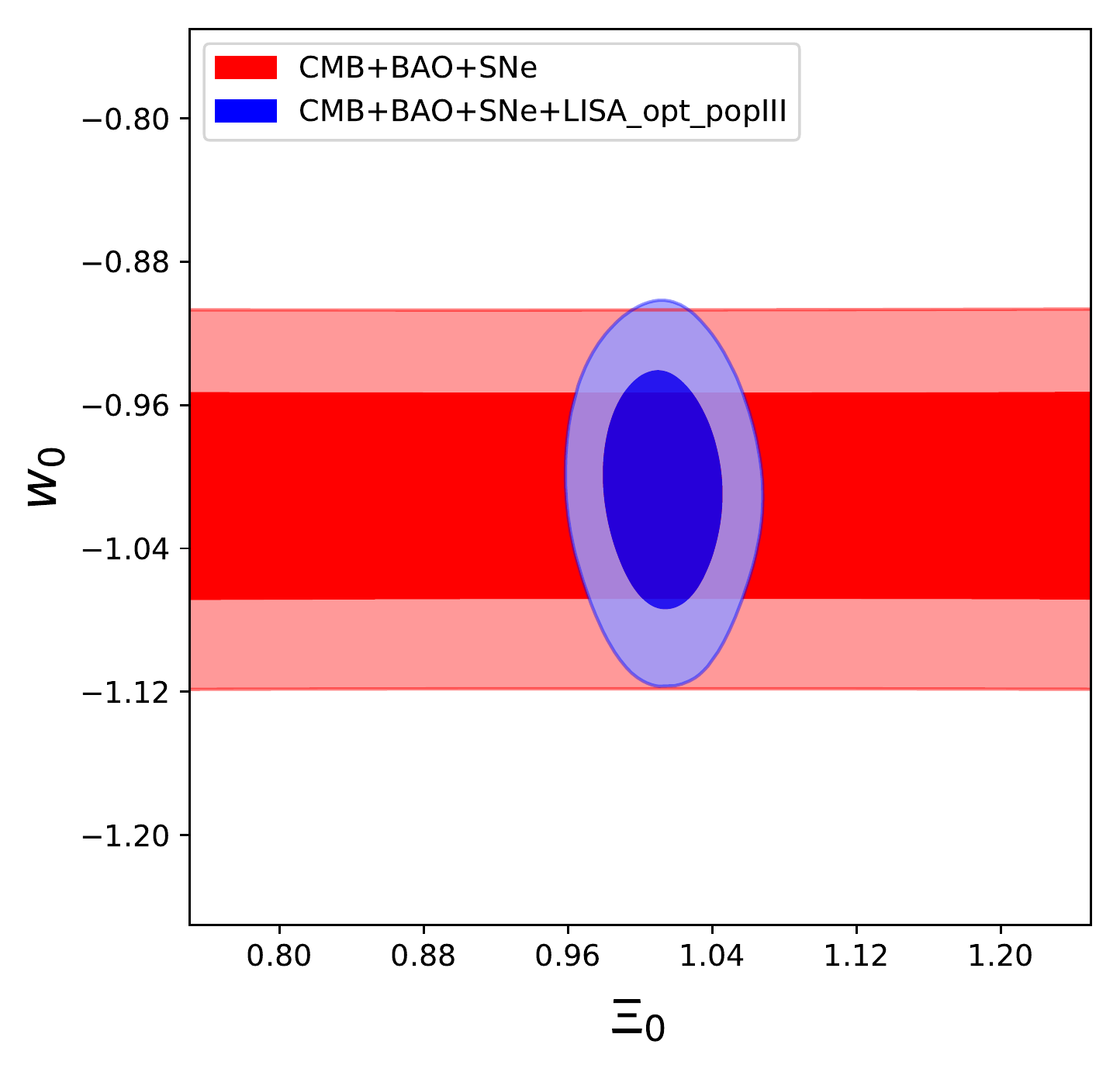}
\end{center}
\caption{As in Fig.~\ref{fig:xi0w0real}, for the scenario (2) (``optimistic") for the error on the luminosity distance.}
\label{fig:xi0w0opt}
\end{figure}

Fig.~\ref{fig:xi0w0real} shows the two-dimensional likelihood in $(\Xi_0,w_0)$ plane, for the realistic scenario for the  error on the luminosity distance, and the three seed  scenarios, while 
Fig.~\ref{fig:xi0w0opt} shows the results obtained using the optimistic scenario for the  error on the luminosity distance.
We plot the  limit from  CMB+BAO+SNe and the combined limit from  CMB+BAO+SNe+LISA standard sirens  while, as before,
using  only standard sirens, with the addition of these new parameters with respect to $\Lambda$CDM,  the MCMC chains fail to converge because of the limited number of sources.
Note that CMB, BAO and SNe, as any other electromagnetic probe, are blind to $\Xi_0$, and therefore the corresponding contour from  CMB+BAO+SNe are flat in the $\Xi_0$ direction. Standard sirens, however, lift this flat direction.
The  errors on $\Xi_0$ and $w_0$ from the corresponding one-dimensional likelihoods are shown in Tables~\ref{tab:Xiw0}.

\begin{table}[t]
\centering
\begin{tabular}{||c|c|c||} 
\hline
{\bf seeds }& $ \Delta\Xi_0$ & {\bf $\Delta w_0$ }  \\ \hline 
hnd    & $0.023$          &   $0.045$            \\ \hline 
hQ3   & $0.036$          &   $0.046$            \\ \hline 
popIII& $0.044$          &   $0.045$            \\ \hline 
\end{tabular}
\hspace*{15mm}
\begin{tabular}{||c|c|c||} 
 \hline 
{\bf seeds } & $\Delta\Xi_0$ & $\Delta w_0$   \\ \hline 
hnd    & $0.011$          &   $0.045$            \\ \hline 
hQ3   & $0.017$          &   $0.044$            \\ \hline 
popIII& $0.022$          &   $0.044$            \\ \hline 
\end{tabular}
\caption{\label{tab:Xiw0} Forecasts for the 1$\sigma$ errors on $\Xi_0$ and $w_0$ from CMB+BAO+SNe+LISA standard sirens,  for the  three seed  scenarios. Left: realistic scenario for the  error on the luminosity distance. Right: optimistic scenario.
}
\end{table}

Exactly  as in the case of $w$CDM discussed above,
we see that the accuracy  that LISA, combined with CMB+BAO+SNe, can reach  on $w_0$  is only of about $4.4\%$ (at least using only MBH binaries with counterpart as standard sirens), which is basically entirely determined  by the current CMB+BAO+SNe observations.

In contrast, even under the current set of assumptions (see the discussion in sect.~\ref{sect:LCDMresults}), that have led to relatively large errors on $H_0$ and on $w_0$ from LISA standard sirens, $\Xi_0$ still turns out to be  an extremely interesting observable for LISA. First of all it can be observed only with GW experiments, and second, as we already anticipated in the discussion in section~\ref{sect:phenopar}, $\Xi_0$ can be measured more accurately than $w_0$. This is confirmed by the results of our MCMC, which shows that, in the best case, $\Xi_0$ can be measured to $1\%$ accuracy, and even in the worst case we still have a $4.4\%$ accuracy, see Table~\ref{tab:Xiw0}. By comparison, for instance, the RT nonlocal model discussed in section~\ref{Sec:nonlocal} predicts a deviation from the GR value $\Xi_0=1$ at the level of $6.6\%$. {\gt  Similar values can be obtained for the scalar-tensor theories discussed
in Section \ref{sect:scalartens}: the $f(R)$ and  the coupled Galileon models  of Section \ref{subsec_hd} and the DHOST system of Section \ref{sub_dhost} can reach deviations of the $\Xi_0$ parameter
of order respectively $5\%$ and $2\%$, see  Figures, \ref{fig:horndeskiparametrisation},  \ref{fig:brans_dicke_and_coupled_gals} and \ref{fig:Luminosity_distance}.} \footnote{In sections \ref{sect:phenopar}, \ref{sub_dhost} we also discussed more complex parametrizations for the quantity $\Xi(z)$, designed
to accommodate the features of specific models:  it should  be straightforward to apply the methods
used so far also to those cases, but we leave this for future works.}
%and similar results can be obtained from the other models discussed.
Thus, the accuracy that LISA can reach on modified GW propagation is extremely interesting for testing modified gravity and dark energy.

It is interesting to compare the results that we have obtained for  LISA with the expectations for a third-generation ground based interferometer such as the Einstein Telescope (ET).  The Einstein Telescope is expected to detect of order $10^5-10^6$ binary neutron stars per year, up to redshift $z\sim 2$. For standard sirens with electromagnetic counterpart, the main uncertainty in the analysis comes from the estimate of the fraction of  coalescences that will have an observed electromagnetic counterpart. This depends not only on details of the electromagnetic emission, that can in principle be modeled, but, given this very high event rate,  also on issues that are currently difficult to predict, such as what will be the network of GRB detectors and of ground-based telescopes that will be available at the time of ET, and what fraction of telescope time will be devoted to the electromagnetic follow up of gravitational events. A rather common working hypothesis is that, out of these $10^5-10^6$ events per years, in a few years of running one could collect of order $10^3$ events with electromagnetic counterpart. Under this hypothesis, it was found in  \cite{Belgacem:2018lbp} that, by combining ET with the same CMB+BAO+SNe dataset that we are using in this paper, and extending the dark energy sector through the parameters $\Xi_0$ and $w_0$ (and fixing $n=2.5$, as in this paper), at ET one can get the accuracy $\Delta \Xi_0=0.008$ and $\Delta w_0=0.032$. Comparing with the results for LISA in Table~\ref{tab:Xiw0} we see that the accuracy at ET would be slightly better  than that in the `hnd'  scenario for LISA, although quite comparable. However, this comparison relies heavily on the assumption of having $10^3$ sources with counterpart, which at the present stage is only a reasonable working hypothesis.~\footnote{A more detailed study of the accuracy on 
$(\Xi_0,w_0)$ obtained with a realistic modeling of the  joint GW-GRB observations between ET and the proposed THESEUS mission will be presented in~\cite{Belgacem:inprep}.} One should also observe that ET and LISA observations are complementary because most of the ET events are  at redshift $z \, \lsim \, 1$, while, as we see from Fig.~\ref{fig:catalog_dl}, LISA has events distributed up to $z\sim 6$. This will allow us to study the ratio $d_L^{\,\rm gw}(z)/d_L^{\,\rm em}(z)$ in different redshift ranges. In terms of the $(\Xi_0,n)$ parametrization, this means that LISA will be able to determine the asymptotic value $\Xi_0$, without much contamination from the value of $n$ (and, in turn, it will not be able to determine $n$ very accurately), while the lower redshift events of ET will rather be sensitive to a combination  of $\Xi_0$ and $n$, so the joint  LISA and ET data would lift this degeneracy.

\section{Conclusions}\label{sect:conclusions}

In this paper we have examined the potential of LISA MBH binaries (with electromagnetic counterparts)  for constraining cosmology, and in particular the {dark energy sector associated with modified gravity theories}.  
As in \cite{Tamanini:2016zlh}, we have generated a population of MBHBs using semi-analytic models for the formation and evolution of galaxies, and we have examined three different scenarios for massive black hole formation at high redshifts, namely pop~III seeds, and heavy seeds with and without a delay between galaxy merger and the merger of the central MBHs.

Compared to earlier work specifically related to LISA, in particular ref.~\cite{Tamanini:2016zlh}, we have used updated estimates for the LISA configuration and sensitivity curve, we examined a scenario with a more realistic error for the redshift determination of the source and delensing, we produced more realistic mock catalogs by including the scattering of the mock data according to the observational error, and we took fully into account the degeneracies between cosmological parameters by performing a series of MCMCs, both with standard sirens only and with standard sirens combined with current CMB+BAO+SNe data.  On the other hand, we have emphasized (see the discussion at the end of sect.~\ref{sect:LCDMresults}) that many more issues must be addressed before giving a final word of the sensitivity of LISA to cosmological parameters, and this paper should be considered as a contribution toward that goal, under the given assumptions that we have discussed in detail.

One result of this analysis is that, under the assumptions that we have used, the estimate for the accuracy that can be obtained for 
the Hubble parameter using LISA MBH  is revised toward higher values. In particular, in the most favorable formation scenario, which turns out to be the heavy-seed no-delay one, 
for the relative error on $H_0$ from standard sirens  we find $3.8\%$ assuming the ``optimistic" scenario for the accuracy of redshift measurement and delensing, and $7.7\%$ with possibly more ``realistic" assumptions on redshift determination and delensing. We have presented similar results  for the DE equation of state, where we found that, under the assumptions we made, the inclusion of MBH LISA standard sirens does not improve significantly the determination of $w_0$ compared to current CMB+BAO+SNe observations.
Apart from the role of the assumptions that we have used, it  is also important to stress that LISA will also see other potential GW standard sirens,  in particular stellar mass BH binaries, and extreme mass ratio inspirals (EMRIs). These sources are not expected to have an electromagnetic counterpart, but can still be used as standard sirens by using the statistical method. Work on this is currently in progress within our LISA Cosmology Working Group, and could lead to a significantly better error on $H_0$ and $w_0$.

The most interesting results concerning the potential of LISA for cosmology studies come however from modified GW propagation, which represents the core of this paper. The fact that a modified friction term in the propagation equation of tensor perturbations over FRW can give rise to a modification of the luminosity distance of standard sirens has been recognized in recent years through the study of explicit models such as scalar-tensor theories and nonlocal gravity \cite{Saltas:2014dha,Lombriser:2015sxa,Nishizawa:2017nef,Belgacem:2017ihm,Belgacem:2018lbp} (although the possibility of a modification of the luminosity distance due to gravitons leaking in the bulk was already observed in the context of the DGP model~\cite{Deffayet:2007kf}). In this paper we analyzed   the explicit predictions of some of the best-motivated and most studied modifications of gravity, such 
as scalar-tensor systems (Horndeski and  the more general DHOST family of scalar-tensor theories), nonlocal gravity, bigravity, and theories with extra and varying
dimensions.
 We have found that modified GW propagation is an absolutely generic phenomenon, that takes place in all the theories that we have considered (in spite  of the fact that all these theories comply with the limit from GW170817 on the deviation of the speed of GWs from the speed of light).

We have seen that, in most cases, a simple and accurate description of the effect is captured by the $(\Xi_0,n)$ parametrization of Eq. \eqref{eq:param} originally introduced in \cite{Belgacem:2018lbp}. This parametrization takes into account the fact that the ratio of gravitational to electromagnetic luminosity distances must go to one as the redshift $z\ra 0$, and in most model it saturates to a constant, $\Xi_0$, at large redshift (because in typical  models dark energy is a recent phenomenon on cosmological scales), and smoothly interpolates between these asymptotic values with a power-law determined by the index $n$. Of course, deviations from this simple behavior can occur in some cases because of  specific physical reasons. Most notably, in bigravity we have found a series of oscillations due to the ``beatings'' between the two metrics.   It follows that in some specific cases
 alternative analytic formulas can be useful, and we have presented some alternative parametrizations.

We have run a series of MCMCs to determine the accuracy that LISA can reach on the parameter $\Xi_0$ in the $(\Xi_0,n)$ parametrization of  eq.~\eqref{eq:param}.\footnote{The parameter $n$ is  less important if the sources are at large redshifts, where the expression for the ratio of gravitational and electromagnetic luminosity distances saturates to its asymptotic value $\Xi_0$. In any case, the accuracy that can be  reached on it can be estimated analytically as
$\Delta n =[\Delta \Xi_0/|1-\Xi_0| ]\,\times [(1+z)^n-1]/[\log (1+z)]$ where $z$ is the typical redshift of the sources~\cite{Belgacem:2018lbp}. In the low-redshift limit this reduces to 
$\Delta n/n =\Delta \Xi_0/|1-\Xi_0|$.}
 There are two reasons that make this observable especially important for advanced GW detectors. First of all, modified GW propagation is an observable accessible only to GW observations, and to which electromagnetic observations are blind.\footnote{More precisely, we have seen that  in some models (but not in general), modified GW propagation is related to a time dependence of the effective Newton constant, and in these models one could access $\Xi_0$ indirectly through the effect of the modified Newton's constant on structure formation.} Second, as discussed in section~\ref{sect:phenopar}, in a generic modified gravity theory, in which  the deviation of $d_L^{\,\rm gw}(z)/d_L^{\,\rm em}(z)$ from  1 is of the same order as the deviation of $\wde(z)$ from $-1$, the effect of 
$d_L^{\,\rm gw}(z)/d_L^{\,\rm em}(z)$ on standard sirens dominates over the effect of the DE equation of state, because the latter is partially compensated by 
{\gt degeneracies with other (fitted) cosmological parameters.} 
%Bayesian parameter estimation. 
 As a consequence, the accuracy expected on $\Xi_0$ is better than that on $w_0$. This argument has been confirmed by our explicit MCMC computations. Combining LISA with CMB+BAO+SNe to reduce the degeneracies with the other parameters, 
   in the best case (heavy-seed no-delay formation scenario and  ``optimistic" scenario for the accuracy of redshift measurement and delensing)  
   we have found that  $\Xi_0$ can be measured to an accuracy that reaches $1.1\%$ (to be compared with $4.5\%$ for $w_0$) and even in the worst scenario still is $4.4\%$ (see Table~\ref{tab:Xiw0}).
Last but not least, in several instances the explicit models that we have considered give predictions for $\Xi_0$ larger or equal than these values. For instance the RT model 
 {\gt of Section \ref{Sec:nonlocal}} 
predicts for $\Xi_0$ a deviation from the GR value of order $6.6\%$. {\gt  Similar values can be obtained for the scalar-tensor theories: the $f(R)$ and coupled Galileon models of Section \ref{subsec_hd} and the DHOST system of Section \ref{sub_dhost} can reach deviations of the $\Xi_0$ parameter
of order  $3\%$, $5\%$ and $2\%$ respectively. {Due to their complex dynamics, the bigravity set-up studied in section \ref{Sec:bigravity} and the dimensionally changing systems of section \ref{sec:QG} cannot be faithfully described} in terms of the $(\Xi_0, n)$ parameterization of  Eq. \eqref{eq:param}.  We nevertheless  quantitatively derived projected constrains on their parameter spaces from standard siren catalogs.  }

At the theoretical level, we have examined the predictions on modified GW propagation of a large number of the best studied modified gravity theories. In particular,
we demonstrated the improved capability of LISA to probe GW oscillations, an analog to neutrino flavour oscillations, present in models with {extra} tensor interactions such as bigravity,
 potentially constraining the parameter space of specific theories. 
 We compute predictions for bigravity using a high frequency expansion, focusing on the high mass regime $m_g\gg H_0$.  {In this region of parameter space GW oscillations occur in the mHz range, but the theory is not generically able to account for cosmic acceleration.
Standard sirens at cosmological distances, as will be observed by LISA, have the potential to constrain the mass range $m_g \gtrsim 10^{-25}$eV for most mixing angles,  which would improve by three orders of magnitudes the result obtained from the current LIGO-Virgo detection of GW170817. 
This is a conservative estimate: including frequency-dependent effects on the waveform will improve these bounds. Notice also that
the improvement in constraining power in GW oscillations (e.g. $m_g$) over LIGO/Virgo will not be matched by standard-siren analyses with third-generation ground detectors, as the extended reach of LISA is driven mainly by the lower frequency range (which increases the sensitivity since the signal is suppressed by $(m_g/f)^2$), in addition to a longer oscillation baseline (higher source redshift).
}

The broad conclusion is that modified GW propagation is a prime observable for cosmological studies with advanced GW detectors, and 
MBH binaries detectable with LISA can be a powerful probe of modified gravity and dark energy. Further significant improvements are expected by the study of LISA standard sirens without  electromagnetic counterpart, such as stellar mass BH binaries and EMRIs, that can be used as standard sirens through the statistical method. We leave this for future work.

\vspace{5mm}

\acknowledgments
 It is a pleasure to thank Chiara Caprini for useful discussions,  the LISA Cosmology Working Group Chairs --  Robert Caldwell, Chiara Caprini, Germano Nardini -- for their input and support during the project, and Vuk Mandic and Alberto Sesana for carefully reading  our manuscript and for their useful suggestions. 
The work  of E.B, S.F. and M.M. is supported by the Fonds National Suisse and  by the SwissMap National Center for Competence in Research.
The MCMCs have been run on the Baobab cluster of the University of Geneva.
This project has received funding (to E. Barausse) from the European
Research Council (ERC) under the European Union  Horizon 2020
research and innovation programme (grant agreement no. GRAMS-815673;
project title ``GRavity from Astrophysical to Microscopic Scales'').
This work was supported by the H2020-MSCA-RISE-2015 Grant No.
StronGrHEP-690904. N.B., D.B., A.G and S. M. acknowledge financial support by ASI Grant 2016-24-H.0.
  G.C.\ is supported by the I+D grant FIS2017-86497-C2-2-P of the Spanish Ministry of Science, Innovation and Universities. M.C. is supported by the Labex P2IO and the Enhanced Eurotalents Fellowship.
 C.D. and L.L. were supported by a Swiss National Science Foundation Professorship grant (No. 170547). J.M.E. is supported by the Spanish FPU Grant No. FPU14/01618, the Research Project FPA2015-68048-03-3P (MINECO-FEDER) and the Centro de Excelencia Severo Ochoa Program SEV- 2016-0597. 
 M.R.F. acknowledges  support from STFC grant ST/N000668/1. 
J.G.B. acknowledges support from the Research Project FPA2015-68048-03-3P [MINECO-FEDER] and the Centro de Excelencia Severo Ochoa Program SEV-2016-0597. 
A.K. was supported by the Centre National d'{\'E}tudes Spatiales: he also acknowledges support from the H2020-MSCA-RISE-2015 Grant No. StronGrHEP-690904.
 M.S. is supported in part by STFC grant ST/P000258/1.  G.T. is partially supported by STFC grant ST/P00055X/1.
 M.Z. is supported by the Marie Sklodowska-Curie Global Fellowship Project NLO-CO.

\begin{appendix}
\section{GW luminosity distance and the flux-luminosity relation}\label{app-A}

In this Appendix we show that the GW luminosity distance in modified gravity theories still obeys the standard relation (\ref{fluxluminosity}) between the energy flux ${\cal F}$  measured in the observer frame and  the intrinsic luminosity ${\cal L}$
measured at the source frame. Using purely kinematic
 consideration discussed for example in \cite{Maggiore:1900zz}, Section 4.1.4, we derive an expression for the gravity wave luminosity distance in a case where the graviton
 is massless, $m_T=0$, and the non-minimal coupling with other fields are such that  $\Pi_{ij}=0$.
 GW are then produced by events at high redshift associated with  standard sirens, and then can propagate freely through space-time. (We examine the effect of a graviton mass and couplings with additional spin-2 fields  
 in Section \ref{Sec:bigravity}.)

 In this case, the 
evolution equation \eqref{gen-ev-eq1} can  be obtained by varying the effective quadratic action
for tensor modes
\be
S_T^{(2)}\,=\,\bar M^2\,\int d \eta\,d^3 x\,a^2(\eta)\,\left(\frac{M_{\rm eff}^2(\eta)}{\bar M^2}\right)\,\left[\frac12 h_{ij}'^2 -\frac12 {c_T^2(\eta)} \left(\vec \nabla h_{ij} \right)^2\right]\,.
\ee
Here $\bar M$ is a reference mass scale, whose value we will discuss later.
Studying the GW dynamics governed by this action is equivalent to studying  the propagation of free massless modes on 
 a homogeneous FRW geometry characterised by the line element
\be
d s^2\,=\,a^2(\eta)\,\frac{M_{\rm eff}^2(\eta)}{\bar M^2}\,{c_T(\eta)}\,\left(- {c_T^2(\eta)}\,d \eta^2+{d \vec x^2}\, \right) \label{eff-back}
\,.
\ee
This   expression for the homogeneous background metric  leads to the following
 formula for the   comoving distance between a source and the observer, computed in terms
 of the tensor
 null geodesics $d s \,=\,0$ (we use the relation $d \eta\,=\,d t/a(t)$ connecting
 physical and conformal time): 
  \be \label{rcom1}
r_{\rm com}(t)\,=\, |\Delta x|\,=\,\int_{t_s}^{t}\,\frac{c_T(\tilde t)\,d  \tilde t}{ a( \tilde t)}\,,
\ee
with $t$ an arbitrary time. We denote by $t_{s}$   the time of emission of GW at the source, while $t_0$ the observation time, which we assume being today. The comoving distance depends on the speed of the GW. The physical distance results:
 \be \label{rphys1}
r_{\rm phys}(t)\,=\,a(t)\,\frac{M_{\rm eff}(t)}{\bar M}\, {c_T^{1/2}}(t)
\,r_{\rm com}(t)\,.
%
%r_{com}\,=\, |\Delta x|\,=\,\int_{t_{s}}^{t_{0}}\,\frac{c_T(t)\,d t}{a(t)}\,,
\ee
We now {\em choose} our reference normalization scale $\bar M$ so that in the limit $t\to t_s$
the ratio between physical  and comoving distance 
 acquires the standard expression 
 \be \label{reque1}
\lim_{t\to t_s}\frac{ r_{\rm phys}(t)}{r_{\rm com}(t)}\,=\,a(t_s)
 \ee
at the time of the emission from the source. This might be motivated by requiring that nearby the source
the effects of modified gravity had not yet have time to develop. Condition \eqref{reque1} leads to the definition
\be \label{defbm}
\bar M\,\equiv\,M_{\rm eff}(t_s) \,{c_T^{1/2}}(t_s)
\ee
that we will use in what follows.

Suppose now that the observer measures GW signals
corresponding to wavecrests emitted at different times
from the source. 
 (They travel through the same comoving distance.) 
Using expression \eqref{rcom1} for $r_{\rm com}(t_0)$, at linear order in $\Delta t_{s}$ we can write the relation
\be
\Delta t_{0}\,=\,\frac{c_T(t_{s})}{c_T(t_0)} \,\frac{a(t_0)}{a(t_{s})}\,
%\frac{M_{\rm eff}(t_{obs})}{M_{\rm eff}(t_{em})}
\,\Delta t_{s} \label{tick1}
\ee
between the time difference of two GW wavecrests as measured at emission and observation times.  
%We choose coordinate such that $t_{\rm today}\,=\,0$. 
Formula  \eqref{tick1}, which we  more conveniently express  in terms of redshift, 
 states that   source and observer clocks tick with different
rates. We assume that the observer makes its measure today at redshift equal to zero, while the  emission occurs at resdshift
$z$. Then,  from eq. \eqref{tick1}, we can write
\be
d t_{0}\,=\,\frac{c_T(z)}{c_T(0)}\,
%\frac{M_{\rm eff}(0)}{M_{\rm eff}(z)}\,
\left(1+z \right)\,d  t_s \,,\label{tick2}
\ee
implying that  frequencies measured in the source and observer frames are related by 
\be \label{freqob1}
f^{(0)}\,=\,\frac{f^{(s)}}{\frac{c_T(z)}{c_T(0)}\,
%\frac{M_{\rm eff}(0)}{M_{\rm eff}(z)}\,
\left(1+z \right)}\,.
\ee

%We now derive the luminosity distance for GW propagating
%in the effective background metric \eqref{eff-back}, using the standard methods explained in Maggiore, Volume 1.

We now define the luminosity distance, following  \cite{Maggiore:1900zz}. We call ${\cal F}$ the GW energy flux measured
by the observer, corresponding to the amount of  GW energy per unit time per unit area.  ${\cal L}$ is the luminosity of the source, defined 
as the power it radiates
\be
{\cal L}\,=\,\frac{dE_s}{dt_s}
\,.
\ee
Then 
we define the luminosity distance $d_L^{\rm gw}$ as
\be\label{defdLgwfromF}
{\cal F}\,\equiv\,
\frac{\cal L}{{\text{Area}}}
\,\equiv\,
\frac{\cal L}{4 \pi\,(d_L^{\rm gw})^2}\,.
\ee
Since we measure the energy flux at the observer position, we need to convert
$dE_s/dt_s$ into the observer frame. The energy scales as the frequency, 
eq. \eqref{freqob1}, while $d t_s$ and $d t_{0}$ are related by eq.  \eqref{tick2}. 
Hence
\be
\frac{dE_{0}}{dt_{0}}\,=\,\frac{\cal L}{\frac{c_T^2(z)}{c_T^2(0)}\,
%\frac{M^2_{\rm eff}(0)}{M^2_{\rm eff}(z)}
\,\left(1+z \right)^2}\,.
\ee
The area crossed  by the flux of  GW which propagate radially from the source is [using eq. \eqref{defbm}]
\bea
{\text{ Area}}
&=&4 \pi r^2_{\rm phys}
\\
&=&4 \pi\,\frac{M^2_{\rm eff}(t_0)}{M_{\rm eff}^2(t_s)}\,a^2(t_0)\,\frac{c_T(t_0)}{c_T(t_s)}\,r^2_{\rm com}\,.
\eea
%where we identify -- as explained above -- $\bar M\,=\,M_{\rm eff}(t_s)$.
 We now 
collect  the various pieces of information, and  obtain the following expression for the luminosity  distance [$a(0)=a(z=0)$ is the value of scale
factor today]
\bea
d_L^{\rm gw}&=&a(0)\,\sqrt{\frac{c_T(z)}{c_T(0)}}\,\frac{M_{\rm eff}(0)}{M_{\rm eff}(z)}
%\frac{M^2_{\rm eff}(0)}{M_{\rm eff}(z)}
\,
\left(1+z \right)\,r_{\rm com}
\\
&=& \sqrt{\frac{c_T(z)}{c_T(0)}}\, \exp{\left[-
\,\int_0^z\,\frac{\delta(\tilde z)}{1+\tilde z}\,d \tilde z 
\right]}
\,\left(1+z \right)\, a(0) r_{\rm com}\,.
\eea
Using the fact that 
\be
\frac{d t}{a(t)}\,=\,-\frac{1}{a(0)}\,\frac{d z}{H(z)}\,,
\ee
and eq. \eqref{rcom1}, 
we can write the following relation for the comoving distance for  the GWs:
\be
a(0)\,r_{\rm com}\,=\,\int_0^z\,\frac{c_T(\tilde z)\,d \tilde z}{%M_{\rm eff}(\tilde z)\,
H( \tilde z)}\,.
\ee
%\item
The luminosity distance for GW defined from \eq{defdLgwfromF} can then be written as
\be
d_L^{\rm gw}\,=\, \sqrt{\frac{c_T(z)}{c_T(0)}}\, \exp{\left[-
\,\int_0^z\,\frac{\delta(\tilde z)}{1+\tilde z}\,d \tilde z 
\right]}
\,\left(1+z \right)\,\int_0^z\,\frac{c_T(\tilde z)\,d \tilde z}{%M_{\rm eff}(\tilde z)\,
H( \tilde z)}\,,
\ee
and therefore agrees with \eq{dLgwdLemcT}.

%BIGRAVITY
\section{Technical details on bigravity}
\subsection{Hassan--Rosen theory of bigravity}\label{app-bigravity}

The theory known as bigravity \cite{Hassan:2011zd} {is the only known theory of two interactive spin-2 fields that is free of ghosts at the fully non-linear level.} See \cite{Hinterbichler:2011tt,deRham:2014zqa,Schmidt-May:2015vnx} for reviews.
It is described by the action
\be
S\,=\,\int d^4 x\,\left\{
\kappa\,M_{\rm Pl}^2\,\sqrt{- \tilde g}\,\tilde R
+\sqrt{-g}\,\left[
M_{\rm Pl}^2 \left( R-2 m^2\,V\right)+{\cal L}_{\rm matt}
\right]
\right\} \,,
\ee
with $g_{\mu\nu}$
and $\tilde g_{\mu\nu}$ the two metric tensors, 
 $M_{\rm Pl}^2$ and $\kappa M_{\rm Pl}^2$ the corresponding squares of 
 Planck masses and $m$ the graviton mass. Matter is coupled only to the first metric. The interaction potential between the two metrics is indicated by $V$, and it takes the form 
 \be \label{bigpot}
 V\,=\,\sum_{n=0}^4\,a_n \,V_n \,,
 \ee
 with $a_n$ dimensionless parameters, and
 \begin{eqnarray}
 V_0&=&1 \,,
 \\
  V_1&=&\tau_1 \,,
  \\
    V_2&=&\tau_1^2-\tau_2 \,, 
      \\
    V_3&=&\tau_1^3-3 \tau_1 \tau_2+2 \tau_3 \,, 
    \\
    V_4&=&\tau_1^4-6 \tau^2_1 \tau_2   +8 \tau_1 \tau_3 3 \tau_2^2-6 \tau_4 \,,
 \end{eqnarray}
 where $\tau_i \,=\, {\rm tr} \left[ Y^i \right] $ 
 with $Y_\mu^{\,\,\nu}\,=\,[\sqrt{X}]_\mu^{\,\,\nu}$.
 
 This theory admits homogeneous FRW configurations described by two independent metrics
\begin{eqnarray}
d s^2&=&a^2(\tau) \left(-d \tau^2+d \vec x^2 \right) \,,
\\
d \tilde s^2&=&\omega^2(\tau) \left(-c^2(\tau)\,d \tau^2+d \vec x^2 \right) \,.
\end{eqnarray}
The ratio of scale factors is denoted with 
\be 
\xi(\tau)\,=\,\frac{\omega(\tau)}{a(\tau)}\,, \label{xibi}
\ee
and from now on $\Hc=a'/a$ is the Hubble parameter corresponding to the first metric. The Friedmann
equation for the first metric reads
\be \label{eq:bigravityFriedmann}
\frac{\Hc^2}{a^2}=\frac{8\pi G}{3}\rho+m^2\lpar2a_3\xi^3+2a_2\xi^2+a_1\xi+\frac{a_0}{3}\rpar\,.
\ee 
The theory admit two branches of solutions, but only one describes physically interesting cosmological configurations \cite{Comelli:2012db}. 
In this branch the Bianchi identities are realised in the form
\be
c(\tau)-1=\frac{1}{\Hc(\tau)}\frac{\xi'(\tau)}{\xi(\tau)}\,,
\ee
and together with Friedmann equations lead to an algebraic equation for $\xi$:
\be \label{eq:bigravityBianchi}
\frac{8a_4}{\kappa}\xi^2+\frac{6a_3}{\kappa}\xi+\frac{2a_2}{\kappa}+\frac{a_1}{3\kappa}\frac{1}{\xi}=\frac{\Hc^2}{m^2a^2}\,,
\ee
where the coefficients $a_i$ are the parameters of the bigravity potential \eqref{bigpot}. This information about homogeneous configurations
is sufficient for the scope of this work, more details can be found in \cite{Volkov:2011an,vonStrauss:2011mq,Comelli:2011zm,Comelli:2012db}. 
%---WKB APPROX---
\subsection{Details on the WKB approximation for bigravity}
\label{app:wkb}

In this Appendix we spell out  details on the calculations using a WKB approximation leading to the solution (\ref{eq:h1sch}--\ref{eq:h2sch}) for the system of equations (\ref{eq:bigravitymodes}) we use in the main text. 
In  order to obtain WKB solutions, it is convenient to work in matrix notation. After absorbing the cosmic friction term and defining a vector containing the two tensor modes,
\be
\Phi\,=\,\bpm ah_1 \\ ah_2 \epm\,,
\ee
we can express the evolution equation (\ref{eq:bigravitymodes}) for $h_1$ and $h_2$ in bigravity as
\be
\lbra \frac{d^2}{d\tau^2} + \nM\frac{d}{d\tau} +\cM k^2 + \mM-(\Hc^2+\Hc')\iM-\Hc\nM\rbra \Phi =0\,, 
\ee
where we have defined
\begin{align}
\nM&=\bpm 0 & 0 \\ 0 & 2\xi'/\xi-c'/c  \epm, & \cM&=\bpm 1 & 0 \\ 0 &  c^2\epm\,, & \mM&=m^2a^2f_1\bpm 1 & -1 \\ -c/(\kappa\xi^2) &  c/(\kappa\xi^2)\epm\,. 
\end{align}
Next, we introduce a dimensionless expansion parameter $\epsilon$, rescaling time as $d\tau\rightarrow d\tau/\epsilon$, and controlling 
different orders in a high frequency WKB  approximation.
We make the following Ansatz for the solution $\Phi$ for tensor modes,
\be \label{ansphi1}
\Phi=\eM\,e^{i\int\frac{\tM}{\epsilon}d\tau}\lpar\Phi_0+\epsilon\Phi_1+\cdots\rpar\,,
\ee
and aim to solve order by order in the expansion parameter $\epsilon$. In eq. \eqref{ansphi1} we expand over a basis of eigenvectors 
 controlled by the matrix $\eM$, associated with  the matrix of eigenfrequencies $\tM$, which appears at the exponent of \eqref{ansphi1}
 and is diagonal by definition.
 At leading order $\epsilon^0$ of our expansion parameter, we obtain the eigenfrequencies 
\be
\begin{split}
\theta^2_{1,2}=\frac{1}{2}\Big((1+c^2)k^2&+m^2a^2f_1\lpar1+\frac{c}{\kappa\xi^2}\rpar-2\Hc^2 \\
&\mp\sqrt{4m^4a^4f_1^2\frac{c}{\kappa\xi^2}+\lpar\lpar1-c^2\rpar k^2+m^2a^2f_1\lpar1-\frac{c}{\kappa\xi^2}\rpar\rpar^2}\Big) \,,
\end{split}
\ee
and the matrix of eigenvectors
\be
\eM=\bpm 1 & \frac{m^2a^2f_1}{k^2+m^2a^2f_1-\Hc^2-\theta_2^2} \\ \frac{m^2a^2f_1c/(\kappa\xi^2)}{c^2k^2+m^2a^2f_1c/(\kappa\xi^2)-\Hc^2-\theta_1^2} & 1\epm\,. 
\ee
At next to leading order, $\epsilon^1$, the amplitude $\Phi_0$ can be obtained solving
\be
2\eM\tM\gM\Phi'_0+\lpar \eM\tM'+2\eM'\tM+i\mathcal{H}'\eM+i\mathcal{H}\nM\eM+\nM\eM\tM\rpar\gM\Phi_0=0\,,
\ee
where we defined for convenience the matrix $\gM\equiv e^{i\int\tM d\tau}$. For general time dependent coefficients, this matrix equation cannot be solved analytically (because the matrices in the parenthesis should commute at any given time). However, within the regime of the WKB, a matrix exponential solution is a very good approximation (that we have checked numerically) 
\be
\Phi_0=\tM^{-1/2}\exp\lbra-\frac{1}{2}\int d\tau\gM^{-1}\tM^{-1/2}\eM^{-1}\lpar2\eM'+i\Hc'\eM\tM^{-1}+i\Hc\nM\eM\tM^{-1}+\nM\eM\rpar\tM^{1/2}\gM\rbra\bar{C}_0\,, \label{Phi0bi}
\ee 
where $\vec{C}_0$ is a vector of constant coefficients to be fixed with the initial conditions. Recall that $\tM$ is a diagonal matrix and thus the term $\tM^{-1/2}$ in (\ref{Phi0bi}) is the usual WKB scaling. If there is time dependence, there can be corrections to this scaling, which corresponds to the matrix exponential. The fact that the matrix exponential works as a solution is because the matrix in the exponent is small in this regime and, as a consequence, corrections arising from commutators of this matrix are further suppressed.  
For higher order corrections in the WKB expansion, one can proceed iteratively and solve $\Phi_1$ at order $\epsilon^2$ using the solution of~$\Phi_0$.

Having an approximate analytical solution allows us to understand the role of each parameter. In particular, we note that the speed of the massive mode has a key role in the mixing. This is better seen in the high-$k$ limit where the phases tend to 
\begin{align} \label{eq:phasehigh-k}
\theta_1^2&=k^2+m^2a^2f_1-\Hc^2+\mathcal{O}(k^{-2})\,, \\
\theta_2^2&=c^2k^2+\frac{m^2a^2f_1c}{\kappa\xi^2}-\Hc^2+\mathcal{O}(k^{-2})\,.
\end{align}
Focusing on the non-diagonal terms of the matrix of eigenvectors,
\begin{align} \label{eq:bigravitymixing}
&\eM_{12}=\frac{m^2a^2f_1}{(c^2-1)k^2+(1-c/(\kappa\xi^2))m^2a^2f_1+\mathcal{O}(k^{-2})}\,, \\
&\eM_{21}=\frac{m^2a^2f_1c/(\kappa\xi^2)}{(c^2-1)k^2+(1-c/(\kappa\xi^2))m^2a^2f_1+\mathcal{O}(k^{-2})}\,,
\end{align}
we notice that in the high-$k$ limit the mode mixing is  suppressed, i.e. $\eM_{12}\ll1$ and $\eM_{21}\ll1$,
if each tensor propagate at a different speed, $c\neq1$, and there is a large hierarchy between the mass term $m$ and the wavevector $k$.

 While in the main text we focussed on a large mass regime, we conclude this
Appendix  considering a small mass regime, with a mass parameter of the order of the Hubble constant $m\sim\Hc_0$,
and show that in this case there is no distinctive observational effect of bigravity
in the GW propagation. In this regime,
 the energy density proportional to $m^2$ in the right hand side of the Friedmann equation (\ref{eq:bigravityFriedmann}) is of the same order of magnitude of the observed vacuum energy. Then, the viable branch of solutions satisfies $\xi\ll1$, which gives the following value for the speed of the second tensor modes \cite{Comelli:2011zm}
\be
c-1=3(w+1)+\mathcal{O}\lpar\frac{m^2}{G\rho}\rpar^2\,.
\ee
This implies that during matter domination $c^2\sim16$,  and during radiation domination $c^2\sim25$. Therefore,
 given that  LISA frequency of  GWs  is  
   much larger than the rate of expansion, $k_\text{LISA}\sim10^{16}\Hc_0$, inequality  \eqref{eq:bigravityscales}
    is not satisfied and mixing among different modes is negligible in
the small mass regime.

Moreover, this mass range is also still far from being constrained through the modified dispersion relation with LISA. In particular, the bound on an effective mass term $m\,f_1^{1/2}$ in this case is \cite{Will:1997bb}
\be
m\,f_1^{1/2}\lesssim10^{-26}\,\frac{\text{eV}}{c^2}\lpar\frac{10\,\text{Gpc}}{D}\,\frac{f}{10^{-2}\,\text{Hz}}\,\frac{100}{\text{SNR}}\rpar^{1/2}\,,
\ee
where we have introduced the expected distance, frequency and signal-to-noise ratio ($\text{SNR}\sim 1/(f\Delta t)$) of a massive BH binary in the LISA band. This is seven orders of magnitude larger than $\Hc_0$, thus far from the small mass regime.

\end{appendix}

%%%%%%%references%%%%%%%%%
\addcontentsline{toc}{section}{References}
\bibliographystyle{utphys}
\bibliography{myrefsMM}
%%%%%%%%%%%%%%%%%%%%%%
\newpage
\section*{Affiliations}

{$^a$ D\'epartement de Physique Th\'eorique and Center for Astroparticle Physics,\\
Universit\'e de Gen\`eve, 24 quai Ansermet, CH-1211 Gen\`eve 4, Switzerland
\\
$^b$ Instituto de Estructura de la Materia, CSIC, Serrano 121, 28006 Madrid, Spain
\\
$^c$ Institut de physique th\'eorique, Universit\'e  Paris Saclay CEA, CNRS, 91191 Gif-sur-Yvette, France
\\
$^d$ AIM, CEA, CNRS, Univ. Paris-Saclay, Univ. Paris Diderot, Sorbonne Paris Cit\'e, F-91191 Gif-sur-Yvette, France
\\
$^e$ Laboratoire de Physique Th\'eorique, CNRS, Univ. Paris-Sud, Universit\'e  Paris-Saclay, 91405 Orsay, France
\\
$^f$ Center for Theoretical Astrophysics and Cosmology,\\
Institute for Computational Science, University of Z\"urich, CH-8057 Z\"urich, Switzerland
\\
$^g$ Instituto de F\'isica Te\'orica UAM-CSIC, Universidad Auton\'oma de Madrid,
Cantoblanco, 28049 Madrid, Spain
\\
$^h$ Institute of Cosmology and Gravitation, University of Portsmouth,\\
Dennis Sciama Building, Portsmouth, PO1 3FX, United Kingdom
\\
$^i$ Dipartimento di Fisica e Astronomia ``G. Galilei", 
	Universit\`a degli Studi di Padova,\\
	and INFN, Sezione di Padova, via Marzolo 8, I-35131 Padova, Italy
	\\
$^j$ Max-Planck-Institut f\"ur Gravitationsphysik, Albert-Einstein-Institut, Am M\"uhlenberg 1,
14476 Potsdam-Golm, Germany	
\\
$^k$ Department of Physics, Swansea University, Swansea, SA2 8PP, UK
\\
$^l$ Berkeley Center for Cosmological Physics, LBNL and University of California at Berkeley, Berkeley, California 94720, USA
\\
$^m$ Institut d'Astrophysique de Paris, CNRS and Sorbonne Universit\'es,
UMR 7095, 98 bis bd Arago, 75014 Paris, France
\\
$^n$ Scuola Internazionale di Studi Superiori Avanzati (SISSA),
via Bonomea 265, 34136 Trieste, Italy and INFN Sezione di Trieste
\\
$^o$ {INAF - Osservatorio Astronomico di Padova,  Vicolo dell’Osservatorio 5, I-35122 Padova, Italy}
\\
$^{p}$ Gran Sasso Science Institute,  Viale F. Crispi 7, I-67100 L'Aquila, Italy
\\
$^{q}$ Theoretical Particle Physics and Cosmology Group, Physics Department, King's College London, University of London, Strand, London WC2R 2LS, United Kingdom\\
$^{r}$ Institute for Gravitational Wave Astronomy and School of Physics and
Astronomy, University of Birmingham, Edgbaston,
Birmingham B15 2TT, United Kingdom
}

\end{document}